\documentclass[aps,prd,superscriptaddress,twocolumn,floatfix,nofootinbib]{revtex4}
\usepackage{amssymb}
\usepackage{amsmath}
\usepackage{dcolumn}
\usepackage{rotating}
\usepackage{color}
\usepackage{hyperref}
\usepackage{appendix}
\newcommand{\ba}{\begin{eqnarray}}
\newcommand{\ea}{\end{eqnarray}}
\newcommand{\be}{\begin{equation}}
\newcommand{\ee}{\end{equation}}
\newcommand{\bi}{\begin{itemize}}
\newcommand{\ei}{\end{itemize}}
\newcommand{\by}{\begin{array}}
\newcommand{\ey}{\end{array}}
\newcommand{\al}{\alpha}
\newcommand{\bt}{\beta}
\newcommand{\ga}{\gamma}
\newcommand{\ta}{\theta}

\newcommand{\da}{\delta}
\newcommand{\la}{\lambda}

\newcommand{\za}{\zeta}
\newcommand{\sa}{\sigma}
\newcommand{\en}{\varepsilon}

\newcommand{\un}{\upsilon}
\newcommand{\vphi}{\varphi}

\newcommand{\Ta}{\Theta}
\newcommand{\Da}{\Delta}

\newcommand{\La}{\Lambda}
\newcommand{\cC}{{\cal C}}
\newcommand{\cD}{{\cal D}}
\newcommand{\cE}{{\cal E}}
\newcommand{\cF}{{\cal U}}

\newcommand{\cI}{{\cal I}}

\newcommand{\cN}{{\cal N}}
\newcommand{\cP}{{\cal P}}
\newcommand{\cR}{{\cal C}}
\newcommand{\cQ}{{\cal Q}}
\newcommand{\cS}{{\cal S}}
\newcommand{\cO}{{\cal O}}

\newcommand{\e}{\overline}

\newcommand{\w}{\widetilde}
\newcommand{\x}{\ast}

\newcommand{\ra}{\rightarrow}
\newcommand{\Ra}{\Rightarrow}

\newcommand{\LF}{\left(}
\newcommand{\RF}{\right)}
\newcommand{\LT}{\left[}
\newcommand{\RT}{\right]}

\newcommand{\Rd}{\right.}

\newcommand{\ch}{\widehat{c}}

\newcommand{\eh}{\widehat{e}}

\newcommand{\nh}{\widehat{u}}
\newcommand{\ph}{\widehat{p}}

\newcommand{\sh}{\widehat{s}}
\newcommand{\uh}{\widehat{u}}
\newcommand{\yh}{\widehat{y}}

\newcommand{\eph}{\hat{\en}}
\newcommand{\sv}{\vec{s}}
\newcommand{\nv}{\vec{n}}
\newcommand{\wv}{\vec{w}}
\newcommand{\yv}{\vec{y}}
\newcommand{\ytv}{\vec{\w{y}}}

\newcommand{\env}{\vec{\en}}
\newcommand{\enh}{\hat{\en}}
\newcommand{\dav}{\vec{\da}}

\newcommand{\eee}{\w{e}}

\newcommand{\2}{\frac{1}{2}}

\newcommand{\mx}{\mbox}
\newcommand{\mt}{\mathtt}
\newcommand{\mand}{\mx{ and }}
\newcommand{\for}{\mx{ for }}
\newcommand{\where}{\mx{ where }}
\newcommand{\with}{\mx{ with }}
\newcommand{\sgn}{\mx{Sgn}}

\newcommand{\appr}{\mt{appr}}
\newcommand{\tot}{\mt{tot}}
\newcommand{\ie}{{\it i.e.}}

\newcommand{\vs}{\vspace{2mm}\\}
\newcommand{\non}{\nonumber\\}

\newcommand{\ycr}{y_{\mt{cr}}}
\newcommand{\ycmax}{y_{\mt{cr,max,0}}}
\newcommand{\ycmin}{y_{\mt{cr,min,0}}}
\newcommand{\ycmaxx}{y_{\mt{cr,max}}}
\newcommand{\ycminn}{y_{\mt{cr,min}}}
\newcommand{\yct}{y_{\mt{cr,max,1}}}
\newcommand{\ycmint}{y_{\mt{cr,min,1}}}
\begin{document}

\title{Geometric framework to predict structure from function in neural networks}

\author{Tirthabir Biswas}
\email{biswast@janelia.hhmi.org}
\affiliation{{Janelia Research Campus, Howard Hughes Medical Institute, Ashburn, VA 20147, USA.}}
\affiliation{{Department of Physics, Loyola University, New Orleans, LA 70118, USA.}}
\author{James E. Fitzgerald}
\email{fitzgeraldj@janelia.hhmi.org}
\affiliation{{Janelia Research Campus, Howard Hughes Medical Institute, Ashburn, VA 20147, USA.}}


\begin{abstract}

Neural computation in biological and artificial networks relies on the nonlinear summation of many inputs. The structural connectivity matrix of synaptic weights between neurons is a critical determinant of overall network function, but quantitative links between neural network structure and function are complex and subtle. For example, many networks can give rise to similar functional responses, and the same network can function differently depending on context. Whether certain patterns of synaptic connectivity are required to generate specific network-level computations is largely unknown. Here we introduce a geometric framework for identifying synaptic connections required by steady-state responses in recurrent networks of threshold-linear neurons. Assuming that the number of specified response patterns does not exceed the number of input synapses, we analytically calculate the solution space of all feedforward and recurrent connectivity matrices that can generate the specified responses from the network inputs. A generalization accounting for noise further reveals that the solution space geometry can undergo topological transitions as the allowed error increases, which could provide insight into both neuroscience and machine learning. We ultimately use this geometric characterization to derive certainty conditions guaranteeing a non-zero synapse between neurons. Our theoretical framework could thus be applied to neural activity data to make rigorous anatomical predictions that follow generally from the model architecture. 

\end{abstract}

\maketitle

\section{INTRODUCTION}
Structure-function relationships are fundamental to biology~\cite{DNA, Milo, Hunter}. In neural networks, the structure of synaptic connectivity critically shapes the functional responses of neurons~\cite{Seung09, Bargmann}, and large-scale techniques for measuring neural network structure and function provide exciting opportunities for examining this link  quantitatively~\cite{Bock, Varshney, Ahrens, Schrodel, Ohyama, Lemon, Naumann, Hildebrand, Scheffer20, Biswas}. The ellipsoid body in the central complex of {\it Drosophila} is a beautiful example where modeling showed how the structural pattern of excitatory and inhibitory connections enables a persistent representation of heading direction~\cite{Ben-Yishai, Skaggs, Kim, Turner-Evans}. Lucid structure-function links have also been found in several other neural networks \cite{Kim14, Kornfeld, Wanner, Vishwanathan}. However, it is generally hard to predict either neural network structure or function from the other~\cite{Marder, Bargmann}. For example, functionally inferred connectivity can capture neuronal response correlations without matching structural connectivity~\cite{Friston, Schneidman, Pillow, Huang}, and network simulations with structural constraints do not automatically reproduce function~\cite{Tschopp, Zarin, LitwinKumar}. Two broad modeling difficulties hinder the establishment of robust structure-function links. First, models with too much detail are difficult to adequately constrain and analyze. Second, models with too little detail may poorly match biological mechanisms, the model mismatch problem. Here we propose a rigorous theoretical framework that attempts to balance these competing factors to predict components of network structure required for function.

Neural network function probably does not depend on the exact strength of every synapse. Indeed, multiple network connectivity structures can generate the same functional responses~\cite{Prinz, Fisher}, as illustrated by structural variability across individual animals~\cite{Marder, Goaillard} and artificial neural networks~\cite{Baldi, Dauphin, Kawaguchi, Tschopp}. Such redundancy may be a general feature of emergent phenomena in physics, biology, and neuroscience~\cite{Machta, Transtrum, O'Leary}. Nevertheless, some important details may be consistent despite this variability, and here we find well-constrained structure-function links by characterizing all connectivity structures that are consistent with the desired functional responses~\cite{Marder}. We also account for ambiguities caused by measurement noise. Our goal is not to find degenerate networks that perform equivalently in all possible scenarios. We instead seek a framework that finds connectivity required for specific functional responses, independently of whatever else the network might do.   

The model mismatch problem has at least two facets. First, neurons and synapses are incredibly complex~\cite{Abbott, Spruston, Zeng, Grant}, but which complexities are needed to elucidate specific structure-function relationships is unclear~\cite{Bargmann, CurtoR, Billeh}. This issue is very hard to address in full generality, and here we seek a theoretical framework that makes clear experimental predictions that can adjudicate candidate models empirically. In particular, we predict neural network structure only when it occurs in all networks generating the functional responses. This high bar precludes the analysis of biophysically-detailed network models, which require numerical exploration of the connectivity space that is typically incomplete ~\cite{Marder, Prinz, Almog, Bittner, Goncalves}. We instead focus on recurrent firing rate networks of threshold-linear neurons, which are growing in popularity because they strike an appealing balance between biological realism, computational power, and mathematical tractability ~\cite{Ben-Yishai, Naumann, Kim, Kim14, Treves, Salinas, Hahnloser, Hahnloser03, Vishwanathan, Morrison, CurtoP, Wanner, Tschopp, Zarin, Kawaguchi}. 

\begin{figure}[!thbp]
	\centering
	\includegraphics[width=0.46\textwidth,angle=0]{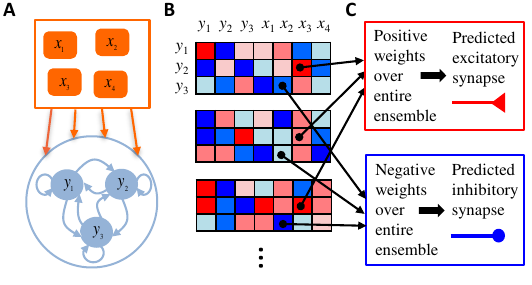}
	\caption{{\bf Cartoon of theoretical framework.} {\small ({\bf A}) We first specify some steady-state responses of a recurrent threshold-linear neural network receiving feedforward input. ({\bf B}) We then find all synaptic weight matrices that have fixed points at the specified responses. Red (blue) matrix elements are positive (negative) synaptic weights. ({\bf C}) When a weight is consistently positive (or consistently negative) across all possibilities, then the model needs a nonzero synaptic connection to generate the responses. We therefore make the experimental prediction that this synapse must exist. We also predict whether the synapse is excitatory or inhibitory.
			\label{fig:abstract}}}
\end{figure}

The second facet of the model mismatch problem is hidden variables, such as missing neurons, neuromodulator  levels, and physiological states~\cite{Bargmann, Marder2012, Aitchison, Mu}. Here we take inspiration from whole-brain imaging in small organisms ~\cite{Biswas}, such as \emph{C. elegans} ~\cite{Schrodel}, larval zebrafish ~\cite{Ahrens, Naumann, Mu}, and larval \emph{Drosophila} ~\cite{Lemon}, and assume access to all  relevant neurons. Our model neglects neuromodulators and other state variables, which would be interesting to consider in the future. Furthermore, many experiments indirectly assess neuronal spiking activity, such as by calcium florescence ~\cite{Grienberger, Wilt, Theis, Aitchison} or hemodynamic responses ~\cite{Friston, Logothetis, Bartolo, Heinzle}. We restrict our analysis to steady-state responses to mitigate mismatch between fast firing rate changes and these inherently slow measurement techniques.  

Our analysis begins with an analytical characterization of synaptic weight matrices that realize specified steady-state responses as fixed points of neural network dynamics (Figs. ~\ref{fig:abstract}A-B). A key insight is that asymmetrically constrained dimensions appear as a consequence of the threshold nonlinearity. Synaptic weight components in these semi-constrained dimensions are completely uncertain in one half of the dimension but well-constrained in the other. We then compute error surfaces by finding weight matrices with fixed points near the desired ones. This error landscape has a continuum of local and global minima, and constant-error surfaces exhibit topological transitions that add semi-constrained dimensions as the error increases. This may help explain the importance of weight initialization in machine learning, as poorly initialized models can get stuck in semi-constrained dimensions that abruptly vanish at nonzero error. By studying the geometric structure of the neural network ensemble that can approximate the functional responses, we derive analytical formulas that pinpoint a subset of connections, which we term certain synapses, that must exist for the model to work (Fig.~\ref{fig:abstract}C). These analytical results are especially useful for studying high-dimensional synaptic weight spaces that are otherwise intractable. Since the presence of a synapse is readily measurable, our theory generates accessible experimental predictions (Fig.~\ref{fig:abstract}C). Tests of these predictions assess the utility of the modeling framework itself, as the predictions hold across model parameters. Their successes and failures can thus move us forward towards identifying the mechanistic principles governing how neural networks implement brain computations.  

The rest of the paper begins in Section II with a toy problem that concretely demonstrates the approach illustrated in Fig.~\ref{fig:abstract} and relates the geometry of the solution space (all synaptic weight matrices that realize a given set of response patterns) to the concept of a certain synapse. In Section III, we explain how the solution space for a limited number of response patterns can be calculated for an arbitrarily large threshold-linear recurrent neural network. Section IV is devoted to three simple toy problems that provide additional insights into how the geometry of the solution space can help us to identify certain synapses. This is followed by Section V, where we explain and numerically test the precise algebraic relation that must be satisfied for a synapse to be certain when the response patterns are orthonormal. Section VI generalizes our analyses to include noise, including numerical tests via simulation. Finally, Section VII concludes the paper by summarizing our main results and discussing important future directions.
\section{An illustrative toy problem}
\begin{figure}[thbp]
	\centering
	\includegraphics[width=0.46\textwidth,angle=0]{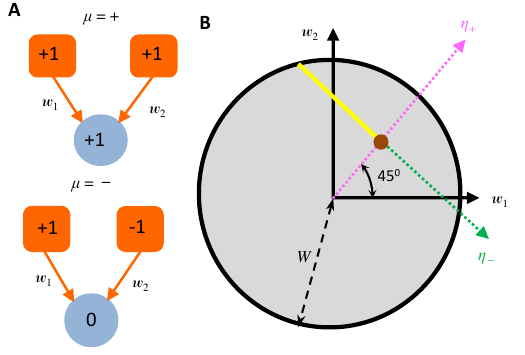}
	\caption{{\bf An illustrative two-dimensional problem.} {\small ({\bf A}) Cartoon depicting two stimulus response patterns in a simple feedforward network with two input neurons and one driven neuron. ({\bf B}) Since the driven neuron in (A) responds in one condition but not the other, we have one constrained dimension (magenta axis) and one semi-constrained dimension (green axis). The yellow ray depicts the space of weights, $(w_1,w_2)$, that generate the stimulus transformation. The weight vector $(\frac{1}{2},\frac{1}{2})$ (brown dot) would uniquely generate the neural responses in a linear network. We assume that the magnitude of the weight vector is bounded by $W$, such that all candidate weight vectors lie within a circle of that radius. A nonzero synapse $x_2\ra y$ exists in all solutions, but the $x_1\ra y$ synapse can be zero because the yellow ray intersects the $w_1=0$ axis.
	\label{fig:toymodel}}}
\end{figure}

\noindent
To gain intuition on how robust structure-function links can be established, including the effects of nonlinearity, we begin by analyzing the structural implications of functional responses in a very simple threshold-linear feedforward network (Fig.~\ref{fig:toymodel}A). We assume that two input neurons, $x_1$ and $x_2$, provide signals to a single driven neuron, $y$, via synaptic weights, $w_1$ and $w_2$. The weights are unknown, and we constrain their possible values using two neuronal response patterns, labeled $\mu = +$ and $\mu = -$. We suppose that steady-state activities of the input neurons and driven neuron are nonlinearly related according to
\be
y=\Phi(w_1x_1+w_2x_2)\ ,
\label{2dtoy}
\ee
where $x_1$, $x_2$, and $y$ denote firing rates of the corresponding neurons, and
\be
\Phi(s) = \max(0,s)
\ee
is the threshold-linear transfer function. The driven neuron responds ($y=1$) when $x_{1}=x_2=1$ in the $\mu = +$ pattern. In contrast, the driven neuron does not respond ($y=0$) when $x_{1}=-x_2=1$ in the $\mu=-$ pattern. If the transfer function were linear, then it is easy to see that there is a unique set of weights, $w_1=w_2=\frac12$, that produces these driven neuron responses, the brown dot in Fig.~\ref{fig:toymodel}B. 

How does the nonlinearity change the solution space of weights that reproduce the driven neuron responses? To answer this question, we define two  linear combinations of weights,
\be
\eta_{\pm}=w_1\pm w_2\ ,
\ee
which correspond to the driven neuron's input drive in patterns $\mu = \pm$. Eq. (\ref{2dtoy}) now yields rather simple algebraic constraints for the two patterns: 
\ba
y_+=1=\Phi(\eta_+)\Ra \eta_+=1,\\
y_-=0=\Phi(\eta_-)\Ra \eta_-\leq 0.
\ea
Note that $\eta_-$ would have had to be zero if $\Phi$ were linear, but because the threshold-linear transfer function turns everything negative into a null response, $\eta_-$ can now also be any negative number. However, sufficiently negative values of $\eta_-$ correspond to implausibly large weight vectors, and hence we focus on solutions with norm bounded above by some value, $W$. The nonlinearity thus turns the unique linear solution (brown dot in Fig.~\ref{fig:toymodel}B) into a continuum of solutions (yellow line segment in Fig.~\ref{fig:toymodel}B). This continuum lies along what we will refer to as a semi-constrained dimension. Indeed, this will turn out to be a generic feature of threshold-linear neural networks: every time there is a null response, a semi-constrained dimension emerges in the solution space\footnote{Assuming that the number of patterns does not exceed the dimensionality of the synaptic weight vector.}.

Although we found infinitely many weight vectors that solve the problem, all solutions to the problem have a synaptic connection $x_2\ra y$, and this connection is always excitatory (Fig.~\ref{fig:toymodel}B). Positive, negative, or zero connection weights are all possible for $x_1\ra y$. However, this reveals why the value of the synaptic weight bound, $W$, has important implications for the solution space. For example, all solutions in Fig.~\ref{fig:toymodel}B with $|\vec w| < 1$ have $w_1 > 0$, whereas larger magnitude weight vectors have $w_1 \le 0$. Therefore, one would be certain that an excitatory $x_1\ra y$ synapse exists if the weight bound were biologically known to be less than $W_{\mt{cr}} =1$. We refer to this weight bound as $W$-critical. Looser weight bounds raise the possibility that the synapse is absent or inhibitory. Note that too tight weight bounds, here less than $W_{\min}=1/\sqrt 2$, can exclude all solutions.

The example of Fig.~\ref{fig:toymodel} concretely illustrates the general procedure diagrammed in Fig. ~\ref{fig:abstract}. First, we specified a network architecture and steady-state response patterns (Figs.~\ref{fig:abstract}A, \ref{fig:toymodel}A). Second, we found all synaptic weight vectors that can implement the nonlinear transformation (Figs.~\ref{fig:abstract}B, \ref{fig:toymodel}B). Finally, we determined whether individual synaptic weights varied in sign across the solution space (Figs.~\ref{fig:abstract}C, \ref{fig:toymodel}B). Section~\ref{Sec: SolutionSpace} will generalize the first two parts of this procedure to characterize the solution space of any threshold-linear recurrent neural network, assuming that the number of response patterns is at most the dimensionality of the weight vectors. Sections~\ref{Sec: CertainCondition3D} and ~\ref{Sec: CertainConditionGen} will then generalize the final part of this procedure to pinpoint synaptic connections that are critical for generating any specified set of orthonormal responses.
\section{Solution Space Geometry}\label{Sec: SolutionSpace}
\begin{figure}[thbp]
\centering
	\includegraphics[width=0.46\textwidth,angle=0]{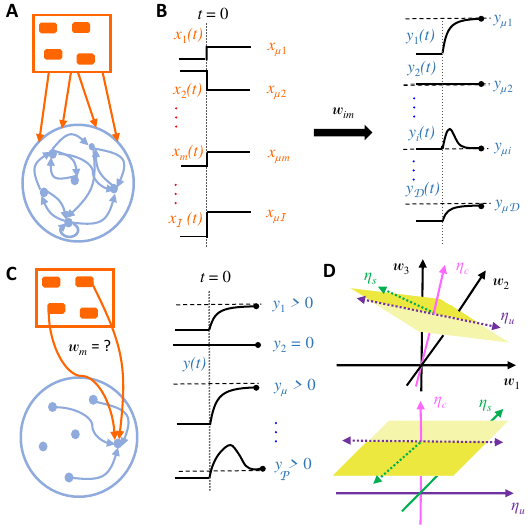}
	\caption{{\bf Finding network structure that implements functional responses.} {\small ({\bf A}) Cartoon depicting a recurrent network of \emph{driven} neurons (blue) receiving feedforward input from a population of \emph{input} neurons (orange). ({\bf B}) The $\mu^\mathrm{th}$ pattern of input neuron activity ($x_{\mu m}$) appears at $t=0$ and drives the recurrent neurons to approach the steady-state response pattern ($y_{\mu i}$) via feedforward and recurrent network connectivity ($w_{im}$). 
	({\bf C}) (\emph{Left}) We focus on one driven neuron at a time, referred to henceforth as the {\it target} neuron, to determine its possible incoming synaptic weights, $w_m$. (\emph{Right}) These weights must reproduce the target neuron's $\cP$ steady-state responses from the steady-state activity patterns of all $\cN$ presynaptic neurons. ({\bf D}) The yellow planes depict the subspace of incoming weights that can exactly reproduce all non-zero responses of the target neuron, and the subregion shaded dark yellow  indicates weights that also reproduce the target neuron's zero responses. The top graph depicts the weight space parametrized by physically meaningful $w$-coordinates, but the solution space is more simply parametrized by abstract $\eta$-coordinates (bottom). The $\eta$-coordinates depend on the specified stimulus transformation ($x_{\mu m}\ra y_{\mu i}$), and $\eta_c$, $\eta_s$, and $\eta_u$ are coordinates in $\cR$-dimensional constrained, $\cS$-dimensional semi-constrained, and $\cF$-dimensional unconstrained subspaces, respectively.
			\label{fig:cartoons}}}
\end{figure}

\noindent
{\bf Neural network structure and dynamics:} Consider a neural network of \(\cI\) input neurons that send signals to a recurrently connected population of \(\cD\) driven neurons (Fig.~\ref{fig:cartoons}A). We compactly represent the network connectivity with a matrix of synaptic weights, \(w_{im}\), where \(i = 1,\cdots,\cD\) indexes the driven neurons, and \(m = 1,\cdots,\cD+\cI\) indexes presynaptic neurons from both the driven and input populations. We suppose that activity in the population of driven neurons dynamically evolves according to 
\begin{align} 
\tau_i \frac{dy_i}{dt} =  -y_i + \Phi\left( \sum_{m=1}^{\cD} w_{im}y_m + \sum_{m=\cD+1}^{\cD+\cI} w_{im} x_{m-\cD}\right), \label{eqn: RateEqn} 
\end{align}
where \(y_i\) is the firing rate of the \(i^\mathrm{th}\) driven neuron, \(x_m\) is the firing rate of the \(m^\mathrm{th}\) input neuron, and \(\tau_i\) is the time constant that determines how long the \(i^{th}\) driven neuron integrates its presynaptic signals.
It is possible that prior biological knowledge dictates that certain synapses appearing in Eq. (\ref{eqn: RateEqn}) are absent. For notational convenience, in this paper we will assume that the number of synapses onto each driven neuron remains the same\footnote{It will become progressively evident that our construction of the solution space and certainty condition can be trivially adapted to the case where the number of presynaptic neurons changes from one driven neuron to another.}, and we will denote this number of the incoming synapses as $\cN$. Note that $\cN =\cI+\cD$ for a general recurrent network, $\cN =\cI+\cD-1$ for recurrent networks without self-synapses, and $\cN=\cI$ for feedforward networks. We suppose that the network functionally maps input patterns, \( x_{\mu m}\), to steady-state driven signals, \(y_{\mu i} \ge 0\), where \(\mu = 1,\cdots,\cP\) labels the patterns (Fig.~\ref{fig:cartoons}B). We assume throughout that \(\cP\le\cN\), as the number of known response patterns is typically small, and the number of possible synaptic inputs is large. Experimentally, different response patterns often correspond to different stimulus conditions, so we will often refer to $\mu$ as a stimulus index and $x_{\mu m}\rightarrow y_{\mu i}$ as a stimulus transformation. 
\vs
{\bf Decomposing a recurrent network into $\cD$ feedforward networks:}
Our goal is to find features of the synaptic weight matrix that are required for the stimulus transformation discussed above. For notational simplicity, let us consider the case where we potentially have all-to-all connectivity, so that $\cN=\cD+\cI$, but we will later explain how our arguments generalize. Since all time-derivatives are zero at steady-state, the response properties provide \(\cD\times \cP\) nonlinear equations for \(\cD\times \cN\) unknown parameters\footnote{A slightly different rate equation, 
$$\tau_i\frac{dv_i}{dt}=-v_i+\sum_{m=1}^\cD w_{im}r_m+\sum_{m=\cD+1}^{\cD+\cI} w_{im}x_{m-\cD},$$ 
with $r_i=\Phi(v_i)$, is also in vogue. While the dynamics of this model are slightly different from Eq.(\ref{eqn: RateEqn}), at steady state they reduce to the same form as Eq.(\ref{steady-state}). In particular, $r_i=\Phi(\sum_{m=1}^\cD w_{im}r_m+\sum_{m=\cD+1}^{\cD+\cI} w_{im}x_{m-\cD})$.}:
\be
y_{\mu i} = \Phi\left( \sum_{m=1}^{\cD} w_{im}y_{\mu m} + \sum_{m=\cD+1}^{\cD+\cI} w_{im} x_{\mu, m-\cD}\right)\ .
\label{steady-state}
\ee
Inspection of the above equation, however, reveals that each neuron's steady-state activity depends only on a single row of the connectivity matrix (Fig.~\ref{fig:cartoons}C); the responses of the $i^{th}$ driven neuron, $\{y_{\mu i}, \mu=1,\dots, \cP\}$, are only affected  by its incoming synaptic weights, $\{w_{im}, m=1,\dots, \cN\}$. Thus, the above equations separate into \(\cD\) independent sets of equations, one for each driven neuron. In other words, we now have to solve $\cD$ feedforward problems, each of which will characterize the incoming synaptic weights of a particular driven neuron, which we term the target neuron. Note that since a generic target neuron receives signals from both the input and the driven populations, the activities of both input and driven neurons serve to produce the presynaptic input patterns that drive the responses of the target neuron in the reduced feedforward problem.  \vs
{\bf Solution space for feedforward networks:} 
We have just seen how we can solve the problem of finding synaptic weights consistent with steady-state responses of a recurrent population of neurons, provided we know how to solve the equivalent problem for feedforward networks. Accordingly, we will now focus on a feedforward network, where a single target neuron, $y$, receives inputs from $\cN$ neurons $\{x_m; m=1,\dots, \cN\}$, to find the ensemble of synaptic weights that reproduce this target neuron's observed responses. The constraint equations are
\begin{align} 
y_{\mu} = \Phi\left( \sum_{m=1}^{\cN} x_{\mu m}w_{m}\right), \label{eqn: SingleSteadyStateEqn} 
\end{align}
where \(y_\mu\) now stands for the activity of the target neuron driven by the $\mu^{th}$ input pattern, and \(\vec w\) is the $\cN$-vector of synaptic weights onto the target neuron. Assuming that the $\cP\times\cN$ matrix \(x\) is rank $\cP$, we let the $\cN\times\cN$ matrix \(X\) be rank $\cN$ with $X_{\mu m}=x_{\mu m}$ for $\mu=1,\dots, \cP$. This implies that the last $\cN-\cP$ rows of $X$ span the null space of $x$, and \(X\) defines a basis transformation on the weight space, 
\begin{align}
\eta_\mu = \sum_{m=1}^\cN X_{\mu m} w_m \Leftrightarrow  w_m =\sum_{\mu=1}^\cN X^{-1}_{m\mu} \eta_\mu \label{eq: omega-coord}.
\end{align} 
The $\cN$ linearly-independent columns of $X^{-1}$ define the basis vectors corresponding to the $\eta$-coordinates, 
\begin{align} X^{-1} = \left(\begin{matrix}\env_1 & \cdots & \env_\mu & \cdots & \env_\cN\end{matrix}\right)\ .
\label{eta-vec}
\end{align}
In other words,
\be
\env_\mu=\sum_{m=1}^{\cN}\eh_m X^{-1}_{m\mu}\ ,
\ee
where $\{\eh_m\}$ is the {\it physical} orthonormal basis whose coordinates, $\{w_m\}$,  correspond to the material substrates of network connectivity. These basis vectors can be obtained from $\{\env_{\mu}\}$ by an inverse basis transformation:
\be
\eh_m=\sum_{\mu=1}^{\cN}\env_\mu X_{\mu m}\ .
\label{ehtoenv}
\ee
We can thus write any vector of incoming weights as 
\begin{align} \vec w = \sum_{m=1}^\cN w_m \eh_m = \sum_{\mu=1}^\cN \eta_\mu \env_\mu. \end{align}

In terms of $\eta$-coordinates, the nonlinear constraint equations take a rather simple form: 
\begin{align} y_\mu =\Phi\left(\eta_\mu\right) \for  \mu=1,\cdots,\cP\ .\label{eqn: SemiFlat} \end{align}
Accordingly, $\eta$-coordinates succinctly parametrize the solution space of all weight matrices that support the specified fixed points (Fig.~\ref{fig:cartoons}D). Each $\eta$-dimension can be neatly categorized into one of three types. First, for each stimulus condition \(\mu\) where  \(y_{\mu}>0\), we must have \(\eta_{\mu}>0\). This in turn implies that $\Phi(\eta_\mu)=\eta_\mu  = y_{\mu}$. Because the coordinate \(\eta_\mu\) must adopt a specific value to generate the transformation, we say that \(\mu\) defines a {\it constrained} dimension. We denote the number of constrained dimensions as \(\cR \le \cP\). Second, note that the threshold in the transfer function implies that \(\Phi(a)=0\) for all \(a\le0\). Therefore, for any stimulus condition such that \(y_{\mu}=0\), we have a solution whenever \(\eta_{\mu}  \le 0\). Because positive values of \(\eta_\mu\) are excluded but all negative values are equally consistent with the transformation, we say that \(\mu\) defines a {\it semi-constrained} dimension. We denote the number of semi-constrained dimensions as \(\cS = \cP - \cR\). Finally, we have no constraint equations for \(\eta_\mu\) if $\mu=\cP+1,\cdots, \cN$. Because all positive or negative values of \(\eta_\mu\) are equally consistent with the stimulus transformation, we say that \(\mu\) defines an {\it unconstrained} dimension. We denote the number of unconstrained dimensions as \(\cF = \cN - \cP\). Altogether, the stimulus transformation is consistent with every incoming weight vector that satisfies
\be
\begin{array}{rl}
\eta_\mu =y_\mu &\mathrm{if}\ y_\mu > 0, \mu \le \cP\\
-\infty<\eta_\mu\leq 0 & \mathrm{if}\ y_\mu = 0, \mu \le \cP\\
-\infty<\eta_\mu<\infty & \mathrm{if}\ \mu > \cP \ .
\label{nature}
\end{array}
\ee
Note that one can enumerate the solutions in the physically meaningful $w$-coordinates by simply applying the inverse basis transformation in Eq. (\ref{eq: omega-coord}) to any solution found in \(\eta\)-coordinates. 

Going forward, it will be convenient to extend the $\cP$-dimensional vector of target neuron activity to an $\cN$-dimensional vector whose components along the unconstrained dimensions are equal to zero, because this will allow us to compactly write equations in terms of dot products between the activity vector and vectors in the $\cN$-dimensional weight space. Rather than introducing a new notation for this extended $\cN$-dimensional vector, we simply write $\yv$ with $y_\mu = 0$ for $\mu = \cP+1,\cdots,\cN$. It is critical to remember that this is merely a notational convenience, and the solution space distinguishes between semi-constrained dimensions and unconstrained dimensions according to Eq. (\ref{nature}). In particular, $y_\mu = 0$ is a constraint equation for semi-constrained dimensions, but $y_\mu = 0$ is a notational convenience for unconstrained dimensions. \vs
{\bf Back to the recurrent network:} 
To understand how the solution space geometry of the feedforward network can be translated back to the recurrent network, it is useful to group together the steady-state activities of all input and driven neurons that are presynaptic to the $i^{th}$ driven neuron as a $\cP\times \cN$ input pattern matrix, $z^{(i)}$~\footnote{In fact, one can easily incorporate the case when the number of presynaptic partners differs from one driven neuron to another. This just means that the $z^{(i)}$ matrices will have dimensions $\cP\times \cN_i$, where $\cN_i$ represents the number of presynaptic partners of the $i^{th}$ neuron.}. The entries of the matrix, $z^{(i)}_{\mu m}$, correspond to the responses of the $m^{th}$ presynaptic neuron to the $\mu^{th}$ stimulus. At this point it is easy to see that when biological constraints dictate that some of the synapses are absent, then one should just exclude those presynaptic neurons when constructing $z^{(i)}$, such that the $m$ index excludes those presynaptic neurons. Similarly, by a suitable reordering, which will depend on the driven neuron, we can always ensure that $m=1,\dots, \cN$ runs only over the neurons that are  presynaptic to the given driven neuron. 

Once the input patterns feeding into the $i^{th}$ neuron are known, we can follow the steps outlined in the previous subsection to define the $\cN\times\cN$ full rank extension of $z^{(i)}$, $Z^{(i)}$, and the $\eta^{(i)}$ coordinates via
\be
\eta^{(i)}_{\mu}=\sum_{m=1}^{\cN}Z^{(i)}_{\mu m}w_{im}\ .
\ee
The nature of the $\eta^{(i)}_{\mu}$ coordinates, that is whether they are constrained, semi-constrained, or unconstrained, is determined by how the $i^{th}$ neuron responded to the stimulus conditions, as in Eq.~(\ref{nature}). Repeating this process for all driven neurons provides a geometric characterization of the entire recurrent network solution space, which involves all elements of the synaptic weight matrix, $w_{im}$.

An important special case is all-to-all network connectivity. In this case, the $Z^{(i)}$ matrices are the same for all driven neurons, and therefore the directions corresponding to the $\eta$-coordinates are also preserved\footnote{Nevertheless, the vector spaces of synaptic weights are fundamentally distinct for different driven neurons, as these vector spaces pertain to the incoming synapses onto different driven neurons. The fact that the $Z^{(i)}$ matrices are the same for all $i$ means that the relative orientation of the $\eta$-directions, with respect to the physical $w$-coordinate axes (labeled by the presynaptic indices), remains the same for all the driven neurons.}. In particular, the orientation of the unconstrained subspace with respect to the physical basis doesn't change from one driven neuron to another. However, how a given driven neuron responds to a particular stimulus determines whether the corresponding $\eta$-direction is going to be constrained or  semi-constrained for the feedforward network associated with that driven neuron.
\begin{figure*}[!htbp]
\centering
\includegraphics[width=0.96\textwidth,angle=0]{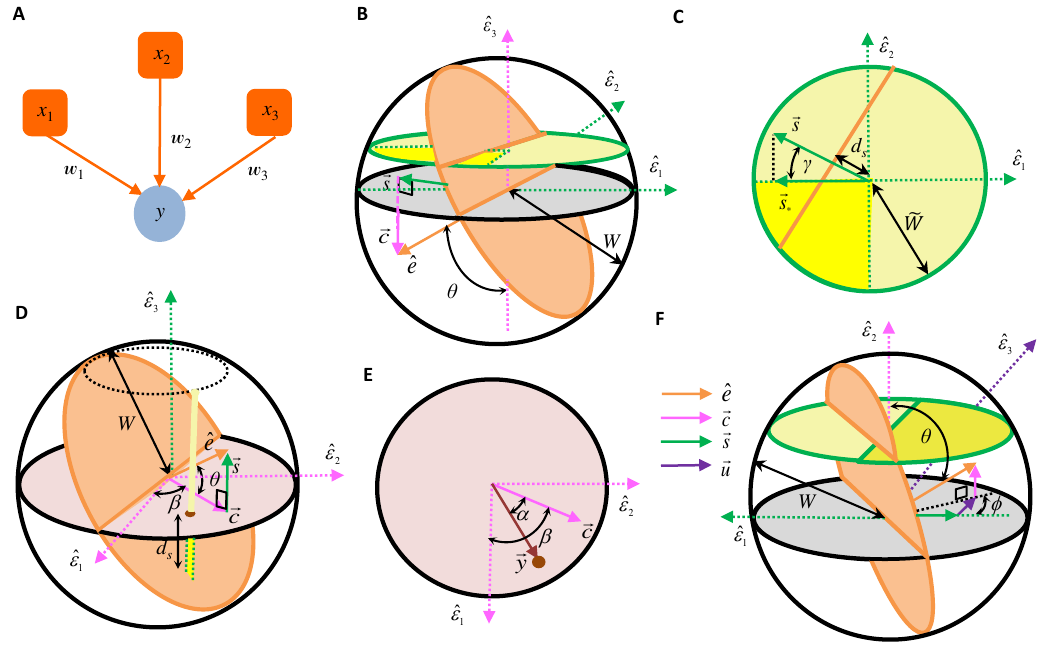}
	\caption{{\bf Geometric quantities determining whether neurons must be synaptically connected in several three-dimensional toy problems. }{\small  ({\bf A}) Cartoon depicting the $\cN=3$ feedforward network corresponding to the toy problems. ({\bf B-C}) Geometrically determining whether a synapse is nonzero when the target neuron responds to one input pattern but doesn't to two other patterns.  
	A synapse can only vanish if the $w_1=0$ plane (orange circle) intersects the solution space (dark yellow wedge) within the weight bounds (bounding sphere). For example, this intersection occurs in panel (B), so the synapse is not required for the responses. For every synapse one can associate   a direction in synaptic weight space (orange arrow) that is normal to the planes with constant synaptic weight. This synapse vector can be decomposed into its projections into the semi-constrained subspace (green arrow, $\sv$) and along the constrained dimension (pink arrow, $\vec c$). In this example, whether the synapse is certain is determined by the size of the bounding synapse space, $W$ (see (B)), the angle $\ta$ between the synapse direction (orange arrow) and the closest axis of the constrained dimension (-$\env_3$) (see (B)), and the angle $\ga$ between $\sv$ and its closest vector in the solution space ($\sv_{\x}$) (see (C)). In (C), $d_s$ depicts the perpendicular distance from the origin of the yellow semi-constrained plane in (B) to its intersection line with the $w_1=0$ orange plane. If this distance is sufficiently large, then the orange line will not intersect the solution space within the yellow plane's circular bound of radius $\w{W}$. ({\bf D-E}) Geometrically determining whether a synapse is nonzero when the target neuron responds to two input patterns but not the third pattern. In panel (D), the orange $w_1=0$ plane intersects the solution space (deep yellow line) within the bounding sphere, so the synapse is not certain. In this example, the factors that determine synapse certainty are $W$ (see (D)), the angle $\ta$ that the synapse vector (orange arrow) makes with its projection along the constrained subspace (pink arrow) (see (D)), and the angle $\al$ between the target response vector (brown arrow) and the pink arrow (see (E)). The angle $\beta$ does not ultimately matter, but it is included in the diagrams to aid the derivation. Here $d_s$ is the distance from the brown dot to the point of intersection between the yellow line and the orange plane. Again this point will lie outside the bounding sphere if $d_s$ is large enough, and this signals a certain synapse. ({\bf F}) Geometrically determining whether a synapse is nonzero when the target neuron responds to one input pattern but doesn't to a second pattern. In the figure shown, the $w_1=0$ orange plane intersects the solution space (deep yellow semi-circle) within the bounding sphere, so the synapse is not certain. In this example, apart from $W$, what determines synapse certainty are the angles $\ta$ and $\phi$, which encode how the synapse vector (orange arrow) can be decomposed into its projections along the constrained direction (pink arrow), semi-constrained direction (green arrow) and unconstrained direction (purple arrow). 
\label{fig:certaintygeometry}}}
\end{figure*}
\section{Certain synapses in illustrative 3D examples}\label{Sec: CertainCondition3D}
\noindent
Although we've found infinitely many weight matrices that produce a given stimulus transformation, it's nevertheless possible that the solutions imply firm anatomical constraints (\emph{e.g.}, {\bf Section II}). In this paper we focus on finding synapses that must be non-zero in order for the response patterns to be fixed points of the neural network dynamics. We refer to such synapses as certain, because the synapse must exist in the model, and its sign is identifiable from the response patterns. It is clear from the geometry of the solution space that the relative orientations between the $\eta$-coordinates and the physical $w$-coordinates are significant determinants of synapse certainty. To build quantitative intuition for how the solution space geometry precisely determines synapse certainty, we begin by first analyzing a few illustrative toy problems. In the next section we will describe the more general treatment of high-dimensional networks. Importantly, we select and parameterize each toy problem to introduce concepts and notations that will reappear in the general solution.  

More specifically, we first consider three feedforward examples with $\cN=3$ (Fig.~\ref{fig:certaintygeometry}A). The first two  examples have $\cP=3$, and the third has $\cP=2$. In the first example, we will assume that the driven neuron doesn't respond  to the first two stimulus patterns, but responds positively to the third pattern. So we have two semi-constrained  and one constrained dimension,
\be
\eta_1\leq 0\ ,\ \eta_2\leq 0\ ,\mand 
\eta_3=y_3>0\ .
\ee
In contrast, in the second example we will have two constrained and one semi-constrained dimension,
\be
\eta_1=y_1>0\ ,\ \eta_2=y_2>0\ ,\mand 
\eta_3\leq 0\ .
\ee
The final example will feature one unconstrained, one semi-constrained, and one constrained dimension,
\be
\eta_1\leq 0,\  \eta_2=y_2>0\ ,\mand 
-\infty < \eta_3 < \infty.\
\ee
For technical simplicity we will consider orthonormal input patterns, $X^{-1}=X^T$, which implies that
\be
\sum_{m=1}^{\cN}X_{\mu m}X_{\nu m}=\da_{\mu\nu}=\env_\mu\cdot\env_\nu\ ,
\ee
where $\da_{\mu\nu}$ is the Kronecker delta function, which equals $1$ if $\mu=\nu$ and $0$ if $\mu\ne\nu$, so $\enh_\mu = \env_\mu$. This trivially implies that the $\eta$-coordinates are related to the synaptic coordinates via a rotation, so the spherical biological bound on the physical coordinates transforms to an identical spherical bound on the $\eta$-coordinates:
\be
\sum_{\mu=1}^{\cN=3}\eta_\mu^2 = \sum_{m=1}^{\cN=3}w_m^2 \le W^2 \ .
\ee
\vs
{\bf Problem 1:} Let us first focus on the example with two semi-constrained and one constrained dimension, whose solution space is depicted in deep yellow in Fig.~\ref{fig:certaintygeometry}B. Suppose we are interested in assessing whether the $w_1$ synapse is certain. Since the $w_1=0$ plane divides the weight space into the positive and the negative halves, the synapse will be certain if this plane doesn't intersect with the solution space, which clearly depends on the orientation of the plane relative to the various $\eta$-directions (Fig.~\ref{fig:certaintygeometry}B). It is thus useful to consider how the $w_1=0$ plane's unit normal vector pointing towards positive weights, $\eh\equiv\eh_1$, is oriented relative to the $\eta$-directions. For ease of graphical illustration, here we assume the specific orientation diagrammed in Figs.~\ref{fig:certaintygeometry}B-C. Using Eq.~(\ref{ehtoenv}) and the orthogonality of $X$, we  can  parametrize $\eh$ as
\ba
\eh&=&\sum_{\mu=1}^{\cN=3}X_{\mu 1}\eph_\mu=\cos\ta\ch+\sin\ta\sh,
\label{ehprojections}
\ea
where
\ba
\ch&=&-\eph_3\ ,\mand \sh=-\cos\ga\eph_1+\sin\ga\eph_2
\ea
(Figs.~\ref{fig:certaintygeometry}B-C). Geometrically, $\ch$ and $\sh$ are unit vectors along the projections of $\eh$ onto the constrained and semi-constrained subspaces (Fig.~\ref{fig:certaintygeometry}B). Thus, $\cos\ta\ge0$ and $\sin\ta\ge0$, making $\ta$ an acute angle. In this example, $\ga$ is also an acute angle, as depicted in Fig.~\ref{fig:certaintygeometry}C.

Note that all solutions lie within the 2-dimensional semi-constrained subspace having $\eta_3 = y_3$. The $w_1=0$ plane intersects this semi-constrained subspace as a line (Figs.~\ref{fig:certaintygeometry}B, C), and its equation in $\eta$-coordinates is
\be
w_1=\eh\cdot \vec{w}=\sin\ta(-\cos\ga\eta_1+\sin\ga\eta_2)-\cos\ta y_3=0\ .
\label{line}
\ee
From the geometry of the problem (Fig.~\ref{fig:certaintygeometry}C), it is clear that if the perpendicular distance, $d_s$, from the origin to this line is large enough, then it will not intersect the all-negative quadrant of the semi-constrained subspace within the weight bound. According to simple trigonometry, this occurs when
\be
d_s>\w{W}\cos\ga=\sqrt{W^2-y_3^2}\cos\ga \ ,
\label{distance}
\ee
where $\w{W}=\sqrt{W^2-y_3^2}$ is the radius of the semi-constrained subspace containing the solutions. The perpendicular distance can be identified from Eq.~(\ref{line}) as
\be
d_s= y_3\cot\ta\ .
\ee
Substituting this expression for $d_s$ into Eq.~(\ref{distance}), one finds through simple algebra that the $w_1=0$ hyperplane doesn't intersect the solution space, and hence the synapse is certain, if the response magnitude exceeds a critical value,
\be{ 
y_3>\ycr= W\sqrt{\sin^2\ta\cos^2\ga\over \cos^2\ta+\sin^2\ta\cos^2\ga}\ ,
\label{2S1C}}
\ee
which we generally refer to as $y$-critical. 

Notice that if $\ta$ increases in Fig.~\ref{fig:certaintygeometry}B, then the orange line in Fig.~\ref{fig:certaintygeometry}C comes closer to the origin, making it intersect with the solution space for more $\ga$ angles. Therefore the synapse is more difficult to identify, and indeed Eq. (\ref{2S1C}) shows that $\ycr$ increases. On the other hand, if $\ga$ increases, the orange line in Fig.~\ref{fig:certaintygeometry}C rotates away from the solution space, making the synapse easier to identify with small $d_s$. Accordingly, $\ycr$ decreases. 

It will turn out that the concept of $y$-critical is general, and $\ycr$ can always be expressed in terms of projections of $\hat e$ along several specific directions. In this example, if we define $e_{s\x}$ and $e_y$ to be projections of $\eh$ along $\sh_{\x}=-\env_1$ and $\yh=\env_3$, respectively, then it is easy to check that one can re-express $\ycr$ as
\be
\ycr = W\sqrt{\frac{e_{s\x}^2}{e_y^2+e_{s\x}^2} }\ . 
\label{SCprojections}
\ee
We will later discover that these projections are closely related to correlations between pre-synaptic and post-synaptic neuronal activity patterns. Thus, the expressions in  Eq.~(\ref{SCprojections}) will provide a deeper understanding of the determinants of synapse certainty. \vs
{\bf Problem 2:} Having identified two key angles, $\theta$ and $\gamma$, that play a role in synapse certainty, let us look at the example of two constrained and one semi-constrained dimensions to uncover other important geometric quantities. In this case, the solution space is a ray defined by $\eta_1=y_1, \eta_2=y_2$, and $-\infty<\eta_3\leq 0$, and the magnitude of $\eta_3$ is at most
\be
\w{W}=\sqrt{W^2-y_1^2-y_2^2}\
\ee
for solutions within the weight bound (Fig.~\ref{fig:certaintygeometry}D). Fig.~\ref{fig:certaintygeometry}D shows a geometry where the $w_1=0$ plane intersects the solution space at the point 
\be
\wv_{\mt{int}}=y_1\env_1+y_2\env_2+\eta_3\env_3\ .
\ee 
Now we must have 
\be
\eh\cdot\wv_{\mt{int}}=0\ ,
\label{w1toy3}
\ee
as the intersection point lies on the $w_1=0$ plane by definition, where we have defined $\eh\equiv\eh_1$ as in the previous toy problem. The projection directions of $\eh$ onto the constrained and semi-constrained subspaces are given by 
\be
\ch=\cos\bt\eph_1+\sin\bt\eph_2\ ,\mand \sh=\eph_3\ ,
\label{toy2}
\ee 
(Fig.~\ref{fig:certaintygeometry}D). Then combining Eqs.~(\ref{ehprojections}) and (\ref{toy2}), we can find an equation to determine $\eta_3$ at the intersection point
\be
\eh\cdot\wv_{\mt{int}}=\cos\ta(y_{1}\cos\bt+y_2\sin\bt)+\eta_3\sin\ta=0\ .
\label{toy2eqn}
\ee
We next introduce $\al$ to represent the angle between $\ch$ and $\vec y$ (Fig.~\ref{fig:certaintygeometry}E), such that 
\be
y_1 = y \cos(\beta-\alpha)\mand y_2 = y \sin(\beta-\alpha),
\ee
where $y=|\yv|$. The first two terms in Eq.~(\ref{toy2eqn}) can then be trigonometrically combined with a difference of angles identity to arrive at
\be
y\cos\ta \cos\al+\eta_3\sin\ta=0 \Longrightarrow \eta_3 = -y\cot\theta\cos\alpha.
\ee 
To be able to identify the sign of $w_1$, this intersection point must lie beyond the weight bounds of the solution line segment, so $\eta_3< -\w{W}$.  After some straightforward algebra we obtain the certainty condition as
\be
y>\ycr\equiv W\sqrt{\sin^2\ta\over \cos^2\ta\cos^2\al+\sin^2\ta}.
\label{2C1S}
\ee

From the geometry of the problem in  Figs.~\ref{fig:certaintygeometry}D-E, one sees that as $\ta$ or $\al$ increases, the point where the orange hyperplane intersects the yellow line is closer to the origin. Indeed $\ycr$ increases, making it  more difficult to identify the synapse sign. Again, one can re-express $\ycr$ as Eq.~(\ref{SCprojections}) in terms of projections, with the role of $\sh_{\x}$ being played by $\env_3$.
\vs
{\bf Problem 3:} Through the two above examples we found three angles, $\ta, \al$, and $\ga$, that determine how large the response of the driven neuron has to be in order for a given synapse to be certain. However in both examples the number of patterns were equal to the number of synapses, $\cP=\cN$. When $\cP<\cN$, we have unconstrained dimensions, and the projection of the $\eh\equiv\eh_1$ vector into the unconstrained subspace will also matter, because it relates to how much we do not know about the response properties of the driven neuron. 

Here we consider a $\cN=3$ example with one constrained, one semi-constrained, and one unconstrained dimension (Fig.~\ref{fig:certaintygeometry}F). In this case, we can express the $\eh$ synaptic direction as a linear combination of its projections along the constrained, semi-constrained and unconstrained dimension as
\begin{align}
\eh= \sum_{\mu=1}^\cN X_{\mu 1} \vec \varepsilon_\mu = \cos\ta \ch + \sin\ta\cos\phi\sh + \sin\ta\sin\phi\uh, \label{Eq: SubspaceDecompEm} 
\end{align}
where we can always choose the directions of the unit vectors to make $\ta$ and $\phi$ acute angles. For the example shown in Fig.~\ref{fig:certaintygeometry}F, this is achieved by choosing
\be
\ch=\env_2\ , \sh=-\env_1\ ,\mx{ and }  \uh=\env_3\ .
\ee
Obtaining the certainty condition again involves ascertaining whether the $w_1=0$ hyperplane intersects the deep yellow solution space (Fig.~\ref{fig:certaintygeometry}F). In the example of Fig.~\ref{fig:certaintygeometry}F, one can see that increasing the driven neuron response moves the yellow plane up, and there will come a critical point when the orange $w_1=0$ plane just touches the solution space at the corner ($\eta_1=0$,   $\eta_2=\ycr$, $\eta_3$). Thus, 
\be
\wv_{\mt{int}}=\ycr\env_2+\eta_3\env_3\ .
\ee 
Since this corner point has a negative $\eta_3$ component and lies on the bounding sphere, we must also have
\be
\eta_3=-\sqrt{W^2-\ycr^2}\ ,
\ee
(Fig.~\ref{fig:certaintygeometry}F). Substituting $\wv_{\mt{int}}$ in the $w_1=0$ plane equation, 
\ba
\eh\cdot \vec{w}_{\mt{int}}=-\sin\ta\sin\phi\sqrt{W^2-\ycr^2}+\cos\ta\ \ycr=0,
\ea
we can then determine $\ycr$ through simple algebra as
\ba
\ycr=W\sqrt{\sin^2\ta\sin^2\phi\over \cos^2\ta+\sin^2\ta\sin^2\phi}\ .
\label{1S1C1U}
\ea

The final result now depends on the two acute orientation angles, $\ta$ and $\phi$. By inspection of Fig.~\ref{fig:certaintygeometry}F or Eq. (\ref{1S1C1U}), it is clear that $\ycr$ increases if either $\ta$ or $\phi$ increases towards $\pi/2$. One therefore needs a larger response ($y_3$) to make the synapse certain. We can again express $\ycr$ in terms of projections
\ba
\ycr=W\sqrt{e_u^2\over e_y^2+e_u^2}\ ,\label{UCprojections}
\ea
where $e_u$ is the projection of $\eh$ along $\uh$, and $e_{s\x}$ does not appear because the intersection occurred at the origin of the semi-constrained subspace. 
\section{Certain synapses, the general treatment}\label{Sec: CertainConditionGen}
\begin{figure}[thbp]
	\centering
	\includegraphics[width=0.46\textwidth,angle=0]{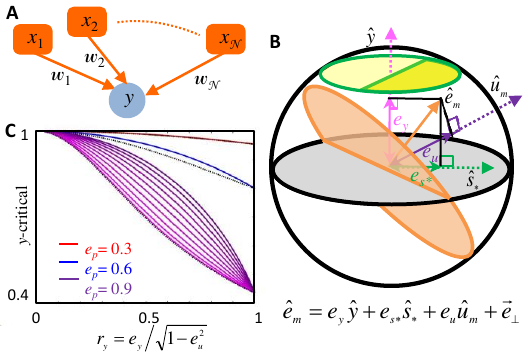}
	\caption{{\bf Identifying certain synapses in high-dimensional networks.} {\small  ({\bf A}) Cartoon depicting the high-dimensional feedforward network under consideration. ({\bf B}) Geometrically determining whether a synapse is nonzero throughout a high-dimensional solution space. A synapse can only vanish if the $w_m=0$ hyperplane (orange circle) intersects the solution space (dark yellow wedge) within the weight bounds (bounding sphere). In the example shown,  this intersection doesn't occur, so the synapse must be present. For orthonormal neural responses, only a few parameters determine whether this intersection occurs (Appendix A). First, the magnitude of the weight bound, $W$, controls the extent of the solution space. Second, there are three projections of the synapse direction (orange arrow) whose lengths are important determinants of the certainty condition: $e_y$, the length of projection along the target response vector (pink arrow); $e_{s\x}$, the length of projection along the closest boundary vector in the semi-constrained solution subspace  (green arrow, see also $\vec s_{\x}$ in Fig.~\ref{fig:certaintygeometry}C); and $e_u$, the length of projection into the unconstrained subspace (purple arrow). Note that the shown example would have had an intersection if the solution space (dark yellow wedge) were moved down (along $\ch$) to lie below the hyperplane (orange circle). The solution space's height is proportional to the magnitude of the postsynaptic responses, $y$. Thus, the solution space does not intersect the hyperplane only if $y$ exceeds a critical value, $\ycr$. 
	({\bf C}) Plots of the certainty condition, Eq. (\ref{ycritical-ind}), for $W=1$. The red, blue, and purple curves plot $\ycr$ as a function of $r_y=e_y/e_p$ for $e_p=$ 0.3, 0.6, and 0.9, respectively. Different purple shades correspond to different values of $r_{s\x}=e_{s\x}/\sqrt{e_p^2-e_y^2}$. As this ratio increases, nonlinear effects increase $\ycr$ and make the sign harder to determine. The red and blue curves are for the maximally nonlinear case when $r_{s\x}=1\Ra e_{s\x}=\sqrt{e_p^2-e_y^2}$. The dashed black curves represent $\ycr$ in a linear model, which cannot exceed the nonlinear $\ycr$. 
			\label{fig:projections}}}
\end{figure}

\noindent
{\bf High-dimensional feedforward networks:} We have seen in the previous section how geometric considerations can identify synapses that must be present to generate observed response patterns in small networks. One can similarly ask when a synapse is required in high-dimensional networks (Fig.~\ref{fig:projections}A). 

Although the rigorous derivation is intricate, this certainty condition is remarkably simple for orthonormal \(X\) (Appendix A). Quantitatively, orthonormal $X$ imply that only a few parameters matter for the certainty condition, each illustrated in the previous section and abstractly summarized in Fig.~\ref{fig:projections}B. 

For any given synapse, its physical basis vector, $\eh_m$, can always be written as a sum of components in the constrained, semi-constrained, and unconstrained subspaces,
\begin{align}\eh_m = \sum_{\mu=1}^\cN X_{\mu m} \env_\mu = \vec c_m + \vec s_m + \vec u_m, \label{Eq: SubspaceDecompEmG} 
\end{align}
where $\vec c_m$, $\vec s_m$, and $\vec u_m$ denote the partial sums over $\mu$ in the constrained, semi-constrained, and unconstrained subspaces, respectively. Note that $\{\env_\mu\}$ are orthogonal unit vectors if and only if $X$ is an orthogonal matrix. In this case, the decomposition of $\eh_m$ is a sum of three orthogonal vectors that can be parameterized by two angles,
\begin{align} \eh_m = \cos\theta\ \ch_m  + \sin\theta\cos\phi\ \sh_m + \sin\theta\sin\phi\ \uh_m , \label{Eq: ThetaPhiDecomp} 
\end{align}
where 
$\ch_m$, $\sh_m$, and $\nh_m$ are unit vectors in the constrained, semi-constrained, and unconstrained subspaces, and $(\ta,\phi)$ are spherical coordinates\footnote{The angles also depend on the synapse but we have dropped the $m$ index for brevity.} specifying the orientation of $\eh_m$ with respect to these subspaces (\emph{e.g.}, Fig.~\ref{fig:certaintygeometry}F). In particular,
\begin{align}\cos\theta& = \sqrt{\sum_{\{\mu|y_\mu > 0\}}^\cP X_{\mu m}^2},\nonumber\\
\sin\theta\cos\phi& = \sqrt{\sum_{\{\mu|y_\mu = 0\}}^\cP X_{\mu m}^2},\nonumber\\
\sin\theta\sin\phi& = \sqrt{\sum_{\mu = \cP+1}^\cN X_{\mu m}^2}. \end{align}
As we have seen in the toy examples, these two orientation angles heavily influence whether the synapse is certain. 

Additionally, because the solution space's height along $\ch_m$ (\emph{e.g.} Fig.~\ref{fig:certaintygeometry}B) is controlled by the angle between $\ch_m$ and $\yh$, the equation for the $w_m=0$ hyperplane that divides the positive and negative synaptic regions in the solution space  depends on
\begin{align} \vec y \cdot \ch_m  = y\cos\alpha, \end{align}
where $y$ is the length of $\vec y$ and $\alpha$ is the angle between $\vec y$ and $\ch_m $ (Fig.~\ref{fig:certaintygeometry}E). Finally, there is another critical angle, which we call $\ga$, that encodes how $\sh_m$ is oriented with respect to the solution space in the semi-constrained subspace. Using a more convenient direction, $\sh\,'_m \equiv  -\sgn(\cos\alpha)\sh_m$, which is either along or opposite to the $\sh_m$ direction, we define $\gamma$ to be the minimal angle between $\sh\,'_m$ and the solution space (\emph{e.g.} Fig.~\ref{fig:certaintygeometry}B). It is generally given by
\begin{align} \cos\gamma = \sqrt{\sum_{\{\mu|\sh\,'_\mu < 0\}} \sh_\mu^{\,'2}} 
\label{gamma-defn}
\end{align}
(Appendix A), where $\sh\,'_\mu$ is the $\mu^{th}$ component of $\sh\,'_m$, and we have suppressed $m$ to avoid cluttered notation. Although this definition and equation for $\ga$ may initially appear opaque, we soon clarify its meaning in terms of interpretable projections of the synapse vector.

Putting all the pieces together, we find that the $m^{th}$ synapse must be present, and its sign is unambiguous, if and only if $y$ exceeds the critical value
\be
\ycr=W\sqrt{\cos^2\ga \sin^2\ta\cos^2\phi+\sin^2\ta\sin^2\phi \over \cos^2\al\cos^2\ta+ \cos^2\ga\sin^2\ta\cos^2\phi+\sin^2\ta\sin^2\phi }
\label{yc-angles}
\ee
(Appendix A).  Intuitively, \(W\) bounds the magnitude of weight vectors, and large $W$ increase $\ycr$ by admitting more solutions. Note that a synapse is certain, for a given $y$, when the weight bound is less than a critical value, 
\be
W_{\mt{cr}} = y \sqrt{\cos^2\al\cos^2\ta+ \cos^2\ga\sin^2\ta\cos^2\phi+\sin^2\ta\sin^2\phi \over \cos^2\ga \sin^2\ta\cos^2\phi+\sin^2\ta\sin^2\phi }.
\ee
Finally, we note that we must have $W \ge y$ for any solutions to exist. One can straightforwardly obtain the special cases Eqs.~(\ref{2S1C}), (\ref{2C1S}) and (\ref{1S1C1U}), by substituting $\al=\phi=0$, $\ga=\phi=0$, and $\al=\cos\gamma=0$ in the general expression given by Eq.~(\ref{yc-angles}).

The geometric description of Eq.~(\ref{yc-angles}) can be written more intuitively as
\begin{align} \ycr = W\sqrt{\frac{e_{s\x}^2+e_u^2}{e_y^2+e_{s\x}^2+e_u^2} } = W\sqrt{\frac{1}{1+e_y^2/(e_{s\x}^2+e_u^2)} } \label{ycritical} 
\end{align}
(Appendix A), where $\sh_{\x}$ is the unit vector in the solution space that is most aligned with $\sh\,'_m$ (\emph{e.g.} Fig.~\ref{fig:certaintygeometry}B), and $e_y$, $e_{s\x}$, and $e_u$ are the projections of $\eh_m$ onto $\yh = \vec y/y$, $\sh_{\x}$, and $\uh_m$ (Fig.~\ref{fig:projections}B). Indeed, Eqs.~(\ref{SCprojections}) and (\ref{UCprojections}) can be readily recognized as special cases of the above general expression. 

Each of these projections is interpretable in light of the fact that $x_{\mu m}$ represents the activity level of the $m^{th}$ presynaptic neuron in the $\mu^{th}$ response pattern. Most simply,
\begin{align} 
e_y = \eh_m \cdot \yh = \frac{\sum_{\mu=1}^{\cP} y_\mu x_{\mu m}}{\sqrt{\sum_{\nu=1}^{\cP} y_{\nu}^2}} \label{Eqn: ey} 
\end{align}
is a normalized correlation of the pre- and postsynaptic activity (note that $\sum_{\rho=1}^\cN X_{\rho m}^2=1$). As expected, synapse certainty is aided by large magnitudes of $e_y$. Moreover, the sign of a certain synapse is the sign of this correlation, or equivalently the sign of $e_y$. Synapse sign identifiability is hindered by large values of
\begin{align} 
e_u = \eh_m \cdot \uh_m  = \sqrt{1-\sum_{\mu=1}^\cP x_{\mu m}^2}, \end{align}
which effectively measures the weakness of the presynaptic neuron's activity, as it is the amount of presynaptic drive for which we do not have any information on the target neuron's response. The more subtle quantity is
\ba e_{s\x} = \eh_m\cdot \sh_{\x} &=& -\sgn(\cos\alpha) \sqrt{\sum_{\{\mu|s'_{\mu} < 0\}}^\cP x_{\mu m}^2},\nonumber \\
&\Longrightarrow& e_{s\x}^2 = \sum_{\{\mu|s'_\mu < 0\}}^\cP x_{\mu m}^2. 
\label{esstar}
\ea
The condition that $s'_\mu < 0$ selects for patterns where the sign of the presynaptic activity is $\sgn(\cos\alpha) = \sgn(e_y)$, but the postsynaptic neuron does not respond. In other words, presynaptic activity should have promoted a response in the target neuron according to the observed activity correlation. That it doesn't generates uncertainty in the sign of the synapse. See Appendix A for a heuristic derivation of $\ycr$ based on this argument. 

We can gain more useful intuition by interpreting our result in relation to what we would obtain in a linear neural network. In the linear problem, there are only constrained and unconstrained dimensions; every dimension that was semi-constrained in the nonlinear problem becomes constrained, with all solutions having $\eta_\mu = y_\mu$ for $\mu = 1,\cdots,\cP$. This implies that
\begin{align}  y_{\mt{cr,lin}} = W\sqrt{\frac{e_u^2}{e_y^2+e_u^2} }  \label{ycriticall} \ .
\end{align}
Returning to the nonlinear problem, recall that the certainty condition finds the largest $y$ for which the solution space and $w_m=0$ hyperplane intersect within the weight bound, and this intersection is simply a point when $y=\ycr$. Importantly, each semi-constrained dimension can either behave like a linear constrained dimension with $\eta_\mu=y_\mu=0$ at this intersection point (toy problems 1 and 3), or like an unconstrained dimension with $\eta_\mu < 0$ at the intersection point (toy problems 1 and 2)\footnote{Since this intersection point depends on $m$, the semi-constrained dimension indexed by $\mu$ can behave as constrained for some synapses and unconstrained for others.}. The first case occurs when $\eh_m\cdot \env_{\mu}$ and $e_y=\eh_m\cdot\yh$ have  opposite signs and $s'_\mu > 0$; the second case occurs when they have the same sign and $s'_\mu < 0$. This means that one could compute the nonlinear theory's $y$-critical from $y_{\mt{cr,lin}}$ by appending the second class of semi-constrained dimensions onto the unconstrained dimensions. Mathematically, this corresponds to the replacement
\ba 
e_u^2\rightarrow e_u^2+e_{s\x }^2=\sum_{\mu=\cP+1}^{\cN}x_{\mu m}^2+\sum_{\{\mu|s'_\mu < 0\}}^\cP x_{\mu m}^2\ ,
\ea
which indeed transforms Eq. (\ref{ycriticall}) to Eq. (\ref{ycritical}). The role of $e_{s\x}^2$ is to quantify the uncertainty introduced by the subset of semi-constrained dimensions that do not behave as constrained at the intersection point.

Since the parameters $e_y, e_{s\x}$, and $e_u$ cannot be set independently, it is convenient to reparameterize Eq.~(\ref{ycritical}) as 
\begin{align} \ycr  = W\sqrt{\frac{1}{1+r_y^2 e_p^2/\left(1 - e_p^2\left(1-r_{s\x}^2\left(1-r_y^2\right)\right)\right)}}, \label{ycritical-ind}\end{align}
where $e_p^2 = 1 - e_u^2$, $r_y^2 = e_y^2/e_p^2$, $r_{s\x}^2 = e_{s\x}^2/(e_p^2 - e_y^2)$, and all three composite parameters can be independently set between 0 and 1. Conceptually, $r_y$ and $r_{s\x}$ merely normalize $e_y$ and $e_{s\x}$ by their maximal values, and $e_p$ is the projection of $\eh_m$ into the activity-constrained subspace spanned by both constrained and semi-constrained dimensions. One could also interpret $r_{s\x}$ as quantifying the effect of threshold nonlinearity. For instance, $r_{s\x}=0$ describes the case where all semi-constrained dimensions are effectively constrained, but $r_{s\x}$ increases as some of the semi-constrained dimensions start to behave like unconstrained dimensions. As expected, $\ycr$ is a decreasing function of $r_y^2$ and $e_p^2$ and an increasing function of  $r_{s\x}^2$ (Fig.~\ref{fig:projections}C). \vs
{\bf Regarding non-orthogonal input patterns:} While a complete treatment of the certainty condition for generally correlated input patterns is beyond the scope of this paper, we could find a conservative bound for $y$-critical that may be useful when patterns are close to being orthogonal. The details of the derivation are discussed in the final subsection of Appendix A. 

The major challenge caused by non-orthogonal patterns is that the spherical weight space becomes elliptical in terms of the $\eta$-coordinates. Thus, the main idea behind the bound is that one can always find the sphere that just encompasses this ellipse. We can then use our formalism to obtain a conservative $y$-critical, such that if the norm of $\yv$ is larger than this value then all solutions within the encompassing sphere have a consistent sign for the synapse under consideration. An interesting insight that emerges from our analysis is that the relative orientations between 
\be
\vec{n}_m\equiv \sum_{\mu=1}^{\cN} X^{-1}_{m\mu}\env_{\mu}\
\ee
and the various important $\eta$-directions play the role of $\ta, \phi, \al$  and $\ga$ (Appendix A). Note that $\vec n_m = \hat e_m$ when $X$ is an orthogonal matrix. We anticipate that $\nv$ will also be an important player in a more comprehensive treatment of non-orthogonal patterns.
\vs
{\bf Application to recurrent networks:} As we explained in Section \ref{Sec: SolutionSpace}, to find the ensemble of all incoming weight vectors onto the $i^{th}$ driven neuron, one can use the results obtained for the feedforward network and just substitute $X$ with the $Z^{(i)}$ matrix. Consequently, identifying certain synapses onto the $i^{th}$ neuron can follow the route outlined for the feedforward scenario as long as $Z^{(i)}$ is orthogonal. So for example, if we want to ascertain whether any incoming synapse to the $i^{th}$ neuron is certain, we have to replace $x_{\mu m}\ra z^{(i)}_{\mu m}$ and $y_{\mu}\ra y_{\mu i}$ in (\ref{ycritical})-(\ref{esstar}) to compute $\ycr$.
\vs
\begin{figure}[!htbp]
\centering
\includegraphics[width=0.48\textwidth,angle=0]{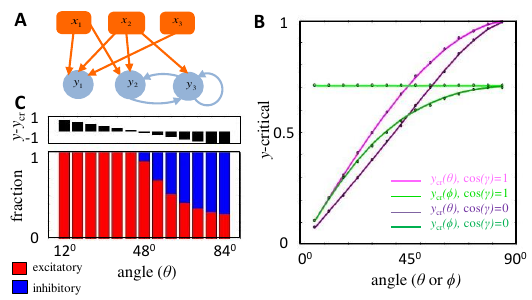}
	\caption{{\bf Testing the certainty condition with exhaustive low-dimensional simulations. }{\small ({\bf A}) A simple recurrent network with three input neurons and three driven neurons (Appendix E). ({\bf B}) We plot the theoretically derived $\ycr$ for feedforward synapses to $y_1$ as we vary $\ta$ (green curves) or $\phi$ (magenta curves), keeping the other angle fixed at $45^\circ$. The lighter shades correspond to $\cos\ga=1\Ra r_{s\ast}=1$. The darker shades correspond to $\cos\ga=0\Ra r_{s\x}=0$, where the predictions from the nonlinear network match those of a linear network. The dots represent $\ycr$ estimated through simulations, and they agree well with the theory. ({\bf C}) (\emph{Bottom}) Bar graph of the fraction of solutions with positive (red) and negative (blue) self-couplings ($y_3\ra y_3$) as a function of $\ta$. (\emph{Top}) As predicted, all solutions have positive $w_{y_3,y_3}$ when $y-\ycr > 0$. 
			\label{fig:zeroerror}}}
\end{figure}
{\bf Numerical illustration of the certainty condition:} To illustrate and test the theory numerically, we first considered a small neural network of three input neurons and three driven neurons (Fig.~\ref{fig:zeroerror}A). This small number of synapses meant that we could comprehensively scan the entire spherical weight-space without relying on a numerical algorithm to find solutions\footnote{Our results for the certainty condition hold for network ensembles that exactly generate the desired responses. For numerical tests, we had to allow for small deviations from the desired responses, but our predictions proved robust.}. This is important because numerical techniques, such as gradient descent learning, potentially find a biased set of solutions that incompletely test the theory. We supposed that each driven neuron has three inputs, and we constrained weights with two orthonormal stimulus responses. We set $W=1$ for all simulations and numerically screened weights randomly. See Appendix F for complete simulation details.

The first driven neuron in Fig.~\ref{fig:zeroerror}A, \(y_1\), receives only feedforward drive, and we suppose that it responds to one stimulus condition with response $y$ ($\mu=2$), but it does not respond to the other ($\mu=1$). Its synapses thus have one constrained, one semi-constrained, and one unconstrained dimension, and all of the terms in Eq.~(\ref{yc-angles}) contribute to $y$-critical. We could thus use $y_1$ to verify Eq.~(\ref{yc-angles}). Moreover, this scenario includes the illustrative example of Fig.~\ref{fig:certaintygeometry}F as a special case, so we could also use $y_1$ to verify Eq.~(\ref{1S1C1U}). 

To these ends, we decided to focus on a two-parameter family of input patterns,
\begin{align}
x=&\LF\by{ccc}
 -\sin\psi \cos\chi& \cos\psi \cos\chi&\sin\chi \\
 \cos\psi        &  \sin\psi  &0    
\ey
\RF\ ,\label{inputpatterns}
\end{align}
where rows correspond to different input patterns and columns correspond to different input neurons, as usual, and we extend $x$ to the full-rank orthogonal matrix
\begin{align}
X=&\LF\by{ccc}
 -\sin\psi \cos\chi& \cos\psi \cos\chi&\sin\chi \\
 \cos\psi        &  \sin\psi  &0 \\
 \sin\psi\sin\chi& -\cos\psi\sin\chi& \cos\chi
\ey
\RF\ .\label{extendedinputpatterns}
\end{align}
By Eq. (\ref{Eq: SubspaceDecompEmG}), the physical basis vector corresponding to the synapse from the first input neuron is thus
\begin{align}
\eh_1= \cos\psi \env_2 + \sin\psi\cos\chi(-\env_1) + \sin\psi\sin\chi\env_3, \label{example}
\end{align}
and it has the same general form as Eqs.~(\ref{Eq: SubspaceDecompEm}) and (\ref{Eq: ThetaPhiDecomp}), where $\env_3$ plays the role of $\uh$.  If $\psi$ and $\chi$ are both acute, then one can identify them with $\ta$ and $\phi$ in Fig.~\ref{fig:certaintygeometry}F, and the roles of $\ch$ and $\sh$ are played by $\env_2$ and $-\env_1$, respectively. In this case $\al=0$, $\cos\ga=0$, and the theoretical dependencies of $\ycr$ on $\ta$ and $\phi$ are given by Eq. (\ref{1S1C1U}). Fig.~\ref{fig:zeroerror}B illustrates these dependencies as the purple and dark green curves. If $\psi$ is acute, but $\chi$ is obtuse, then according to our conventions, $\ta=\psi$, $\phi=\pi-\chi$, $\ch=\env_2$, and $\sh=\env_1$. Now $\al=0$ and $\cos\ga=1$, and our general formula, Eq.~(\ref{yc-angles}), implies
\be
\ycr=W\sin\ta=W\sin\psi\ .
\label{SCUprime}
\ee
These dependencies are plotted as the pink and the light green curves in Fig.~~\ref{fig:zeroerror}B.  We do not plot cases where $\psi$ is obtuse, because obtuse and acute $\psi$ result in equivalent $\ycr$ formulae. Whether $\psi$ is acute or obtuse nevertheless matters because it determines the sign of the $w_1$ synapse when it is certain.

The black dots in Fig.~\ref{fig:zeroerror}B show the largest response magnitude, $y$, for which we numerically found solutions with both positive and negative $w_1$ (see Appendix F for numerical methods), thereby providing a numerical estimate of $\ycr$. The theoretical curves and numerical points precisely aligned in all cases. The differences between the light and dark theoretical curves illustrates the effect of nonlinearity. When $\chi$ is obtuse, the semi-constrained dimension effectively behaves as unconstrained, and the mixing angle between the semi-constrained and unconstrained dimension is irrelevant to $y$-critical. When $\chi$ is acute, the semi-constrained dimension effectively behaves as constrained, as if its coordinate were set to zero. Moreover, these results confirmed that stronger responses were needed to make synapses fixed sign when the synaptic direction was less aligned with the constrained dimension (Fig.~~\ref{fig:zeroerror}B, purple and pink). Furthermore, smaller $y$-critical values occurred when the synaptic direction anti-aligned with the semi-constrained dimension (Fig.~\ref{fig:zeroerror}B, purple vs. pink, dark green vs. light green).

We next wanted to check the validity of our results for the recurrently connected neurons in Fig.~\ref{fig:zeroerror}A. We therefore needed to  tailor the steady-state activity levels of the recurrent network to result in orthogonal presynaptic input patterns for each driven neuron. In mathematical terms, $Z^{(i)}$ must be an orthogonal matrix for $i=1,2,3$. We achieved this by considering a two-parameter family of driven neuronal responses in which the activity patterns of $y_2$ and $y_3$ were matched to those of $x_1$ and $x_3$, respectively. This construction means that all three driven neurons receive the same input patterns. To ensure positivity of driven neuronal responses, we set $\chi$ as an acute angle and $\psi$ as the negative of an acute angle. 

Although \(y_2\) has both feedforward and recurrent inputs, we can analyze its connectivity in exactly the same way as \(y_1\). Recurrence only complicates the analysis for neurons that synapse onto themselves, like \(y_3\), since changing the output activity also changes the input drive. So \(|\yv_3|\) and \(\ycr\) are not independent. Here we focused on the certainty condition for the self-synapse, $w_{y_3,y_3}$, for which
$\ycr=\cos\chi$, and $|\yv_3|=\sin\chi$. Therefore, the synapse should be certain if $45^0<\chi\le 90^0$. Since $\ta=\pi/2-\chi$ according to our conventions\footnote{For $y_3$,  $\mu=1,2$ are constrained and semi-constrained respectively. Accordingly, $$\eh_{y_3y_3}=\sin\chi\env_1+\cos\chi\env_3=\cos\ta\env_1+\sin\ta\env_3=\cos\ta\ch+\sin\ta\uh$$.}, this is equivalent to $0\le\ta<45^0$ (Fig.~\ref{fig:zeroerror}C, \emph{top}). Our numerical results precisely recapitulated these theoretical expectations (Fig.~\ref{fig:zeroerror}C, \emph{bottom}), as the self-connection was consistently positive across all simulations whenever this condition on $\theta$ was met. See Appendix E for certainty condition analyses for other synapses onto $y_3$ and Appendix F for complete simulation details.

\section{Accounting for noise}
\noindent
{\bf Finding the solution space in the presence of noise:} So far we have only considered exact solutions to the fixed point equations. However, it's also important to determine weights that lead to fixed points near the specified ones. For example, biological variability and measurement noise generally make it infeasible to specify exact biological responses. Furthermore, numerical optimization typically produces model networks that only approximate the specified computation. We therefore define the \(\cE\)-{\it error surface} as those weights that generate fixed points a distance \(\cE\) from the specified ones,
\begin{align} \mathcal{V}_\cE = \left\{w \Big | \sum_{\mu = 1}^\cP \sum_{i = 1}^\cD \left(y_{\mu i} - \w{y}_{\mu i}(w) \right)^2 = \cE^2 \right\}, 
\label{error}
\end{align}
where \(y_{\mu i}\) is the specified activity of the \(i^{th}\) driven neuron in the \(\mu^{th}\) fixed point, and \(\w{y}_{\mu i}\) is the corresponding activity level in the fixed point approached by the model network when it's initialized as \(y_i(t=0)=y_{\mu i}\). If the network dynamics do not approach a fixed point, perhaps oscillating or diverging instead~\cite{Morrison}, we say \(\cE = \infty\). 

\begin{figure}[!htbp]
\centering
		\includegraphics[width=0.46\textwidth,angle=0]{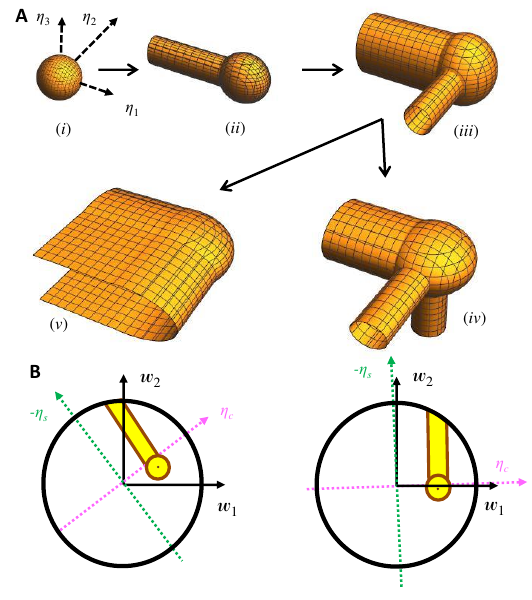}
	\caption{{\bf The solution space geometry changes as the allowed error increases.} {\small ({\bf A}) Error surface contours in a three-dimensional subspace corresponding to $\eta_1$,  $\eta_2$, and  $\eta_3$. Several topological transitions occur as the error increases. (\emph{i}) We consider the case where all responses are positive, so the contours are spherical for small errors, just like in a linear neural network. (\emph{ii})-(\emph{iii}) Two cylindrical dimensions sequentially open up when the error is large enough for some $\eta$-coordinates to become negative. (\emph{iv})-(\emph{v}) After that, either a third cylindrical dimension can open up, or the two cylindrical axes can join to form a plane. Which transition occurs at lower error depends on the pattern of neural responses. ({\bf B}) (\emph{Left}) We illustrate a case where there is a unique exact solution to the problem (brown dot). Allowing error but neglecting topological transitions would expand the solution space to an ellipse (here, brown circle), but the signs of $w_1$ and $w_2$ remains positive. Including topological transitions in the error surface can cap the ellipse with a cylinder (full yellow solution space). Now we can say with certainty that the sign of $w_2$ is positive, but negative values of $w_1$ become possible. (\emph{Right}) Graphical conventions are the same. However, in this case all solutions inside the cylinder have $w_2>0$. Therefore the topological transition breaks a near symmetry between positive and negative weights. 
			\label{fig:transitions}}}
\end{figure}

Each \(\cE\)-error surface can be found exactly for feedforward networks. For illustrative purposes, let us first consider the $\cD=1$ feedforward scenario in which the driven neuron is active in every response pattern. This means that $y_\mu>0$ for all $\mu=1,\cdots,\cP$, and we can reorder the $\mu$ indices to sort the driven neuron responses in ascending order, \(0 < y_{1} < y_{2} < \cdots < y_{\cP}\). Here we assumed that no two response levels are exactly equal, as is typical of noisy responses. Since all responses are positive, the zero-error solution space has no semi-constrained dimensions, and the only freedom for choosing $w$ is in the $\cF=\cN-\cP$ unconstrained dimensions. Therefore, the zero-error surface of exact solutions, \(\mathcal{V}_0\), is a \(\mathcal{U}\)-dimensional linear subspace, and \(\mathcal{V}_0\) is a point in the \(\mathcal{P}\)-dimensional activity-constrained  subspace. 

How does this geometry change as we allow error? For \(0<\cE<y_1\), we must have \(\w{y}_\mu > 0\) for all \(\mu\). Therefore, the nonlinearity is irrelevant, and \(\cE\)-error surfaces are spherical in the activity-constrained $\eta$-coordinates (Eq.~\ref{error}, Fig.~\ref{fig:transitions}Ai). However, once $\cE=y_1$ it becomes possible that \(\w{y}_1=0\), and suddenly a semi-infinite line of solutions appears with $\eta_1\leq 0$.  As \(\cE\) further increases, this line dilates to a high-dimensional cylinder (Fig.~\ref{fig:transitions}Aii). A similar transition happens at $\cE=y_2$, whereafter two cylinders cap the sphere (Fig.~\ref{fig:transitions}Aiii). Things get more interesting as $\cE$ increases further because two transitions are possible. A third cylinder appears at $\cE'=y_{3}$. However, at $ \cE'' = \sqrt{(y_1)^2+(y_2)^2}$ it's possible for both \(\w{y}_1\) and \(\w{y}_2\) to be zero, and the two cylindrical axes merge into a semi-infinite hyperplane defined by $\eta_1 \le 0, \eta_2 \le 0$. Thus, when \(\cE' < \cE''\) the error surface grows to attach a third cylinder (Fig.~\ref{fig:transitions}Aiv), and when \(\cE'' < \cE'\) the two cylindrical surfaces merge to also include planar surfaces in between  (Fig.~\ref{fig:transitions}Av). These topological transitions continue by adding new cylinders and merging existing ones, and the sequence is easily calculable from $\{y_\mu\}$. Note that we use the terminology ``topological transition'' to emphasize that the structure of the error surface changes discontinuously at these values of error. The geometric transitions we observe here also relate to topological changes in a formal mathematical sense. For instance, while there are no incontractible circles in Fig.~\ref{fig:transitions}A(ii), one develops as we transition to Fig.~\ref{fig:transitions}A(iii). 

In general, $y_\mu$ may also be zero or negative in the presence of noise. Whenever $y_\mu=0$, the $\mu^{th}$ response pattern generates a semi-constrained dimension in $\mathcal{V}_0$. On the other hand, if some response levels are negative, then there are no exact solutions at all. However, it becomes possible to find solutions when $\cE=\sqrt{\sum_{\{\mu|y_{\mu}<0\}}y_{\mu}^2}$, and each response pattern associated with a negative $y_\mu$ acts as a semi-constrained dimension in $\mathcal{V}_\cE$. As illustrated above, more semi-constrained dimensions open up as more error is allowed in each of these cases. 

This geometry only approximates \(\cE\)-error surfaces for recurrent networks (Appendix C). For instance, displacing $y_{\mu i}$ from its specified value changes the input pattern that define the $\env_\mu$-directions for downstream driven neurons, but this effect is neglected here. We will nevertheless find that this feedforward approximation to \(\cE\)-error surfaces is practically useful for predicting synaptic connectivity in recurrent networks as well.\vs
{\bf Predicting connectivity in the presence of noise:} The threshold nonlinearity and error-induced topological transitions can have a major impact on synapse certainty (Fig.~\ref{fig:transitions}B). For example, one might model a neuronal dataset with a linear neural network and find that models with acceptably low error consistently have positive signs for some synapses. However, if measured neural activity was sometimes comparable to the noise level, then semi-constained dimensions could open up that suddenly make some of these synapse signs ambiguous (Fig.~\ref{fig:transitions}B, left). Although semi-constrained dimensions can never make an ambiguous synapse fully unambiguous, semi-constrained dimensions can heavily affect the distribution of synapse signs across the model ensemble by providing a large number of solutions that have consistent anatomical features (Fig.~\ref{fig:transitions}B, right). 

We therefore generalized the certainty condition to include the effects of error, including topological transitions in the error surface (Appendix C). As before, finding the certainty condition amounts to determining when the $w_m=0$ hyperplane intersects the solution space within the weight bound, but to account for noise of magnitude $\en$, we must now check whether an intersection occurs with any \(\cE\)-error surface with $\cE \le \en$. No intersections will occur if and only if every non-negative $\vec{\tilde y}$ within $\en$ of the provided $\cP$-vector of noisy target neuron activity  (Fig.~\ref{fig:cartoons}C) satisfies its zero-error certainty condition, and each $\vec{\tilde y}$ is a possible denoised version of it (Eq. \ref{error}). We thus define $y$-critical in the presence of noise as the maximal $\ycr$ (Eq. \ref{ycritical}) amongst this set of $\vec{\tilde y}$.

Although we lack an exact expression for $y$-critical in the presence of noise, we derived several useful bounds and approximations (Appendix C). We usually focus on a theoretical upper bound for $y$-critical, $y_{\mt{cr},\max}$. Note that this upper bound suffices for making rigorous predictions for certain synapses, because $y>y_{\mt{cr},\max}\Longrightarrow y>y$-critical. In the absence of topological transitions, this formula is
\begin{widetext}
	\ba
	&y_{\mt{cr},\max}&=
	W\LT\sqrt{\LF{e_{s\x}^2+e_u^2 \over e_{y}^2+ e_{s\x}^2+e_u^2 }\RF+ {\en^2\over W^2}\LF1+{e_y^2(e_p^2-e_y^2)\over (e_y^2+ e_{s\x}^2+e_u^2)^2}\RF}+ {\en\over W}\sqrt{1+{e_y^2(e_p^2-e_y^2)\over (e_y^2+ e_{s\x}^2+e_u^2)^2}}\RT.\label{ycmaxM}
	\ea
\end{widetext}
We also computed a lower bound, $y_{\mt{cr},\min}$, to assess the tightness of the upper bound. This bound is 
\ba
&y_{\mt{cr},\min}&= W\LT\sqrt{e_{s\x}^2+e_u^2 \over e_{y}^2+ e_{s\x}^2+e_u^2 }+{\en\over W}\RT\ 
\label{ycminM}
\ea
without topological transitions. Both bounds increase with error and should be considered to be bounded above by $W$. As expected, both expressions reduce to Eq. (\ref{ycritical}) as $\en/W\ra 0$. We also note that the two bounds coincide, to leading order in $\en/W$, if $e_y\ll \max(e_{s\x},e_u)$ and $e_p/\max(e_{s\x},e_u)=\cO(1)$, and we argue in Appendix B that this is typical when the network size is large.

\begin{figure*}[!htbp]
\centering
	\includegraphics[width=0.7\textwidth,angle=0]{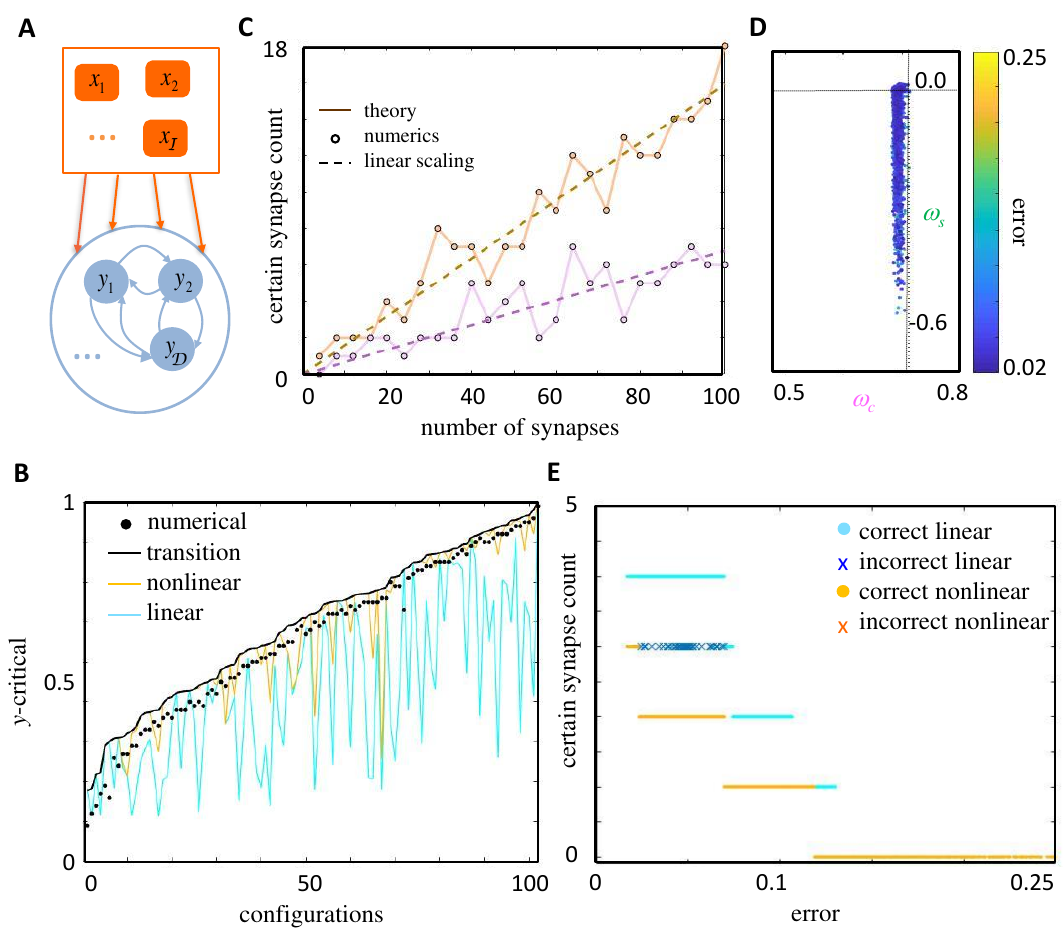}
	\caption{{\bf The theory accounting for error explains numerical ensembles of feedforward and recurrent networks.} {\small ({\bf A}) Cartoon of a recurrent neural network. We disallow recurrent connectivity of neurons onto themselves throughout this figure. $\cD=1$ corresponds to the feedforward case, and $W=1$ for all panels. ({\bf B}) Comparison of numerical and theoretical $y$-critical values for 102 random configurations of input-output activity (Appendix F). We considered a feedforward network with $\cI=6$, $\cP=5$, $\cR=2$. For each configuration and postsynaptic activity level $y$, we used gradient descent learning to numerically find many solutions to the problem with $\cE \approx 0.1$. The black dots correspond to the maximal value of $y$ in our simulations that resulted in an inconsistent sign for the synaptic weight under consideration. The continuous curves show theoretical values for $y$-critical that upper bound the true $y$-critical ($y_{\mt{cr},\max}$, black), that neglect topological transitions in the error surface (yellow), or that neglect the threshold nonlinearity (cyan). Only the black curve successfully upper bounded the numerical points. Configurations were sorted by the $y_{\mt{cr},\max}$ value predicted by the black curve. ({\bf C}) The number of certain synapses increased with the total number of synapses in feedforward networks. Purple and brown correspond to $\cN=2\cP=4\cR$ and $\cN=\cP=4\cR$, respectively. The solid lines plot the predicted number of certain synapses. The circles represent the number of correctly predicted synapse signs in the simulations. The dashed brown and purple lines are best-fit linear curves with slopes  $0.16 (\pm 0.01)$  and $0.07 (\pm 0.01) $ at 95\% confidence level, significantly less than the zero error theoretical estimates of 0.28 and 0.18 (Appendix B). ({\bf D})-({\bf E}) Testing the theory in a recurrent neural network with $\cN=10, \cI = 7, \cD=4, \cP=8$ and $\cR=3$. Each dot shows a model found with gradient descent learning. (D) $x$- and $y$-axes show two $\eta$-coordinates predicted to be constrained and semi-constrained, respectively, and the color axis shows the model's root mean square error over neurons, $\cE/\sqrt{\cD}$. Although our theory for error surfaces is approximate for recurrent networks, the solution space was well explained by the constrained and semi-constrained dimensions. Note that the numerical solutions tend to have constrained coordinates smaller than the theoretical value (vertical line) because the learning procedure is initialized with small weights and stops at nonzero error. (E) The $x$-axis shows the model's error, and the $y$-axis shows the number of synapse signs correctly predicted by the nonlinear theory (yellow dots or red crosses) or linear theory (cyan dots or blue crosses). Dots denote models for which every model prediction was accurate, and crosses denote models for which some predictions failed.
		\label{fig:highDsims}}}
\end{figure*}

The effect of topological transitions is that $y_{\mt{cr},\max}$ and $y_{\mt{cr},\min}$ become the maximums of several terms, each corresponding to a way that constrained dimensions could behave as semi-constrained within the error bound (Appendix C). We compute each term from generalizations of Eqs. (\ref{ycmaxM}) and (\ref{ycminM}) that account for the amount of error needed to open up semi-constrained dimensions.
\vs
{\bf Testing the theory with simulations:} To examine our theory's validity, we assessed its predictions with numerical simulations of feedforward and recurrent networks (Fig.~\ref{fig:highDsims}A). Each assessment used gradient descent learning to find neural networks whose late time activity approximated some specified orthogonal configuration of input neuron activity and driven neuron activity (Appendix F). We then used our analytically-derived certainty condition with noise to identify a subset of synapses that were predicted to not vary in sign across the model ensemble ($W=1$), and we checked these predictions using the numerical ensemble. We similarly checked predictions from simpler certainty conditions that ignored the nonlinearity or neglected topological transitions in the error surface (Appendix C). Note that we expected gradient descent learning to often fail at finding good solutions in high dimensions, as our theory predicts that each semi-constrained dimension induces local minima in the error surface (Fig. \ref{fig:transitions}A). Since we did not want the theory to bias our numerical verification of it, we focused our simulations on small to moderately-sized networks, where we could reasonably sample the initial weight distribution randomly. Future work will consider more realistic neural network applications.

We first considered feedforward network architectures, for which our analytical treatment of noise is exact. To illustrate how nonlinearity and noise affect synapse certainty, we calculated the magnitude of postsynaptic activity needed to make a particular synapse sign certain (Fig.~\ref{fig:highDsims}B). We specifically considered 102 random input-output configurations of a small feedforward network with 6 input neurons ($\cP=5$, $\cC=2$), which were tailored to have orthonormal input patterns and generate one topological error surface transition at small errors. In particular, we generated random orthogonal matrices by exponentiating random anti-symmetric matrices, we set one element of $\hat y$ to a small random value to encourage the topological transition, and we ensured that the other non-zero random element of $\hat y$ was large enough to preclude additional transitions (Appendices C, F). For each input-output configuration, we then systematically varied the magnitude of driven neuron activity, $y$, finding $10^5$ synaptic weight matrices with moderate error, $\cE^2\approx \en^2$, for each magnitude $y$. Since randomly screening a 6-dimensional synaptic weight-space is not numerically efficient, we applied gradient descent learning. Nevertheless, the small network size meant that we could comprehensively sample the solution space and numerically probe the distinct predictions made by each bound or approximation used to estimate $y$-critical. 

As expected, the maximum value of $y$ that produced numerical solutions with mixed synapse signs (Fig.~\ref{fig:highDsims}B, black dots) was always below the theoretical upper bound for $y$-critical (Fig.~\ref{fig:highDsims}B, black line). In contrast, mixed-sign numerical ensembles were often found above theoretical $y$-critical values that neglected topological transitions in the error surface (Fig.~\ref{fig:highDsims}B, yellow line) or that neglected the nonlinearity entirely (Fig.~\ref{fig:highDsims}B, cyan line). This means that these simplified calculations for estimating $y$-critical make erroneous predictions, because the synapse sign is supposed to be exclusively positive or negative whenever $y$ exceeds $y$-critical, by definition. Therefore, we were able to accurately assess synapse certainty, and this generally required us to include both the nonlinearity and noise-induced topological transitions in the error surface.

We next asked how often we could identify certain synapses in larger networks. For this purpose, we generated 25 random input-output configurations in the feedforward setting (Appendix F), again with orthonormal input patterns, but this time we increased the number of input neurons from 4 to 100 across the configurations (Fig.~\ref{fig:highDsims}C). As we increased the size of the network, we kept $\cC/\cN$ fixed at 0.25 and $\cP/\cN$ fixed at 1 (Fig.~\ref{fig:highDsims}C, brown) or 0.5 (Fig.~\ref{fig:highDsims}C, purple). These scaling relationships put our simulations in the setting of high-dimensional statistics \cite{Advani}, where both the number of parameters and the number of constraints increase with the size of the network. In this high-dimensional regime, a simple heuristic argument suggests that the number of zero-error certain synapses should scale linearly with the number of synapses (Appendix B), because $\ycr$ and the typical magnitude of $y$ scale equivalently with $\cN$. Here we tested this prediction by setting $\hat y$ randomly, setting $y=1-\ln 2/\cR$ to approximate the median norm of vectors in the unit $\cC$-ball (Appendix B), and numerically finding a small error solution for each configuration ($\cE^2/\cP\approx 10^{-6}$). 

As expected, we empirically found that the number of certain synapses predicted by the theory (Fig.~\ref{fig:highDsims}C, solid lines) scaled with the network size linearly (Fig.~\ref{fig:highDsims}C, dashed lines). The jaggedness of the solid curves  reflect the fact that each point is specific to the random input-output configuration constructed for that value of $\cN$. The purple curve corresponds to the case when $\cN=2\cP=4\cR$ and the brown curve when $\cN=\cP=4\cR$. Furthermore, for every certain synapse predicted, we verified that its predicted sign was realized in the numerical solution we found (Fig.~\ref{fig:highDsims}C, circles). These results suggest that the theory will predict many synapses to be certain in realistically large neural systems. 

Finally, we empirically tested our theory for a recurrent network (Figs.~\ref{fig:highDsims}D,~\ref{fig:highDsims}E), where our treatment of noise is only approximate. For this purpose, we considered networks without the self-coupling terms, $\cN = \cI+\cD-1$. We constructed a single random configuration with non-negative driven neuron responses and orthogonal presynaptic patterns for one of the driven neurons\footnote{We performed this numerical experiment with several random configurations to confirm that the results did not qualitatively depend on the random sample.} (Appendix F). This driven neuron could thus serve as the target neuron for our analyses. Note that it is sometimes possible to orthogonalize the input patterns for more than one driven neuron, but this is irrelevant to our analysis and is not pursued here. We then used gradient descent learning to find around 4500 networks that approximated the desired fixed points with variable accuracy. For technical simplicity, we first found connectivity matrices using a proxy cost function that treated the network as if it were feedforward. We then simulated the neural network dynamics with these weights and correctly evaluated the model's error as prescribed by Eq. (\ref{error}). 

This network ensemble revealed that constrained and semi-constrained dimensions accurately explained the structure of the solution space for recurrent networks with non-zero error. Fig.~\ref{fig:highDsims}D shows the projection of the corresponding solution space along two $\eta$-directions, one predicted to be constrained by the feedforward theory and the other predicted to be semi-constrained. As predicted, the extension of the solution space along the negative semi-constrained direction was clearly discernible. However, recurrence implies that the exact solution space is not perfectly cylindrical around the semi-constrained axes (Appendix C), because the driven neuron inputs to the target neuron can themselves vary due to noise. Here this effect was empirically insignificant, and the geometric structure of the solution space conformed rather well to our feedforward prediction. One might have expected the error (color in Fig.~\ref{fig:highDsims}D) to increase monotonically as one moves away from the center of semi-constrained cylinder, but this expectation is incorrect for two reasons. First, we are visualizing the error surface as a projection along two dimensions, yet variations in other $\eta$-coordinates add variation to the error\footnote{For example, imagine projecting the 3D surfaces in Fig.~\ref{fig:transitions}A along two dimensions}. Second, we are visualizing the solution space for one target neuron, but other driven neurons in the recurrent network contribute to the summed error represented by the color. 

Moreover, the theory correctly predicted how the number of certain synapses would decrease as a function of \(\en\) (Fig.~\ref{fig:highDsims}E), and we never found a numerical violation of the theoretical certainty condition that included nonlinearity and noise. In Fig.~\ref{fig:highDsims}E, the yellow circles represent the number of certain synapses that were predicted by the theory and verified to have synapse signs that agreed with the theoretical prediction. Here accurate predictions did not require us to account for topological error surface transitions. In contrast, although our simulations usually agreed with the predictions of the linear theory (Fig.~\ref{fig:highDsims}E, cyan circles), they could also disagree. In Fig.~\ref{fig:highDsims}E, the blue crosses indicate configurations where the linear theory incorrectly predicted some synapse signs. The absence of red crosses reiterates the consistency of predictions coming from the nonlinear treatment. 
\section{DISCUSSION}
In summary, we enumerated all threshold-linear recurrent neural networks that generate specified sets of fixed points, under the assumption that the number of candidate synapses onto a neuron is at least the specified number of fixed points. We found that the geometry of the solution space was elegantly simple, and we described a coordinate transformation that permits easy classification of weight-space dimensions into constrained, semi-constrained, and unconstrained varieties. This geometric approach also generalized to approximate error-surfaces of model parameters that imprecisely generate the fixed points. We used this geometric description of the error surface to analyze structure-function links in neural networks. In particular, we found that it is often possible to identify synapses that must be present for the network to perform its task, and we verified the theory with simulations of feedforward and recurrent neural networks.

Rectified-linear units are also popular in state of the art machine learning models ~\cite{Tschopp, Nair, Krizhevsky, Xu15}, so the fundamental insights we provide into the effects of neuronal thresholds on neural network error landscapes may have practical significance. For example, machine learning often works by taking a model that initially has high error and gradually improving it by modifying its parameters in the gradient-direction \cite{Rumelhart}. However, error surfaces with high error can have semi-constrained dimensions that abruptly vanish at lower errors (Fig.~\ref{fig:transitions}). Local parameter changes typically cannot move the model through these topological transitions, because models that wander deeply into semi-constrained dimensions are far from where they must be to move down the error surface. The model has continua of local and global minima, and the network needs to be initialized correctly to reach its lowest possible errors. This could provide insight into deep learning theories that view its success as a consequence of weight subspaces that happen to be initialized well~\cite{Frankle, Zhou}. 

The geometric simplicity of the zero-error solution space provides several insights into neural network computation. Every time a neuron has a vanishing response, half of a dimension remains part of the solution space, which the network could explore to perform other tasks. In other words, by replacing an equality constraint with an inequality constraint, simple thresholding nonlinearities effectively increase the computational capacity of the network ~\cite{Cover, Gardner}. The flexibility afforded by vanishing neuronal responses thereby provides an intuitive way to understand the impressive computational power of sparse neural representations ~\cite{Marr, Treves, Olshausen, Glorot}. Furthermore, the brain could potentially use this flexibility to set some synaptic strengths to zero, thereby improving wiring efficiency. This would link sparse connectivity to sparse response patterns, both of which are observed ubiquitously in neural systems. 

Our theory could be extended in several important ways. First, we only derived the certainty condition to identify critical synapses from orthonormal sets of fixed points. Although our orthogonal analysis also provides a conservative bound for a general set of fixed points (Appendix A), a more precise analysis will be needed to pinpoint synapses in realistic biological settings where stimulus-induced activity patterns may be strongly correlated. Since our error surface description made no orthonormality assumptions, this analysis will only require more complicated geometrical calculations to discern whether the synapse sign is consistent across the space of low-error models. Furthermore, we could use the error surfaces to identify multi-synapse anatomical motifs that are required for function, or to estimate the fraction of models in which an uncertain synapse is excitatory versus inhibitory. It would also be interesting to relax the assumption that the number of fixed points is small. This would allow us to consider scenarios where the fixed points can only be generated nonlinearly. We could also consider cases where no exact solution exists at all. Here we assumed that we knew the activity level of every neuron in the circuit. This is not always the case, and it will be important to determine how unobserved neurons alter the error landscape for synaptic weights connecting the observed neurons. The error landscape geometry will also be affected by recurrent network effects that we ignored here (Appendix C). It will be interesting to see whether the geometric toolbox of theoretical physics can provide insights into the nontrivial effects of unobserved neurons and recurrent network dynamics. Finally, we note that it will sometimes be important to analyze networks with alternate nonlinear transfer functions. Our analyses already apply exactly to recurrent networks with arbitrary threshold-monotonic nonlinear transfer functions (Appendix D). Moreover, our analyses can approximate any nonlinearity by treating its departures from threshold-linearity as noise (Appendix D). An extension to capped rectified linear units ~\cite{Krizhevsky}, which saturate above a second threshold, would also be straightforward. In particular, semi-constrained dimensions would emerge from any condition where the target neuron is inactive or saturated.

Our primary motivation for undertaking this study was to find rigorous theoretical methods for predicting neural circuit structure from its functional responses. This identification can be used to corroborate or broaden circuit models that posit specific connectivity patterns, such as center-surround excitation-inhibition in ring attractors \cite{Ben-Yishai, Skaggs, Kim} or contralateral relay neuron connectivity in zebrafish binocular vision ~\cite{Naumann, Kubo}. More generally, if an experimental test violates the certainty conditions we derived using our ensemble modeling approach, it will suggest that some aspect of model mismatch is important. We could then move on to the development of qualitatively improved models that might modify neuronal nonlinearities, relax weight bounds, incorporate sub-cellular processes or neuromodulation, or hypothesize hidden cell populations. On the other hand, we hope that our focus on predictions that follow with certainty from simple network assumptions will enable predictions that are relatively insensitive to minor mismatches between our abstract model and the real biological brain. More nuanced predictions may require more nuanced models.

An important parameter of the theory is the weight bound. In particular, $W$ bounds the magnitude of synaptic weight vectors in biological networks, and our certainty condition declares a synapse to be necessary when the ratio $y/W$ exceeds a critical value. It is not \emph{a priori} clear how to set this scale parameter without additional biological data. Nevertheless, one could use the neuronal activity data to compute each synapse's $W$-critical value, below which the certainty condition is satisfied, and rank-order the synapses according to decreasing $W$-critical values. Until we know the value of $W$, we do not know where to draw the line between certain synapses and uncertain synapses. However, our theory predicts that all of the certain synapses will be at the top of the list, which specifies a sequence of experimentally testable predictions and may already provide biological insights into the important synaptic connections. Testing these predictions can help constrain the theory's biological bound parameter.

Our theory describes function at the level of neural representations. This description is useful because many systems neuroscience experiments measure representations directly, and it is important to build mechanistic models that explain these data in terms of neural network interactions \cite{Naumann, Biswas, Kim, Kubo}. However, it would also be interesting to link structure to function at the higher levels of behavior and cognition. This is a significantly different problem because multiple representations can support the same high-level functions, and both neural network structure and representation can change over time \cite{Trachtenberg, Ziv, Attardo, Driscoll, Rule, Schoonover, Marks, Deitch}. Consequently, experimental tests of our current framework must measure network structure and representation on timescales shorter than the network's representational dynamics, and certain synapses may be most biologically meaningful in innate circuits with limited plasticity. Extensions to our framework may also be useful for relating structural and representational dynamics in circuits for learning \cite{Kappel}.

An exciting prospect is to explore how our ensemble modeling framework can be combined with other theoretical principles and biological constraints to obtain more refined structure-function links. For instance, we could refine our ensemble by restricting to stable fixed points. Alternatively, once the sign of a given synapse is identified, Dale's principle might allow us to fix the signs of all other synapses from this neuron~\cite{Burnstock}. This would restrict the solution space and could make other synapses certain. Utilizing limited connectomic data to impose similar restrictions might also be a fruitful way to benefit from large-scale anatomical efforts ~\cite{Varshney, Hildebrand, Ohyama, Scheffer20}. Finally, rather than restricting the magnitude of the incoming synaptic weight vector, we could consider alternate biologically relevant constraints, such as limiting the number of synapses, minimizing the total wiring length, or positing that the network operates at capacity ~\cite{Chen, Brunel}. These changes would modify the certainty conditions in our framework, as well as our experimental predictions. We could therefore assess candidate optimization principles and biological priors experimentally. While the base framework developed here was designed to identify crucial network connections required for function, we hope that our approach will eventually allow us to assess theoretical principles that determine how neural network structure follows from function. 
\vs

\section*{ACKNOWLEDGMENTS}
The authors thank Tianzhi (Lambus) Li, Srini Turaga, Andrew Saxe, Ran Darshan, and Larry Abbott for helpful discussions and comments on the manuscript. This work was supported by the Howard Hughes Medical Institute and the Janelia Visiting Scientist Program. 
\onecolumngrid
\renewcommand{\theequation}{A.\arabic{equation}}
\setcounter{equation}{0}
\section*{APPENDICES}
\noindent
{\bf A. A Certainty Condition to Pinpoint Synapses Required for Specified Response Patterns}\vs
{\bf Preliminaries}\vs
For completeness, we begin by briefly reviewing a few central concepts from the main manuscript.\vs
{\it From recurrent to feedforward networks: } Let us consider a neural network of \(\cI\) input neurons that send  signals to an interconnected population of \(\cD\) driven neurons governed by dynamical equations (\ref{eqn: RateEqn}), as described in the main manuscript.  At steady-state, since all time-derivatives are zero, (\ref{eqn: RateEqn}) yields
\begin{align} 
y_{\mu i} = \Phi\left( \sum_{m=1}^{\cD} w_{im}y_{\mu m} + \sum_{m=\cD+1}^{\cD+\cI} w_{im} x_{\mu ,m-\cD}\right)=\Phi\left( \sum_{m=1}^{\cN} w_{im}z_{\mu m}\RF\ , \label{eqn: SteadyStateEqn} 
\end{align}
where, as prescribed in the main manuscript,  \(y_{\mu i}\) and \(x_{\mu m}\) denote steady-state activity levels of the driven and input neurons to the \(\mu^{th}\) stimulus, which we have combined into $z_{\mu m}$, and $\cN$ is the number of incoming synapses onto each of the driven neurons. ({\ref{eqn: SteadyStateEqn}}) provides \(\cD\times \cP\) nonlinear equations for \(\cD\times \cN\) unknown parameters. However, we immediately notice that the steady-state activity of neuron $i$ depends only on the $i^{th}$ row of the connectivity matrix, so these equations separate into \(\cD\) independent sets of \(\cP\) equations with \(\cN\) unknowns, the weights onto a given driven neuron. In other words, the recurrent network involving $\cD$ driven  and $\cI$ input neurons decomposes into $\cD$ feedforward networks with $\cN=\cD+\cI$ feedforward inputs. The steady-state equations for these feedforward networks are given by,
\begin{align} 
y_{\mu} = \Phi\left( \sum_{m=1}^{\cN} z_{\mu m}w_{m}\right)\ , \label{eqn: SingleSteadyStateEqn} 
\end{align}
where we have now suppressed the $i$ index in $y_{\mu i}$ and in $w_{im}$. For this feedforward network we will refer the $i^{th}$ neuron as the target neuron, and it is as if that all the neurons (driven and input) are providing feedforward inputs to it. As long as we only consider exact solutions to the fixed point equations, the problem of identifying synaptic connectivity in a recurrent network reduces to solving the problem for feedforward networks. Thus in the rest of this appendix we will focus on identifying $w_m$'s satisfying (\ref{eqn: SingleSteadyStateEqn}). 

Note that the main text used the notation $z_{\mu m}^{(i)}$ to emphasize that the set of presynaptic neurons may depend on the target neuron, but we simply write $z_{\mu m}$ throughout the Appendices with the understanding that the formalism applies to a specified target neuron whose index is suppressed. Furthermore, for conceptual simplicity the main text first stated many results in a feedforward setting with a single driven neuron, but the Appendices immediately treat the general case where presynaptic partners may come from either the input or driven populations of neurons. 
\vs
{\it A convenient set of variables:} In all our discussions in this section the input neuronal response matrix, $z_{\mu m}$, will be assumed to be fixed. Note that $z_{\mu m}$ connects synaptic weight vectors to the target response vector and can be used to define $\cP$ weight combinations, the $\eta$-coordinates. Each $\eta$-coordinate controls the target response to a single stimulus condition:
\be
y_\mu=\Phi(\eta_\mu)\ ,\where\ \eta_{\mu}\equiv \sum_{m = 1}^{\cN}z_{\mu m} w_m\ .
\label{omega-defn}
\ee
It is rather convenient to extend this set of $\cP$ $\eta$-coordinates to a basis set of $\cN$ $\eta$-coordinates, such that all synaptic weights can be uniquely expressed as a linear combination of these $\eta$-coordinates, and vice versa. To see how this can be done, we will henceforth make the simplifying assumption that the $\cP\times\cN$ matrix has the maximal rank, $\cP$, although we anticipate that much of our framework, results, and insights will apply more generally. If $z$ has maximal rank, its kernel will be an \((\cN-\cP)\)-dimensional linear subspace spanned by \((\cN-\cP)\) orthogonal basis vectors, denoted by $\env_\mu$ for $\mu=\cP+1\dots \cN$. We can now extend $z$  to an $\cN\times\cN$ matrix, $Z$, as follows
\ba
Z_{\mu m}&=&z_{\mu m}\ \for\ \mu=1\dots \cP,\mand\ \forall\ m,\non
Z_{\mu m}&=&\en_{\mu m}\ \for\ \mu=\cP+1\dots \cN,\mand\ \forall\ m,
\ea
where $\en_{\mu m}$ is the $m^{th}$ component of the null vector $\env_\mu$. With this construction, it is easy to see that the new  $\eta$-coordinates,
\be
\eta_\mu\equiv\sum_{m = 1}^{\cN} Z_{\mu m}w_m\ , \for\ \mu =\cP+1,\dots,\cN\ ,
\ee
remain completely unconstrained by the specified response patterns, as these linear combinations do not contribute to any of the target responses. In contrast, the original $\eta$-coordinates,
\be
\eta_\mu=\sum_{m = 1}^{\cN} Z_{\mu m}w_m=\sum_{m = 1}^{\cN} z_{\mu m}w_m\ ,\for\ \mu =1,\dots,\cP\ ,
\label{omega-gen}
\ee
are all constrained by the data:
\be
\eta_\mu  \left\{ \by{ll}= y_\mu &\for\ \mu=1,\dots,\cR, \mx{the constrained dimensions}\\
\leq 0 &\for\ \mu=\cR+1,\dots,\cP, \mx{the semi-constrained dimensions.}
\ey
\Rd\ ,
\label{eqn: SemiFlat} 
\ee
where for notational simplicity we have ordered the response patterns such that $y_\mu\neq0$ only for $\mu=1,\dots,\cC$. Also, we extend the $y_\mu$'s to an $\cN$-dimensional vector, $\yv$, by assigning $y_\mu=0$ for $\mu=\cP+1\dots \cN$.  

The extended response matrix $Z$ defines a basis transformation connecting physical synaptic directions, $\eh_m$, with directions 
\be
\env_\mu\equiv \sum_{m = 1}^{\cN}\eh_m Z^{-1}_{m\mu}\ ,
\ee
along which the $\eta$-coordinates change. These $\env$ vectors clearly differentiate   directions in the weight space that are activity-constrained by neuronal responses ($\mu =1,\dots,\cP$) from those that are not ($\mu =\cP+1,\dots,\cN$). We can  express any weight vector in either the $\{\eh_m\}$ basis or the $\{\env_{\mu}\}$ basis:
\be
\vec{w}=\sum_{m = 1}^{\cN}w_m \eh_m=\sum_{\mu = 1}^{\cN}\eta_\mu \env_\mu\ ,  \where w_m =  \sum_{\mu = 1}^{\cN}  Z^{-1}_{m\mu}\eta_\mu\ , \ \eta_\mu =  \sum_{m = 1}^{\cN}  Z_{\mu m}w_m \ , \env_\mu\equiv \sum_{m = 1}^{\cN}\eh_m Z^{-1}_{m\mu}\ ,\ \ \eh_m= \sum_{\mu = 1}^{\cN} \env_\mu Z_{\mu m}\ .
\label{transformation}
\ee
For later convenience we also define the number of semi-constrained and unconstrained dimensions as, $\cS=\cP-\cR$, and $\cF=\cN-\cP$, respectively.\vs
{\bf Derivation of the certainty condition for orthogonal input patterns:} \vs
Our goal here is to use the solution space (\emph{i.e.} ensemble of weights that are precisely able to recover the specified target responses) to derive a condition for when we can be certain that a given synapse must be nonzero. For technical simplicity, we will specialize to the case when all the response patterns are orthonormal, \ie
\be
\sum_{m=1}^{\cN}z_{\mu m}z_{\nu m}=\da_{\mu \nu}\Leftrightarrow z z^T=I\ ,
\label{orthogonality}
\ee
where $I$ is the identity matrix. Then we can always choose the extended $Z$ matrix  to be an $\cN \times \cN$ orthogonal matrix, such that $Z^{-1}=Z^T$ and the $\env_\mu$ vectors now form an orthonormal basis. Motivated by biological constraints, we will impose a bound on the magnitude of the synaptic weight vector. For orthonormal response patterns, this translates into a spherical bound on $\eta$-coordinates as well (see Fig.~\ref{fig:projections}B)
\be
|\vec{w}|^2=\sum_{m=1}^{\cN} w_m^2=\sum_{\mu=1}^{\cN} \eta_\mu^2\leq W^2\ .
\label{sphere}
\ee
We refer to this $\cN$-dimensional ball, in which all admissable synaptic weights reside, as the weight-space.  
\vs
{\it A heuristic argument for $y$-critical:} Before diving into the rigorous and technical derivation, in this subsection we first try to intuitively understand how the certainty condition (\ref{ycritical}) can arise. For this purpose, let us start with a linear theory with no unconstrained dimension, so $\cS=\cF=0$. In this case, there is a unique set of weights  that can precisely reproduce the observed responses:
\be
w_m=\sum_{\mu=1}^\cN Z^{-1}_{m\mu}y_{\mu}=\sum_{\mu=1}^\cN Z_{\mu m}y_\mu\ .
\label{total-corr}
\ee
Since $Z_{\mu m}=z_{\mu m}$ represents the responses of the $m^{th}$ presynaptic neuron, the solution for the $m^{th}$ synaptic weight (\ref{total-corr}) is simply the correlation between the pre and post synaptic activity. In a linear theory, the sign of the synapse is thus dictated by the sign of the correlation between the pre and post synaptic neuron. 

Let us now allow a single ($\cN^{th}$) unconstrained direction. One can think of this situation as if we do not have the information on how the target neuron would respond to the  unconstrained stimulus pattern. If we knew that this response was say, $y_u$, then we would have been able to determine the sign of $w_m$:
\be
\sgn(w_m)=\sgn\LF\sum_{\mu=1}^{\cN-1} Z_{\mu m}y_\mu+Z_{\cN  m}y_u\RF\ .
\label{A13}
\ee
However, since we do not know what the last term is, if it can cancel the first term for some allowed value of $y_u$ then the overall sign becomes ambiguous. Conversely, $w_m$ becomes certain if 
\be
\left|\sum_{\mu=1}^{\cN-1} Z_{\mu m}y_\mu\right|>|Z_{\cN  m}y_u|\ \forall\ y_u\ .
\label{heuristic-cond}
\ee
Now, it is easy to recognize that the first term is just $\eh\cdot\yv=ye_y$, where we have suppressed the $m$ index on $\eh_m$ here to reduce notational clutter and will continue to do so while referring to the synapse direction whose sign we are considering\footnote{We do want to point out that in the main manuscript since we were introducing the various concepts and relevant quantities,  for clarity we did explicitly keep track of the $m$ index.}. $e_y=\eh\cdot\yh$ refers to the projection of $\eh$ along $\yh$. Also, note that in this simple case with one unconstrained direction, the projection of $\eh$ along the unconstrained subspace is just given by  $e_u=\eh\cdot\env_{\cN}=Z_{\cN m}$. Further, since $Z$ is orthogonal, in order to have any solution at all 
\be
y^2+y_u^2\leq W^2\ .
\label{yu}
\ee
Substituting the maximum $|y_u|$ from (\ref{yu}) into (\ref{heuristic-cond}), after some algebra we get the condition for sign certainty as
\be
y>\ycr=W\sqrt{e_u^2\over e_u^2+e_y^2}\ .
\label{heuristic}
\ee

The same argument applies if the $\cN^{th}$ direction is  semi-constrained instead of unconstrained, with  one notable difference. If the $\cN^{th}$ pattern was semi-constrained that means $y_\cN=0$, and the nonlinear thresholding is masking how the target neuron would have responded in a linear model\footnote{Note that our relation between the sign of the synapse and the sign of the correlation is based on a linear response.}. However, the ambiguity in sign can only arise if the second term has a sign opposite to the first term, or a sign opposite to $\sgn(e_y)$. Moreover, for the thresholding to act, the target response  for the semi-constrained pattern must be negative in the linear theory, so $Z_{\cN m}$ has to have the same sign as $e_y$ to generate the  ambiguity. And, if it is indeed so, then we obtain a certainty condition that is identical to (\ref{heuristic}) except that $e_u\ra e_s$, the projection of $\eh$ along the semi-constrained direction:
\be
y>\ycr=W\sqrt{e_s^2\over e_s^2+e_y^2}\ .
\label{heuristic-s}
\ee
If $Z_{\cN m}$ and $e_y$ have opposite signs, then the synapse always has the same sign throughout the solution space.

While this derivation of $\ycr$ is heuristic and only deals with a single semi-constrained or unconstrained dimension, it provides intuition for the general result (\ref{ycritical}). Essentially, whether the sign of a given synapse is constant across the solution space depends on two competing quantities: the correlation between the pre- and postsynaptic responses; and the strength of the postsynaptic drive for patterns where the target response is either unknown or masked by the thresholding nonlinearity. 
\vs
{\it Hyperplane dividing excitatory and inhibitory synaptic regions:} Having gained some intuition about the certainty condition, let us now proceed to a rigorous derivation of the result. Since the constrained coordinates are fixed for the weight vectors that belong to the solution space (the deep yellow wedge in Fig.~\ref{fig:projections}B, we must have
\be
y^2\equiv\sum_{\mu=1}^{\cR} y_\mu^2=\sum_{\mu=1}^{\cR} \eta_\mu^2 \ ,
\ee
so that the solution space  resides within an $(\cF+\cS)$-dimensional ball with radius 
\be
\overline{W}\equiv \sqrt{W^2-y^2}\ ,
\label{oa-sphere}
\ee
as depicted in Fig.~\ref{fig:projections}B, by the yellow region. We refer to this semi-constrained plus unconstrained subspace as the flexible subspace.

Now, the synaptic direction of interest, $\eh$, can be decomposed into its projections along constrained, semi-constrained and unconstrained subspaces. For notational simplicity, let us denote $e_\mu\equiv(\eh\cdot\env_\mu)= Z_{\mu m}$ as the component of $\eh$ along $\env_\mu$. Note that the second equality follows from the orthogonality assumption and (\ref{transformation}). In general, in this manuscript we will use subscripts on $e$ to denote projections of $\eh$ along different directions or subspaces.  We can now write\footnote{Again, we remind the readers that in the main manuscript these projected vectors were denoted by $\ch_m, \sh_m$ and $\uh_m$.}
\be
\eh=\sum_{\mu=1}^{\cN}e_{\mu}\env_\mu =\sum_{\mu=1}^{\cR}e_{\mu}\env_\mu +\sum_{\mu=\cR+1}^{\cP}e_{\mu}\env_\mu +\sum_{\mu=\cP+1}^{\cN}e_{\mu}\env_\mu =\cos\ta\ch+\sin\ta\cos\phi\sh+\sin\ta\sin\phi\nh\ ,
\label{decomposition}
\ee
where 
\ba 
\ch\equiv {\sum_{\mu=1}^{\cR}e_\mu\env_\mu\over\sqrt{\sum_{\mu=1}^{\cR}e_\mu^2 }}\ ,\ 
\sh\equiv {\sum_{\mu=\cR+1}^{\cP}e_\mu\env_\mu\over\sqrt{\sum_{\mu=\cR+1}^{\cP}e_\mu^2 }}\ ,
\nh\equiv {\sum_{\mu=\cP+1}^{\cN}e_\mu\env_\mu\over\sqrt{\sum_{\mu=\cP+1}^{\cN}e_\mu^2 }}\ ,
\ea
are unit vectors that lie within the constrained, semi-constrained and unconstrained subspaces, and
\ba
e_c\equiv\ch\cdot\eh=\cos\ta&=&\sqrt{\sum_{\mu=1}^{\cR}e_\mu^2 }\geq 0\ ,\non
e_s\equiv\sh\cdot\eh=\sin\ta\cos\phi&=& \sqrt{\sum_{\mu=\cR+1}^{\cP}e_\mu^2 }\geq 0\ ,\non
e_u\equiv\nh\cdot\eh=\sin\ta\sin\phi&=& \sqrt{\sum_{\mu=\cP+1}^{\cN}e_\mu^2 }\geq 0\ 
\label{theta-phi}
\ea
are the projections of $\eh$ along these directions. One could think of $\ta,\phi$ as representing a spherical coordinate system where the role of $x, y$ and $z$ axes are played by $\sh, \nh$ and $\ch$ respectively, and our definitions (\ref{decomposition} - \ref{theta-phi}) imply  the convention, $0\leq \{\ta,\phi\}< \pi/2$. For later convenience, let us also introduce the projection of $\eh$ onto the activity-constrained subspace:
\be
e_p\equiv \ph\cdot\eh=\sqrt{\cos^2\ta+\sin^2\ta\cos^2\phi}=\sqrt{\sum_{\mu=1}^{\cP}e_\mu^2 }\geq 0\ ,\where 
\ph\equiv {\sum_{\mu=1}^{\cP}e_\mu\env_\mu\over\sqrt{\sum_{\mu=1}^{\cP}e_\mu^2 }}\ .
\ee
We would also like to emphasize that we can compute $\ta,\phi$ just from the knowledge of the neuronal responses, $z_{\mu m}=e_\mu$, which is particularly useful for numerical calculations:
\be
\ta\equiv\cos^{-1}\LF\sqrt{\sum_{\mu=1}^{\cR}z_{\mu m}^2 }\RF\ ,\mand\ \phi\equiv\cos^{-1}\LF \sqrt{\sum_{\mu=\cR+1}^{\cP}z_{\mu m}^2 }\middle/\sqrt{1-\sum_{\mu=1}^{\cR}z_{\mu m}^2}\RF\ . 
\ee

\begin{figure}[!thbp]
	\includegraphics[width=0.96\textwidth,angle=0]{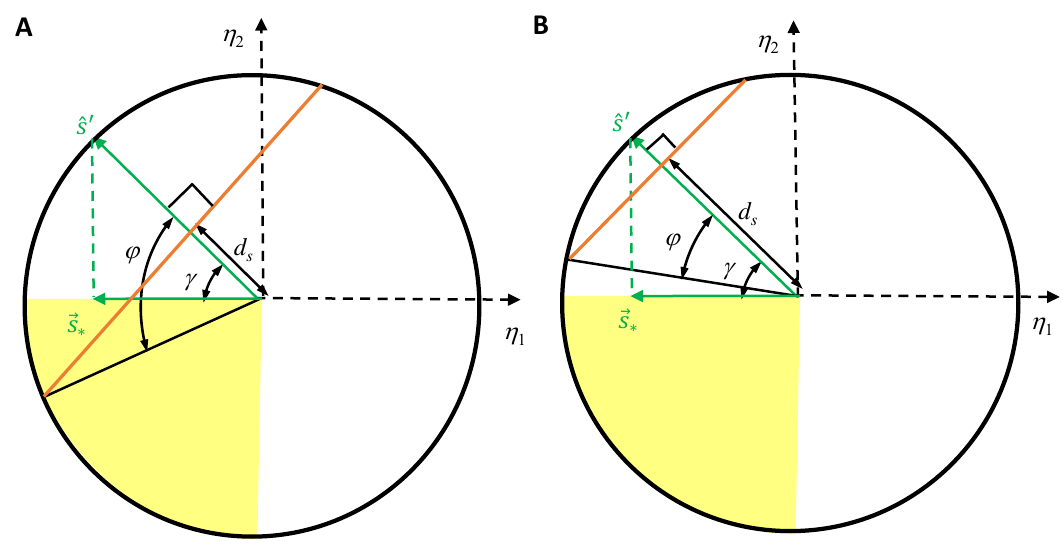}
	\caption{{\small {\bf Cartoons depicting the orientation of the semi-constrained projection of a given synaptic weight direction ($\sh\,'$) within the semi-constrained subspace and its impact on determining the sign of the given weight.} In these plots, the yellow wedges represent the solution space, $\eta_1,\eta_2\leq0$.  $d_s$ is the distance of the $w_m=0$ orange line  (hyperplane in higher dimension) from the origin. If $d_s$ is small, as in the left plot ({\bf A}), the projection angle $\ga$ is smaller than  $\vphi$, half of the angle subtended by the orange line to the origin, and therefore the orange line and the yellow cone intersect. This means that solutions with both positive and negative $w$'s are present. In the right plot ({\bf B}), $d_s$ is sufficiently large such that $\ga>\vphi$ and consequently, all the solutions must have consistent sign.
			\label{fig:semiflat}}}
\end{figure}

Now, any weight vector in the  solution space can be written as
\be
\wv=\yv+\wv_s+\wv_u\ ,
\label{basis-dec}
\ee
where $\vec{w}_s$ and $\wv_u$ are the projections of $\vec{w}$ onto the semi-constrained and unconstrained subspaces, and the constrained part of $\wv$ is fixed at $\vec y$. Using (\ref{decomposition}) and (\ref{basis-dec}), one then finds that the $w_m=0$ hyperplane dividing the excitatory and inhibitory regions in the flexible subspace satisfies the equation
\ba
w=\wv\cdot\eh=y \cos\ta\cos\al +\sin\ta\cos\phi\ \sh\cdot\vec{w}_s+\sin\ta\sin\phi\ \nh\cdot\wv_u=0\ ,
\label{flat-plane}
\ea
where we have now also suppressed the index $m$  in $w_m$. Also, we have defined $\al\in [0,\pi]$ to be the angle between $\vec{y}$ and $\ch$. We now notice that the origin of the flexible subspace, $\wv_s=\wv_u=0$, is in the solution space and the sign of $w$ for this solution point is given by
\be
\sgn(w)=\sgn(\cos\al)=\sgn(e_y)=\sgn\LF \sum_{\mu=1}^{\cR}z_{\mu m}y_\mu\RF\ .
\ee
In other words, if the sign of the synapse is certain, this certain sign must be $\sgn(\cos\al)$, which corresponds to the sign of the correlation between the target neuron and the  presynaptic neuron. Intuitively, positive correlations point to an excitatory connection, and negative correlations point to an inhibitory connection.
\vs
{\it Special case without unconstrained dimensions:} To derive the certainty condition, let's start by looking at the case when $\cP=\cN$, so that there are no unconstrained directions, or equivalently, $\phi=0$. In this case, the solution space is just the all-negative orthant in the $\cS$-dimensional semi-constrained hypersphere (Fig.~\ref{fig:semiflat}), and the equation for the $w=0$ hyperplane can be written as 
\be
\sin\ta\ (\sh\,'\cdot\vec{w}_s)=y \cos\ta|\cos\al|\ ,
\label{s-plane}
\ee
where the right hand side is positive, and we have introduced 
\be
\sh\,'\equiv -\mx{Sgn}(\cos\al)\sh\ ,
\ee
which flips the direction of $\sh$ if $\cos\al>0$, or equivalently, if $e_y>0$.
Now, if the  $w=0$ hyperplane (orange lines in Fig.~\ref{fig:semiflat})  is far enough along $\sh\,'$ from the origin that it does not intersect with the all-negative orthant within the weight bounds, then we can be certain that $w$ is nonzero and always has a consistent sign. To check this, we need to compare the cone angle that the orange hyperplane  subtends at the center, $\vphi$, with the minimum angle, $\ga$, that the $\sh\,'$ vector makes with the all-negative orthant. 

First, $\vphi$ can easily be inferred from trigonometry:
\be
\cos\vphi={d_s\over \overline{W}}={y \cot\ta|\cos\al|\over\sqrt{W^2-y^2} }\ ,
\label{varphi}
\ee
where $d_s=y \cot\ta|\cos\al|$ represents the distance from the center of the semi-constrained sphere to the hyperplane. The expression for $d_s$  follows from the general mathematical result that if,
\be
\vec\bt\cdot \vec x+\bt_0=0\ ,
\ee
is an equation for a hyperplane, where $\vec x$ denotes the coordinate vector and $\vec\bt, \bt_0$ are constants, then the perpendicular distance, $d_\perp$,  to it from a point $\vec x=\vec{\za}$ is given by
\be
d_\perp={|\vec\bt\cdot \vec\za+\bt_0|\over |\vec\bt|}\ .
\label{dperp}
\ee
Note, we are interested in the distance from the origin, $\vec{\za}=0$, to the hyperplane satisfying the equation (\ref{s-plane}), so $\vec\bt=\sin\ta\ \sh\,'\Ra |\vec\bt|=\sin\ta$ and $\bt_0=-y \cos\ta|\cos\al|$.

To provide a geometric intuition for $\ga$, let us first assume that $\sh\,'$ doesn't point into the all-negative orthant. If we can find the projection of $\sh\,'$ on the correct boundary of the solution space, then $\ga$ will be given by the angle between $\sh\,'$ and the appropriate semi-constrained boundary vector, $\vec s_\x$ (Fig.~\ref{fig:semiflat}). Since all the components in the solution space (all-negative orthant) have to be negative or zero, to find the appropriate projection vector of  $\sh\,'$ onto the boundary of solution space, we essentially have to set all the positive components  to zero:
\be
\vec{s}_\x= \sum_{\mu=\cR+1}^{\cP}s'_{\mu}\Ta(-s'_{\mu})\env_{\mu}=\mp\sum_{\mu=\cR+1}^{\cP}s_\mu\  \Ta(\pm s_{\mu})\env_{\mu}\ ,
\label{vecsx}
\ee
depending upon whether $\sgn(e_y)=\pm$. Here $s'_\mu, s_\mu$ are just  the $\mu^{th}$ components of $\sh\,'$ and $\sh$ vectors, and $\Ta(x)$ is the Heaviside step function, which is one if $x$ is positive and zero otherwise. Then, $\ga$ is given by
\be
\cos\ga=\sh\,'\cdot\widehat{s}_\x={\sh\,'\cdot\vec{s}_\x\over |\vec{s}_\x|}=\sqrt{ \sum_{\mu=\cR+1}^{\cP}s_{\mu}^{'2}\ \Ta(-s'_{\mu})}=\sqrt{\sum_{\mu=\cR+1}^{\cP}s_{\mu}^2\ \Ta(\pm s_{\mu})}=|\vec s_\x|\ ,
\label{gamma}
\ee
where again the sign in $\Ta$ is determined by the sign of $e_y$.

A formal way to see that $\ga$ is indeed given by (\ref{gamma}) is to start with any unit vector, $\wv_s$, lying in the solution space. Then, the angle, $\ga$, between $\sh\,'$ and $\wv_s$ is given by
\be
\cos\ga=\sum_{\mu=\cR+1}^\cP s'_\mu w_{s\mu}=\sum_{\mu\in A_+} s'_\mu w_{s\mu}+\sum_{\mu\in A_-} s'_\mu w_{s\mu}\ ,
\ee
where we have defined $A_{\pm}$ as the set of all $\mu$ indices for which $s'_\mu$ is positive/negative, respectively. Since $\wv_s$ is in the solution space, $w_{s\mu}\leq 0$, and therefore the second term sums positive quantities while the first term subtracts. Thus, 
\be
\cos\ga\leq \sum_{\mu\in A_-} s'_\mu w_{s\mu}=\vec s_\x\cdot \wv_s\leq |\vec s_\x|\ ,
\ee
where both the equalities are achieved when $\wv_s$ is aligned with the boundary semi-constrained vector, $\vec s_\x$, or $\wv_s=\sh_\x$, as  argued previously. Note also, that this formal proof didn't assume any restrictions on $\sh\,'$ direction and thus (\ref{gamma}) turns out to be a general result that also holds if $\sh\,'$ points into the all-negative orthant. 

Combining (\ref{gamma}) and (\ref{varphi}), the certainty condition now reads 
\be
\vphi<\ga\Ra{y^2\cot^2\ta \cos^2\al\over W^2-y^2}> \cos^2\ga\Ra y>\ycr \equiv W\sqrt{\cos^2\ga\sin^2\ta\over \cos^2\al\cos^2\ta+\cos^2\ga\sin^2\ta}\ .
\label{cc-semiflat}
\ee 
\vs
{\it General case with unconstrained dimensions:} We can extend the above analysis to the case when we have unconstrained dimensions by noting that, for a given set of unconstrained coordinates, the solution space is again the all-negative orthant in a semi-constrained hypersphere. Isometry along unconstrained dimensions ensures that it is always possible to make one of the null directions, lets say $\env_{\cN}$,  align with $\nh$. Then, the $w=0$ hyperplane equation (\ref{flat-plane}) reads
\be
w=\sin\ta\cos\phi\ \sh\cdot\vec{w}_s+y \cos\ta\cos\al+\eta_{u}\sin\ta\sin\phi=0\ ,
\ee
which can be rewritten as
\be
\sin\ta\cos\phi\ \sh\,'\cdot\vec{w}_s=y \cos\ta|\cos\al|-\eta'_{u}\sin\ta\sin\phi\ ,
\label{semi-ds}
\ee
where we have introduced $\eta'_{u}= -\sgn(\cos\al)\eta_{u}$. To have a certain synapse, the $w=0$ hyperplane cannot intersect the solution space for any allowed value of $\eta'_{u}$. 

The direction of $\sh\,'$ is independent of the unconstrained coordinates and hence the value of $\ga$ remains unchanged. However, the cone-angle, $\vphi$, does depend on the unconstrained coordinates in two ways. Firstly, the radius, $\w{W}$, of the $\cS$-dimensional spherical subspace containing admissible solutions is now:
\be
\w{W}=\sqrt{W^2-y^2-\eta_\perp^2-\eta_{u}^{'2}}
\label{semi-radius}
\ee
where $\eta_\perp$ is the magnitude of the weight-vector in the $(\cF-1)$ dimensional  subspace that is perpendicular to $\env_{\cN}=\nh$. We note in passing that  (\ref{semi-radius}) implies, $\eta_\perp,\eta'_{u}\leq \sqrt{W^2-y^2}=\e{W}$. Secondly, the distance of the hyperplane from the origin that follows from (\ref{semi-ds}) is now a function of $\eta'_{u}$: 
\be
d_s={y |\cos\ta\cos\al|-\eta'_{u}\sin\ta\sin\phi\over\sin\ta\cos\phi}\ .
\label{ds-eqn}
\ee

Strictly speaking, this expression for the distance is only valid as long as the numerator in the $d_s$ expression stays positive. However, if there exists an allowed $\eta'_{u}\leq \e{W}$ (let's call it $\eta_{u0}$) for which the numerator can vanish, that would mean that the synapse cannot have a certain sign, because at that point $d_s=0$, the hyperplane intersects the origin, and the weight can vanish even for a linear theory. In fact, the  $d_s=0$ condition provides us with the $y$-critical value below which the synapse sign becomes uncertain in a linear theory: 
\be
\eta_{u0}={y \cos\ta|\cos\al|\over \sin\ta\sin\phi}\leq \overline{W}\Ra y_{\mt{cr,lin}}=W\sqrt{\sin^2\ta\sin^2\phi\over \cos^2\ta\cos^2\al+\sin^2\ta\sin^2\phi}\ .
\label{yc-linear}
\ee
So we will now look into cases when $y\geq y_{\mt{cr,lin}}$ which also means that  (\ref{ds-eqn}) will remain valid.

Combining (\ref{semi-radius}) and (\ref{ds-eqn}) we get
\be
\cos\vphi={d_s\over \w{W}}={y |\cos\ta\cos\al|-\eta'_{u}\sin\ta\sin\phi\over\sin\ta\cos\phi\sqrt{W^2-y^2-\eta_\perp^2-\eta_{u}^{'2} }}\ .
\label{cosvphi}
\ee
In order for us to be certain that $w$ is nonzero, we have to make sure that even the largest $\vphi$ that one can obtain by varying $\eta_\perp$ and  $\eta'_{u}$ is still smaller than $\ga$. Clearly, to make $\vphi$ large it is best to make $\eta_\perp=0$. Also, it is clear from inspection that $\cos\vphi$ starts to initially decrease as $\eta'_{u}$ increases from zero, being dominated by the linear term. However, as the quadratic term in $\eta'_{u}$ in the denominator becomes more and more important, $\cos\vphi$ reaches a minimum and starts to increase.  
Imposing $d\cos\vphi/d\eta'_{u}=0$, we can find that this minimum  is reached at
\be
\eta'_{u} = {\sin\ta\sin\phi\ (W^2-y^2)\over y \cos\ta|\cos\al|}=\e{W}\sqrt{\LF W/y\RF^2-1\over \LF W/ y_{\mt{cr,lin}}\RF^2-1}\leq \e{W}\ ,
\ee
where we substituted $y_{\mt{cr,lin}}$ from (\ref{yc-linear}) and used the fact that $W\geq y\geq y_{\mt{cr,lin}}$ to obtain the inequality. This proves that the minimum $\cos\vphi$ indeed occurs at an allowed positive value of $\eta'_{u}\leq\e{W} $. 
Substituting the above $\eta'_{u}$ in (\ref{cosvphi}),  after some algebra we  find that this minimum value of $\cos\vphi$, or equivalently the maximum $\vphi$, is given by 
\be
\cos\vphi=\sqrt{y^2\cos^2\ta\cos^2\al- (W^2-y^2)\sin^2\ta\sin^2\phi\over (W^2-y^2)\cos^2\phi\sin^2\ta}\ .
\ee
The certainty condition then requires 
\be
\cos^2\vphi={y^2\cos^2\ta\cos^2\al- (W^2-y^2)\sin^2\ta\sin^2\phi\over (W^2-y^2)\cos^2\phi\sin^2\ta}>\cos^2\ga\ ,
\label{cc-gen}
\ee 
which can be recast as 
\ba 
y>\ycr \equiv W\sqrt{\cos^2\ga\sin^2\ta\cos^2\phi+\sin^2\ta\sin^2\phi \over \cos^2\al\cos^2\ta+ \cos^2\ga\sin^2\ta\cos^2\phi+\sin^2\ta\sin^2\phi }\ ,
\label{ycrappendix-angles}
\ea

It is illuminating to express $y$-critical in terms of the projections, $e_y, e_u, e_{s\x}$, of the synaptic direction, $\eh$, respectively along the data vector, $\yh$, the unconstrained unit vector, $\nh$, and the  semi-constrained boundary vector, $\vec{s}_\x$:
\ba 
\ycr =W\sqrt{e_{s\x}^2+e_u^2 \over e_y^2+ e_{s\x}^2+e_u^2 }\ ,
\label{ycrappendix}
\ea
where 
\be
e_y\equiv \eh\cdot\yh= {\sum_{\mu=1}^{\cR}y_\mu e_\mu \over \sqrt{\sum_{\mu=1}^{\cR}y_\mu^2 } }=\cos\ta\cos\al\ ,\  e_{s\x}\equiv \eh\cdot\sh_{\x}=  \sqrt{\sum_{\mu\in A_-}e_\mu^2 }=-\sgn(\cos\al)\sin\ta\cos\phi\cos\ga\ ,
\label{es-star}
\ee
and $e_u$ is given by (\ref{theta-phi}). 
We note that setting $\phi=0$ precisely reproduces the correct limit with no unconstrained directions (\ref{cc-semiflat}). 
\vs
{\bf Regarding orthogonal input patterns in recurrent networks}
\vs
While our analysis of the solution space and the certainty condition  (\ref{ycrappendix}) translate directly to recurrent networks, the requirement of orthogonality for the derivation of our certainty condition imposes certain technical restrictions on its scope  when it comes  to recurrent neural networks.  

The certainty condition we derived for feedforward networks can be applied to two different recurrent neural network set ups. First, let us consider networks where neurons have self-couplings. A consequence of having orthogonal response patterns in this case is that the certainty condition can only be satisfied for self-couplings $w_{ii}$, as long as $W\geq 1$. 
This is because the imposition of orthogonality in response patterns also restricts the correlation between the target neuron and the other neurons: 
\be
\sum_{m=1}^\cN Z_{\mu m}Z_{\nu m}=\da_{\mu\nu}\Ra \sum_{\mu=1}^\cN Z_{\mu m}Z_{\mu n}=\da_{m n}\ .
\ee
However, for the synapse-sign to be certain, the responses of pre and postsynaptic neurons need to be correlated.  To see the problem more quantitatively, suppose we are interested in constraining the synapse from the $m^{th}$ neuron onto the $i^{th}$ neuron, as before. Now,  the first $\cP$ elements of the unit vectors, $\eh_i$ and $\eh_m$ contain the responses of the $i^{th}$ and the $m^{th}$ neuron in the $\cP$ patterns. We have already derived a decomposition of $\eh_m$ in terms of its projections onto the constrained, semi-constrained and unconstrained subspaces (\ref{decomposition}). Similarly, $\eh_i$ can be decomposed as 
\ba
\eh_i= y \yh+\sqrt{1-y^2}\yh_{\perp} \,
\label{ei-decomposition}
\ea  
where $\yh$ lies entirely along the constrained directions, and $\yh_{\perp}$ is orthogonal to it and only has  components along unconstrained directions. Then, orthogonality implies
\be
\eh_i\cdot\eh_m=y\cos\ta\cos \al+\sqrt{1-y^2}\sin\ta\sin\phi\ (\yh_{\perp}\cdot\nh)=0\Ra \sin\ta\sin\phi=-{y\cos\ta\cos \al\over (\yh_{\perp}\cdot\nh)\sqrt{1-y^2}}\ .
\label{ortho}
\ee
Starting from the certainty condition (\ref{ycrappendix}), we can now go through a sequence of (in)equalities:
\ba
y^2&>&W^2{\sin^2\ta\sin^2\phi+\sin^2\ta\cos^2\phi\cos^2\ga \over\sin^2\ta\sin^2\phi+\sin^2\ta\cos^2\phi\cos^2\ga+\cos^2\ta\cos^2 \al} 
\geqslant  {W^2\sin^2\ta\sin^2\phi \over\sin^2\ta\sin^2\phi+\cos^2\ta\cos^2 \al} \non
\label{inequality}
&=& {W^2y^2\cos^2\ta\cos^2\al \over y^2\cos^2\ta\cos^2\al+(1-y^2)(\yh_{\perp}\cdot\nh)^2\cos^2\ta\cos^2\al}= {W^2y^2 \over y^2+(1-y^2)(\yh_{\perp}\cdot\nh)^2}\ \geq W^2 y^2 ,
\ea
where we substituted $\sin\ta\sin\phi$ from (\ref{ortho}). Note that the RHS is minimized when $\nh$ and $\yh_{\perp}$ are either aligned or anti-aligned. Even in this case, RHS = $W^2y^2$, and thus the certainty condition cannot be satisfied if $W\geq1$. One can check that when $i=m$, because the RHS in the first equation of (\ref{ortho}) is one and not zero, no similar constraints appear. Indeed, the certainty condition may be satisfied depending upon the specific response patterns. 

As a second possibility, suppose that no self-couplings are present. Then to be able to apply our framework and determine the couplings $w_{im}$ for a given $i$, we only need the truncated row vectors of $z$ whose $i^{th}$ column entry is absent, to be orthonormal. Therefore, the response of the $i^{th}$ driven neuron, which consists of the entries of the $i^{th}$ column, can now be chosen independently from the responses of its input neurons. In other words,  $\eh_i$  and   $\eh_m\ , m\neq i$,  no longer need to satisfy orthogonality constraint of (\ref{ortho}). Consequently, the  $w_{im}$ weights can indeed satisfy the certainty condition, just as in the feedforward case. 
\vs
{\bf Implied conservative bound on the certainty condition for non-orthogonal input patterns}\vs
A complete treatment of the certainty conditions for non-orthogonal fixed point patterns is beyond the scope of this work. However, here we provide some preliminary results and insights by explaining how our formalism for analyzing orthogonal fixed-point patterns can be simply adapted to derive an exact, but conservative, upper bound for $y$-critical that applies to general sets of patterns. 

Conceptually speaking, deriving the certainty condition amounts to determining when the $w_m = 0$ hyperplane intersects the solution space within the sphere of weight vectors with norm at most $W$. Because the solution space is exceedingly simple in $\eta$-coordinates, our orthogonal analysis used $\eta$-coordinates to conveniently recast the equations for the bounding sphere and $w_m = 0$ hyperplane. To adapt this analysis to the non-orthogonal case, it's important to account for three important changes to the geometry of the problem. Most fundamentally, the $\env_\mu$ directions corresponding to $\eta$-coordinates are no longer orthonormal. However, the mathematical notions of orthogonality and normality are implicitly defined with respect to the inner-product structure imposed on the vector space, and our geometrical calculations from the orthogonal case easily carry over to the non-orthogonal case if we redefine the inner product structure of the weight space to give an orthonormal coordinate system with respect to the $\eta$-coordinates rather than the physical coordinates\footnote{In particular, for an orthogonal $Z$ matrix, the two inner product structures defined via $\eh_m\cdot\eh_n=\da_{mn}$ and $\env_{\mu}\cdot\env_{\nu}=\da_{\mu\nu}$ are equivalent, but this is not the case when $Z$ is non-orthogonal. Although one would conventionally adopt the first inner product structure, both the derivation and interpretation of the conservative $y$-critical formula is easier in terms of the latter inner product structure, which makes all of the response pattern directions orthonormal by definition.}. In practice, all that this will entail is interpreting the $\eta$-coordinates as if they define coordinates along orthogonal axes, and we will never need to explicitly write down the associated inner product. Second, the equation for the weight bound in the orthogonal system of $\eta$-coordinates is elliptical around the origin, rather than spherical. However, for any ellipse one can find a sphere that just encompasses it. If we can find the radius of this bounding sphere, then one can look for an intersection anywhere within this sphere and our geometrical approach for deriving (\ref{ycrappendix}) will carry over and provide a conservative bound for $y$-critical. This bound will poorly approximate the true $y$-critical when some axes of the ellipse are much longer than others. Third, the normal vector to the $w_m=0$ hyper-plane is no longer $\hat e_m$ in the orthogonal system of $\eta$-coordinates. Therefore, the projections of $\hat e_m$ in (\ref{ycrappendix}) must be generalized to become projections of the hyperplane's normal vector. 

To obtain the radius of the bounding sphere, consider the SVD decomposition of the data-matrix:
\be
z=L \La R^T\ ,
\ee
where $L$ and $R$ are $\cP\times\cP$ and $\cN\times\cN$ orthogonal rotation matrices, and $\La$ is a $\cP\times \cN$ rectangular diagonal matrix whose only nonzero entries are given by
\be
\La_{a a}=\la_a \ge 0 \ ,\ a=1\dots \cP\ .
\ee
Note that $\la_1,\cdots,\la_\cP$ are called the singular values of $z$. We can now define rotated coordinates:
\be
\eta'=L^T\eta\ ,\mand w'=R^Tw\ .
\ee
so that
\be
\eta'=L^T L \La R^T w=\La w'\Ra \eta'_a =\la_a w'_a\ \forall \ a=1\dots \cP.
\ee
Note that $\eta$ and $\eta'$ are $\cP$-vectors in the current notation. We also note that since $w'$ is just a rotation of the original synaptic coordinates, the biological bound doesn't change as we go from $w$ to $w'$ coordinates:
\be
\sum_a w_a'^2=\sum_m w_m^2\leq W^2\ .
\ee
This makes it possible to find an inequality in terms of the $\eta'$ coordinates:
\be
W^2\geq \sum_{a=1}^{\cN} w_a'^2=\sum_{a=1}^{\cP} {\eta^{'2}_a\over \la_a^2} +\sum_{a=\cP+1}^{\cN} w_a'^2\geq {1\over \la_{\max}^2}\sum_{a=1}^{\cP} \eta^{'2}_a +\sum_{a=\cP+1}^{\cN} w_a'^2={1\over \la_{\max}^2}\sum_{\mu=1}^{\cP} \eta^{2}_\mu +\sum_{a=\cP+1}^{\cN} w_a'^2\ ,
\ee
where in the last step we have used the fact that the orthogonal matrix $L$ doesn't change the L2-norm as one goes from $\eta'$ to $\eta$ coordinates, and $\la_{\max}$ is defined as the maximal singular value. To obtain spherical symmetry, we thus define unconstrained coordinates via
\be
\eta_\mu=\la_{\max}w'_\mu \ ,\forall\ \mu>\cP\ ,
\ee
such that
\be
W^{'2}\equiv \la_{\max}^2W^2\geq \sum_{\mu=1}^{\cN} \eta^{2}_\mu\ .
\ee
We now realize that the problem of finding $y$-critical using this conservative bound can be recast into the problem of the orthogonal case: As we just described, the $\eta$-coordinates satisfy the conservative spherical bound. Our goal can then be to find the minimum value of $y$ for which the hyperplane satisfying, $w_m=0$, does not intersect the solution space. Now, if we define $Z$ to be, as in the orthogonal case, a full rank extension\footnote{One can equivalently obtain $Z$ as 
	\be
	Z=\w{L}\w{\La}R\ ,
	\ee
	where $\w{\La}$ is now an $\cN\times\cN$ diagonal matrix with same entries as $\La$ and $ \w{\La}_{\mu\mu}=1/\la_{\max}$ for $\mu>\cP$, and $\w{L}$ is the an $\cN$-dimensional extension of the rotation matrix $L$ where the $(\cN-\cP)$-dimensional block is an identity and there is no mixing between the unconstrained and the activity-constrained coordinates.}  of $z$:
\ba
Z_{\mu m}= 
\left\{\by{lr}z_{\mu m} \ ,&\forall\ \mu\leq \cP\\ 
 \la_{\max}R^T_{\mu m} \ ,&\forall\ \mu> \cP
\ey\Rd\ ,
\ea
so that for all value of $\mu$, 
\be
\eta_\mu= \sum_{m=1}^\cN Z_{\mu m}w_m\ ,
\ee
then the hyperplane equation can be re-written as
\be
w_m=\sum_{\mu=1}^{\cN} Z^{-1}_{m\mu}\eta_\mu=0\ \ .
\label{nonortho-plane}
\ee
From (\ref{nonortho-plane}) it is clear that
\be
\nv_m=\sum_{\mu=1}^{\cN}Z^{-1}_{m\mu}\env_{\mu}\ ,
\ee
is perpendicular to the $w_m=0$ hyperplane, where we remind the readers that $\env_\mu$'s were defined by (\ref{transformation}). $\hat{n}_m$ thus plays the role of $\eh_m$ whose orientation with respect to the constrained, semi-constrained and unconstrained dimensions determines the certainty condition. Specifically, 
\ba 
\ycr =W'\sqrt{n_{s\x}^2+n_u^2 \over n_y^2+ n_{s\x}^2+n_u^2 }=W\la_{\max}\sqrt{n_{s\x}^2+n_u^2 \over n_y^2+ n_{s\x}^2+n_u^2 }\ ,
\label{ycr-nonortho}
\ea
where the various projections of $\hat{n}_m$ are given by
\be
n_y\equiv  {\sum_{\mu=1}^{\cR}y_\mu Z^{-1}_{m\mu} \over \sqrt{\sum_{\mu=1}^{\cR}y_\mu^2 } \sqrt{\sum_{\mu=1}^{ \cN}(Z^{-1}_{m\mu})^2 }}\ ,\  n_{s\x}\equiv {\sqrt{\sum_{\mu\in A_-}(Z^{-1}_{m\mu})^2} \over \sqrt{\sum_{\mu=1}^{ \cN}(Z^{-1}_{m\mu})^2 }}\ ,\mand n_{u}\equiv {\sqrt{\sum_{\mu=\cP+1}^{ \cN}(Z^{-1}_{m\mu})^2}\over \sqrt{\sum_{\mu=1}^{ \cN}(Z^{-1}_{m\mu})^2  }}\ .
\label{nonortho-proj}
\ee
We remind the readers that in the orthogonal case the set $A_-$ contained each semi-constrained $\mu$ index along which the component of $\eh_m$ had the same sign as $e_y$.  Similarly, here $A_-$, contains those semi-constrained $\mu$ indices along which the component of $\nv_m$ have the same sign as $n_y$. To summarize, the above analysis suggests that both $Z$ and $Z^{-1}$ will play an important role in generalizing  the certainty condition to nonorthogonal patterns, and especially the relative orientation of  $\hat{n}_m$ (defined by $Z^{-1}$) with respect to the $\eta$-directions.
\vs
{\bf B. Estimating the Probability that a Synapse is Certain in Large Feedforward Networks:}
\renewcommand{\theequation}{B.\arabic{equation}}
\setcounter{equation}{0}
\vs
For given values of $\cN(=\cI), \cR, \cP$ in a feedforward setting we will here try to assess how likely is it that noiseless orthonormal neuronal responses require a given synapse to be nonzero. As we have seen (\ref{yc-angles}), whether a synapse is certain to exist depends on six parameters, $\ta, \phi, \ga, \al, W,$ and $y$. The first four quantities depend on how $\eh$ is oriented with respect to various  directions in the weight space. Since $\eh$ is a unit vector, typically we expect its component along any given direction to be $\sim\cO(1/\sqrt{\cN})$. Thus, we typically expect
\be
e_y^2=\cos^2\ta\cos^2\al\sim {1\over \cN}\ ;\ e_s^2=\sin^2\ta\cos^2\phi\sim {\cS\over \cN}\ ;\ e_u^2=\sin^2\ta\sin^2\phi\sim {\cF\over \cN}\ \mand\ e_{s\x}^2 =\cos^2\ga\ e_s^2\sim {\cS\over 2\cN}\ .
\label{average}
\ee
Hence we approximate the typical $\ycr$ as
\be
\ycr=W\sqrt{\cS/(2\cN)+\cF/\cN \over 1/\cN+ \cS/(2\cN)+\cF/\cN }=W\sqrt{\cS+2\cF \over 2 +\cS+2\cF }\ .
\label{ycr-appr}
\ee 
Let us now suppose that all the dimensions scale with the network size, such that
\be
\cS=\sa\cN,\mand \cF=\un\cN\ .
\ee
Then, we find that as the network size increases $\ycr$ behaves as
\be
{\ycr\over W}\approx 1-\LF{1\over \sa+2\un}\RF{1\over \cN}\ ,
\ee
and $\ycr$ is essentially pushed up towards $W$. 

However, the typical scale of $y$ behaves similarly as the dimensions increase. To see this concretely, let us define $\yv_{cons}\equiv \{y_{\mu}|\mu=1\dots \cR\}$ as a $\cR$-dimensional vector, and assume that every possible $\yv_{cons}$ is equally likely within a sphere of radius $W$ (larger activity levels of the target neuron admit no solutions). Then the average and median values of $y=|\yv_{cons}|$ are given by
\ba
&<y>&={\int_0^{W} dy y^{\cR}\over \int_0^W dy y^{\cR-1}}=W\LF{\cR\over \cR+1}\RF\non
\Ra&{<y>\over W}&=\frac{1}{1+1/(\eta \cN)}\approx 1-{1\over\eta\cN}\ ,\where \eta\equiv{\cR\over \cN}\ ,\non
\mand& &{\int_0^{y_M} dy y^{\cR-1}\over \int_0^W dy y^{\cR-1}}={1\over 2}\Ra \frac{y_M}{W}=\LF{1\over 2}\RF^{1\over \cR}\non
\Ra&{y_M\over W}&\approx 1-{\ln 2\over\eta\cN}\ ,
\label{avg-median}
\ea
respectively. Since $y$ and $\ycr$ scale similarity as one increases the network size, the probability of a synapse being certain should not change as the network size increases. In Fig.~\ref{fig:highDsims}C, we show that if we choose, $y=1-\ln 2/\cR$, as the approximate median value in simulations with random input-output configurations (see Appendix~F for details), then the number of certain synapses does indeed increase linearly with $\cN$.

To quantitatively estimate the probability of finding a certain synapse, we can compute the fraction of volume of $\yv_{cons}$'s for which the synapse is certain for the typical projections (\ref{ycr-appr}), as compared to the volume of $\yv_{cons}$'s for which solutions to the steady-state equations exist. We know that  $\yv_{cons}$ has to lie within a $\cR$-dimensional sphere of radius $W$ in order for there to be any solutions to the problem\footnote{The allowed $\yv_{cons}$ must also lie in the all positive orthant, but as we will compute the ratio of two spherical volumes the reduction factor will cancel out.}. On the other hand, for the synapse sign to be certain, we need $W\geq |\vec{y}_{cons}|=y>\ycr$, where $\ycr$ is given by (\ref{ycr-appr}). So, we need to compare the spherical shell volume, $V_{W\geq y>\ycr}$, with the volume of the $\cR$-dimensional sphere, $V_{y\leq W}$. In order to  find $V_{W\geq y>\ycr}$, we have to subtract the $\cR$-dimensional spherical volume with radius $\ycr$ from the spherical volume  with radius $W$. Since $n$-dimensional spherical volumes scale as the $n$th power of the radius, the probability, $P$, of ascertaining the sign of the synapse is approximately given by
\be
P\approx {V_{W\geq y>\ycr}\over V_{y\leq W}} = \frac{W^\cR-\ycr^\cR}{W^\cR}= 1-\left(\frac{\ycr}{W}\right)^\cR. 
\ee
Now, when $\cS,\cF\gg 1$ we can approximately evaluate the RHS as follows:
\ba
\left(\frac{\ycr}{W}\right)^\cR&=&\LF{\cS+2\cF \over 2+ \cS+2\cF }\RF^{\cR/2}=\LF1+{2 \over \cS+2\cF }\RF^{-\cR/2}\non
\Ra \ln\LF\left(\frac{\ycr}{W}\right)^\cR\RF &=&-{\cR\over 2}\ln\LF1+{2 \over \cS+2\cF }\RF={-\cR\over \cS+2\cF }\LT1+\cO\LF{1\over \cS+2\cF }\RF\RT  \ .
\ea
Thus we get
\be
 P\approx 1-e^{-{\cR\over \cS+2\cF} }\ .
\label{random-prob}
\ee
The most prominent feature of (\ref{random-prob}) is that the probability only depends on the ratios of the various dimensions. Hence it doesn't change as we increase the size of the network as long as the ratios are kept constant. 

For the purpose of illustration and numerically testing this feature we assessed how certainty predictions changed when the network size is increased while holding the ratios between $\cR$, $\cS$ and $\cF$ fixed. In Fig.~\ref{fig:highDsims}C we have plotted the number of certain synapses in simulations generated from random data as we scale up $\cN$ maintaining the ratios between $\cR, \cS$ and $\cF$ (see Appendix F for more details). We illustrate two cases. In the first example, no unconstrained directions were present, and $\cS=3\cR$. Then $P=1-e^{-1/3}\approx 0.28$, so one has a 28\% chance of being able to determine the sign of the connections. This answer incidentally is the same as an example with $\cR=\cS=\cF$. As another example, Fig.~\ref{fig:highDsims}C considered the case when $\cF=2\cR=2\cS$. According to (\ref{random-prob}), then  $P=1-e^{-1/5}\approx 0.18$, so the chance of determining the sign drops to about 18\%. We only expect these numbers to be approximate. For example, our arguments relied on the assumption that all target responses admitting solutions are equally likely, an assumption that definitely needs to be revisited for realistic networks. However, the scaling behavior should hold for other probabilistic distributions as long as the scale of $\vec{y}_{cons}$ behaves similarly to (\ref{avg-median}) with increasing $\cN$.  
\renewcommand{\theequation}{C.\arabic{equation}}
\setcounter{equation}{0}
\vs
{\bf C. Nonzero-error Certainty Conditions}\vs
There are various reasons why we may want to not only consider weights that exactly reproduce the specified neuronal responses, but also weights that do so approximately. For instance, we are always limited by the accuracy of the measurement apparatus. More importantly, there are  various sources of biological noise that typically lead to uncertainties in observed values of neuronal responses. For the purpose of this paper we will consider any set of weights to be part of the $\en$-error solution space if it is able to reproduce the specified neuronal responses with an error $\cE\leq \en$ (see (\ref{error}) for definition of $\cE$). We will neglect uncertainties in the input responses to the target neuron, but we will comment on their possible effects towards the end of this appendix.\vs
{\bf Errors in feedforward networks}
\vs
Let us first focus on feedforward networks. Allowing for error increases the value of $\ycr$ by expanding the solution space. One way to think about this is to realize that we have to now make sure that (\ref{ycritical}) is satisfied for any non-negative $\ytv=\yv +\dav$, where $\yv $ and $\dav$ are vectors in the $\cP$-dimensional activity-constrained subspace, the former representing the observed responses, and the latter coming from noise. We will initially assume that all the observed responses are non-negative, so a zero-error solution is possible and the noise is bounded by $|\dav|\leq \en$. Our strategy will be to first seek the minimum $y $ needed to have a certain synapse for a given $\dav$. We then find the maximum among these $y$-critical values as we let $\dav$ vary within the $\en$-ball. Since this procedure will guarantee that the $w=0$ hyperplane doesn't intersect the entire solution space with $\cE\leq \en$, this means that the synapse must exist for the network to generate the specified responses patterns. The synapse's sign will match the zero-error analysis. We will first estimate $y$-critical when the error is small enough to not induce topological transitions in the error surface. In the subsequent sections, we will include the effects of topological transitions, as well as explain how to deal with situations where some of the observed responses are negative, which is possible due to noise.
\vs
{\it When all observed responses are non-negative and no topological transitions occur:}
To understand how errors affect the certainty conditions, let us consider the case where the observed responses are non-negative and the allowed error satisfies $0<\en<\min\{y_{\mu}\}_{\mu=1,\dots, \cR}$, so that no topological transitions can occur. If some responses that were zero in $\vec y $ are now nonzero in $\ytv$, then both $e_y$ and $e_{s\x}$ can change due to the noise. Without loss of generality, let us assume that $\dav$ only has nonzero components along  $\mu=\cC+1,\dots,\cQ$ semi-constrained dimensions\footnote{The components of $\dav$ along these semi-constrained directions are all positive since $\vec y$ must be non-negative.}, as well as along some (or all) of the constrained dimensions. Then $e_y$ changes to
\be
\w{e}_y={\eh\cdot(\yv +\dav)\over |\yv +\dav|}\ .
\ee
Furthermore, if some previously semi-constrained components that contributed to $\sh_\x$ have now become constrained\footnote{This will happen if a component of $\yv $ that was zero now has a nonzero component.}, then $\sh_\x$ no longer has those components. This means that we have to subtract these components from $e_{s\x}$:
\be
\widehat{\w{s}}_\x=\sh_\x-\sum_{\mu=\cR+1}^{\cQ}A_\mu e_{\mu}\env_\mu\ \Ra 
\eee _{s\x}^2= e_{s\x}^2-\sum_{\mu=\cR+1}^{\cQ}A_\mu e_{\mu}^2\ ,
\ee
where $A_\mu$ is 1 if $(\yh\cdot\ch)$ and $e_\mu$ have the same sign and 0 otherwise. This follows from the definition of the boundary projection vector (\ref{vecsx}) and $e_{s\x}$  (\ref{es-star}). Thus, for a given $\dav$ the certainty condition (\ref{ycrappendix}) yields
\be
|\ytv|^2=|\yv +\dav|^2 > W^2\LF{\eee_{s\x}^2+e_u^2 \over \eee_y^2+ \eee_{s\x}^2+e_u^2 }\RF\Ra 
|\yv +\dav|^2\LF {\eee_y^2 \over \eee_{s\x}^2+e_u^2 }+1\RF > W^2 \Ra
{|\eh\cdot(\yv +\dav)|^2\over \eee_{s\x}^2+e_u^2}+|\yv +\dav|^2 > W^2.
\label{yexpcr}
\ee
As before, one can interpret the above inequality as equivalently specifying either $y$-critical or $W$-critical. For a fixed $\yv $, one can obtain a minimum value of the left hand side (LHS) of the latter inequality by varying $\dav$ within the $\en$-ball. The square root of this is $W$-critical. Then as long as $W$ is less than $W$-critical, we will have a certain synapse. Inverting the relation, one finds $y$-critical as the minimum $y $ needed to make the synapse sign certain for all $\dav$ and given $\yh $ and $W$. More explicitly, equating the two sides of the inequality for any given $\yh $, $\dav$, and $W$, we  get a minimal-$y $ that depends on $\dav$. To find $y$-critical, we have to take the maximum of the minimal-$y $ as we vary over all possible $\dav$ in the $\en$-ball. 

Let us first obtain a lower bound on $y$-critical. By inspection of the LHS of the above inequality, it is clear that the more the $\dav$-dependent terms can cancel the $\yv $-dependent terms, the harder it is to satisfy the certainty condition. We observe that in (\ref{yexpcr}), the second term is minimized when $\dav=-\en\yh $~\footnote{This assumes that $y  > \en$. Smaller values of $y $ permit $\vec y = \vec 0$ and all weights can be set to zero}.  Accordingly, one can obtain  a lower bound on $y$-critical by substituting $\da=-\en\yh $ in (\ref{yexpcr}):
\be
(\eh\cdot\yh )^2 (y -\en)^2+(e_{s\x}^2+e_u^2)(y -\en)^2 > W^2(e_{s\x}^2+e_u^2)\ ,\mx{ or, } y  > y_{\mt{cr},\min,0}\equiv W\sqrt{e_{s\x}^2+e_u^2 \over e_{y}^2+ e_{s\x}^2+e_u^2 }+\en\ ,
\label{ycmin}
\ee
where we have used $e_y = \eh\cdot\yh $ and $\eee_{s\x}^2=e_{s\x}^2$, since $\dav$ has no components along the semi-constrained directions. We will see later that this simple lower bound can approximate the actual $y$-critical very well in many situations. Notice that we used a subscript ``0'' to denote this lower bound. This is because, as we will soon see, when noise allows for topological transitions, one may be able to obtain stricter lower bounds by allowing some constrained dimensions to behave as semi-constrained. This ``0'' emphasized that no constrained indices behave as semi-constrained.

\begin{figure}[thbp]
	\includegraphics[width=1\textwidth,angle=0]{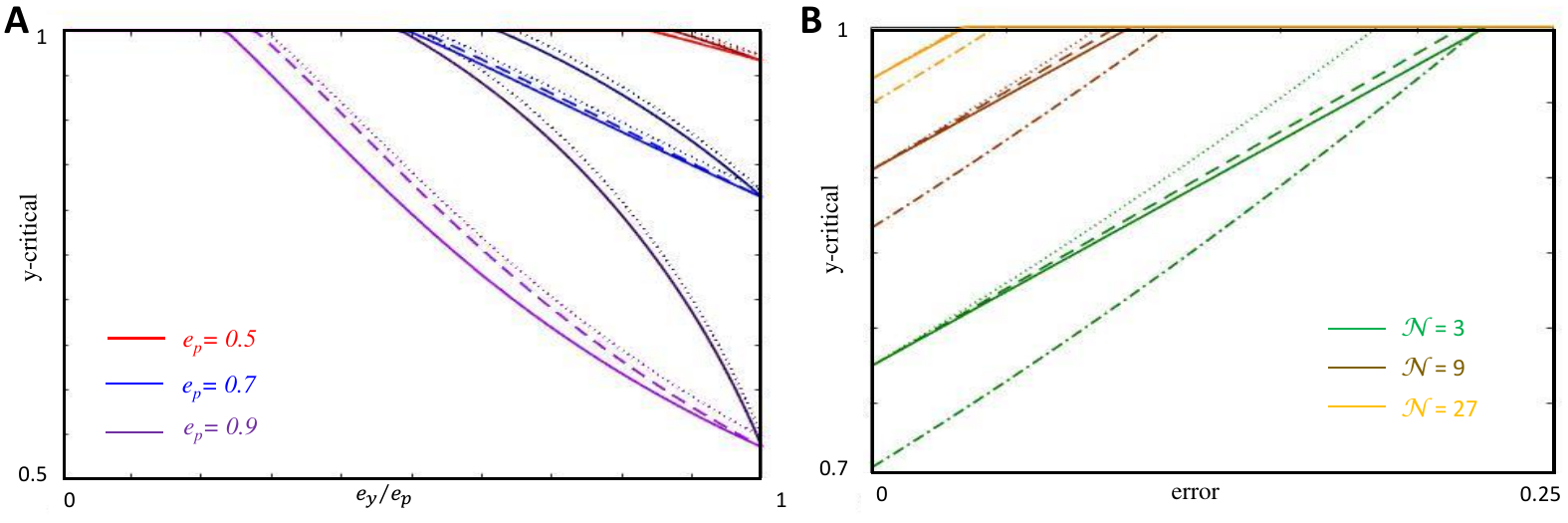}
	\caption{{\small {\bf Dependence of $y$-critical on various parameters for nonzero errors.} ({\bf A}) The red, blue and purple curves track $y$-critical as a function of $e_y/e_p$ for $e_p= 0.5, 0.7$, and $0.9$, respectively. The dotted, dashed and bold curves represent the lower bound, leading-order and upper bound $y$-critical curves for a fixed error, $\en=0.1W$. The darker shade correspond to the most nonlinear case when $e_{s\x}/\sqrt{e_p^2-e_y^2}=1$, while the lighter shade correspond to $e_{s\x}=0$. These latter curves are also the ones that one obtains in a linear theory. Clearly, the difference between the linear and nonlinear theory increases as $e_p$  increases. In all these cases $y$-critical decreases  with increase of $e_p$, and for a given $e_p$, as $e_y/e_p$ increases. Also, as $e_{s\x}$ increases and the  semi-constrained dimensions become more important, it becomes harder to constrain the synapse sign, and therefore $y$-critical  increases. ({\bf B}) The green, brown and orange curves again track $y$-critical, but this time as a function of $\en$, for networks with $\cN=3, 9$, and $27$ input neurons, respectively. The dotted, dashed and bold curves plot the lower bound, leading-order and upper bound on $y$-critical for typical values of $e_p, e_y$ and $e_{s\x}$ that one expects in these networks (\ref{average}). We see that these curves come closer together as the network size increases. The dot-dashed curves correspond to the linear theory ($e_{s\x}=0$), which remains clearly separated from the nonlinear curves. In each of these networks, $\cP/\cN = 2/3$ and $\cR/\cP = 1/2$.  
			\label{fig:theory-error}}}
\end{figure}
Next, we can find an upper bound for $y$-critical by noting
\be
{|\eh\cdot(\yv +\dav)|^2\over \eee_{s\x}^2+e_u^2}+|\yv +\dav|^2\geq{|\eh\cdot(\yv +\dav)|^2\over e_{s\x}^2+e_u^2}+|\yv +\dav|^2\geq{(\eh\cdot\yv )^2+2(\eh\cdot\yv )(\eh\cdot\dav)\over e_{s\x}^2+e_u^2}+y_{0}^2+2\ \yv \cdot\dav\ ,
\label{y0-inequalities}
\ee
where the first inequality is true because $\eee_{s\x}^2\leq e_{s\x}^2$, and the second inequality because we have dropped positive $\cO(\da^2)$ terms. 
Then we can obtain an upper bound on $y$-critical by finding a $y $ such that even the last expression on the RHS is greater than $W^2$. Specifically, 
\be
(\eh\cdot\yv )^2+2(\eh\cdot\yv )(\eh\cdot\dav)+(e_{s\x}^2+e_u^2)(y_{0}^2+2\ \yv \cdot\dav)> W^2(e_{s\x}^2+e_u^2)\ .
\label{intermediate-yc}
\ee
So, let us try to find the $\dav$ that minimizes the LHS: 
\ba
\mx{LHS}&=&y ^2e_y^2+2y  e_y(\eh\cdot\dav)+(e_{s\x}^2+e_u^2)(y ^2+2y \ \yh \cdot\dav)\non
&=&y ^2(e_y^2+e_{s\x}^2+e_u^2)+2y ( e_y\eh+(e_{s\x}^2+e_u^2)\yh )\cdot\dav\non
&=&y ^2(e_y^2+e_{s\x}^2+e_u^2)+2y \ \vec\xi\cdot\dav\,
\ea
where $\vec\xi\equiv e_y\sum_{\mu=1}^\cP e_\mu\env_\mu+(e_{s\x}^2+e_u^2)\yh $, and we have noted that $\uh\cdot\dav=0$ because $\dav$ must be in the activity-constrained subspace. It is now clear that LHS is minimized  if $\dav$ anti-aligns with $\vec\xi$. Then  (\ref{intermediate-yc}) yields
\be
y ^2(e_y^2+e_{s\x}^2+e_u^2)-2y |\vec\xi|\en > W^2(e_{s\x}^2+e_u^2)\ .
\label{inter2}
\ee
Equating the two sides of (\ref{inter2}) and solving for $y $~\footnote{This quadratic equation obviously has two solutions. The correct one can easily be identified, for instance, by taking the $\en\ra 0$ limit.}, we now get an upper bound for $y$-critical:
\be
y_{\mt{cr},\max,0}\equiv
\sqrt{W^2\LF{e_{s\x}^2+e_u^2 \over e_{y}^2+ e_{s\x}^2+e_u^2 }\RF+ {\en^2\xi^2\over (e_y^2+ e_{s\x}^2+e_u^2)^2}}+ {\en\xi\over e_{s\x}^2+e_u^2+e_{y}^2},
\label{ymax-inter}
\ee
where $\xi$ is the norm of $\vec\xi$ and can be simplified as
\ba
\xi^2&=&\sum_{\mu=1}^\cP [e_ye_\mu+(e_{s\x}^2+e_u^2)\yh_{\mu}]^2=e_y^2\sum_{\mu=1}^\cP e_\mu^2+(e_{s\x}^2+e_u^2)^2\sum_{\mu=1}^\cR\yh_{\mu}^2+2(e_{s\x}^2+e_u^2)e_y\sum_{\mu=1}^\cR\yh_{\mu} e_\mu\non
&=&e_y^2e_p^2+(e_{s\x}^2+e_u^2)^2+2(e_{s\x}^2+e_u^2)e_y^2=(e_y^2+e_{s\x}^2+e_u^2)^2+e_y^2(e_p^2-e_y^2)\ .
\ea
Thus, we have
\be
y_{\mt{cr},\max,0}\equiv
\sqrt{W^2\LF{e_{s\x}^2+e_u^2 \over e_{y}^2+ e_{s\x}^2+e_u^2 }\RF+ \en^2\LF1+{e_y^2(e_p^2-e_y^2)\over (e_y^2+ e_{s\x}^2+e_u^2)^2}\RF}+ \en\sqrt{1+{e_y^2(e_p^2-e_y^2)\over (e_y^2+ e_{s\x}^2+e_u^2)^2}}\ .
\label{ycmax}
\ee
As with lower bound, we will see that to obtain the correct upper bound  in presence of topological transitions, one has to maximize over several upper bounds. Hence we refer the above upper bound  that doesn't include any effects from topological transitions with an index ``0''. 

Finally, we would like to point out that for small errors one can also obtain an approximate correction to $y$-critical that lies in between $y_{\mt{cr},\min,0}$ and $y_{\mt{cr},\max,0}$. To  obtain this estimate, let us first write down the bound on $y $ that one would obtain  from (\ref{yexpcr}) as $\da\ra 0$:
\be
y  > W \sqrt{\eee_{s\x}^2+e_u^2 \over \eee_y^2+ \eee_{s\x}^2+e_u^2 }\ .
\label{ynot}
\ee
If $\dav$ has components along any  semi-constrained direction that contributes towards the original $\eh_{s\x}$ vector, then $\eee_{s\x}^2<e_{s\x}^2$, and comparing (\ref{ycrappendix}) and (\ref{ynot}) we see that,  as $\da\ra 0$, the bound on $y $ will be less than the zero-error $\ycr$. In other words, for sufficiently  small errors, if $\dav$ explores directions that contribute to $e_{s\x}$, then the corresponding bound on $y $ is going to be smaller than even the zero-error $\ycr$. Thus, for these small errors the leading order corrections to (\ref{ycrappendix}) is obtained only if $\dav$  do not have any components along these semi-constrained directions. This means $\eee_{s\x}^2=e_{s\x}^2$, and we can reorder the indices such that the semi-constrained directions along which excursions of $\dav$ will be  considered range from $\cR+1,\dots,\cQ$, \ie,   $\sgn(\sh\,'_\mu)$ is negative for these, and only these, semi-constrained indices. To obtain the certainty condition, one can then follow steps (\ref{y0-inequalities})~\footnote{Since we are only interested in the leading order correction, we could also drop the $\cO(\da^2)$ terms needed to arrive at an expression such as the RHS of (\ref{y0-inequalities}).} through (\ref{ymax-inter}) except that
 $\dav$  is restricted to only have nonzero components along constrained directions and those semi-constrained directions that do not contribute to $\eh_{s\x}$, \ie, for $\mu=1\dots \cQ$.  In other words, it can at the most anti-align with a truncated $\vec\xi$,
\be
\vec\xi_{\mt{trunc}}\equiv \sum_{\mu=1}^{\cQ}\xi_\mu\env_\mu=\sum_{\mu=1}^{\cQ}[e_ye_\mu+(e_{s\x}^2+e_u^2)\yh_{\mu}]\env_\mu\ ,\mand (\vec\xi\cdot\dav)_{\min}=-\en\xi_{\mt{trunc}}\ ,
\ee
where $\xi_{\mt{trunc}}$ is the norm of $\vec\xi_{\mt{trunc}}$ and can be simplified as
\ba
\xi_{\mt{trunc}}^2&=&\sum_{\mu=1}^\cQ [e_ye_\mu+(e_{s\x}^2+e_u^2)\yh_{\mu}]^2=e_y^2\sum_{\mu=1}^\cQ e_\mu^2+(e_{s\x}^2+e_u^2)^2\sum_{\mu=1}^\cR\yh_{\mu}^2+2(e_{s\x}^2+e_u^2)e_y\sum_{\mu=1}^\cR\yh_{\mu} e_\mu\non
&=&e_y^2(e_p^2- e_{s\x}^2)+(e_{s\x}^2+e_u^2)^2+2(e_{s\x}^2+e_u^2)e_y^2=(e_y^2+e_{s\x}^2+e_u^2)^2+e_y^2(e_p^2- e_{s\x}^2-e_y^2)\ .
\ea
Substituting $\xi=\xi_{\mt{trunc}}$ into the counterpart of (\ref{ymax-inter}), and keeping only the linear terms in $\en$, we thus get the leading order correction to (\ref{ycrappendix}):
\be
y_{\mt{cr,appr,0}}\approx W\sqrt{{e_{s\x}^2+e_u^2 \over e_{y}^2+ e_{s\x}^2+e_u^2 }}+ {\en\xi_{\mt{trunc}}\over e_{y}^2+e_{s\x}^2+e_u^2}=W\sqrt{{e_{s\x}^2+e_u^2 \over e_{y}^2+ e_{s\x}^2+e_u^2 }}+ \en\sqrt{1+{e_y^2(e_p^2-e_{s\x}^2-e_y^2)\over (e_y^2+e_{s\x}^2+e_u^2)^2}}\ .
\label{ycr-lead-en}
\ee
We will see later how $y_{\mt{cr,appr,0}}$ can be generalized to provide an approximation,  $y_{\mt{cr,appr}}$, to $y$-critical that accounts for topological transitions.

Reassuringly, we see that at $\en=0$, $y_{\mt{cr,appr,0}}$, $y_{\mt{cr},\max,0}$ and $y_{\mt{cr},\min,0}$, all reduce to the zero-error $\ycr$ (\ref{ycrappendix}). Also it is obvious that the coefficient of $\en$ in  $y_{\mt{cr,appr,0}}$ is greater than that of $y_{\mt{cr},\min,0}$ but less than that of $y_{\mt{cr},\max,0}$. Finally, note that $y_{\mt{cr,appr,0}}$ coincides with $y_{\mt{cr},\min}$ in the maximally nonlinear case where $e_{s\x}^2 = e_p^2 - e_y^2$. In Fig.~\ref{fig:theory-error}A, we have plotted how these different quantities depend on $e_p, e_y$ and $e_{s\x}$. In particular we note that as the network size increases, these curves typically come closer together (Fig.~\ref{fig:theory-error}B), so that they  provide a good approximation for $y$-critical. Finally, for future reference we point out that for a given set of input patterns, $z_{\mu m}$, the various $y$-criticals that we have computed above depend on the orientation of the target response vector, or $\yh$, and the total noise budget, $\en$. In other words,  $y_{\mt{cr},\min,0}=y_{\mt{cr},\min,0}(\yh,\en)$, $y_{\mt{cr},\max,0}=y_{\mt{cr},\max,0}(\yh,\en)$, and $y_{\mt{cr},\appr,0}=y_{\mt{cr},\appr.0}(\yh,\en)$.
\vs
{\it Comparing predictions from linear and nonlinear models:}
To assess the effects of nonlinearity it is useful to compare the predictions for certain-synapses between the linear and nonlinear theory. In a linear theory, there are no semi-constrained directions, and therefore, a lower bound, leading order and upper bound on  $y$-critical can be obtained from (\ref{ycmin}), (\ref{ycr-lead-en}) and (\ref{ycmax}) respectively by setting $e_{s\x}=0$:
\ba
y_{\mt{cr,min,lin}}&=&W\sqrt{{e_u^2 \over e_{y}^2+e_u^2 }}+ \en \ ,\\
y_{\mt{cr,appr,lin}}&=&W\sqrt{{e_u^2 \over e_{y}^2+e_u^2 }}+ \en\sqrt{1+{e_y^2(e_p^2-e_y^2)\over (e_y^2+e_u^2)^2}}\ ,\\
y_{\mt{cr,max,lin}}&=&
\sqrt{W^2\LF{e_u^2 \over e_{y}^2+e_u^2 }\RF+ \en^2\LF1+{e_y^2(e_p^2-e_y^2)\over (e_y^2+e_u^2)^2}\RF}+ \en\sqrt{1+{e_y^2(e_p^2-e_y^2)\over (e_y^2+e_u^2)^2}}\ .
\label{ycr-lin}
\ea
We note that since all these quantities are increasing function of $e_{s\x}$, the linear values  are always less than or equal to the nonlinear counterparts. Since no toplogical transitions are possible in a linear theory, these expressions don't need a qualifying ``0'' index. In Fig.~\ref{fig:theory-error}, we show a comparison between the upper bound on $y$-critical obtained in the linear and the nonlinear theories.
\vs
{\it $\ycmaxx$, when one and only one non-zero response is smaller than noise:} In the previous section we have considered responses which are either zero, or positive and greater than the noise bound, $\en$. In this section, we consider a situation where one and only one of the observed responses is smaller than the noise, $|y_1 |<\en$. 

Note that once we admit noise, it is possible for the small observed response to be negative. In this case, there is no zero-error solution as $\Phi(\eta_1 )$ cannot be negative, and therefore we need a minimum noise, and incur a minimum error:
\be
\da_1 =-y_1  \Ra \cE_{\min}=y_1 ^2\ .
\label{minnoise}
\ee
In fact, since the noise for this observation has to be positive, we must have
\be
\Phi(\eta_1)= y_1 +\da_1 \equiv\da'_1 \geq 0\ .
\ee
Then using, 
\be
\da_1 ^2=\da_1 ^{'2}+y_1 ^2-2y_1 \da'_1 \ ,
\ee
we obtain a modified bound on the noise:
\be
\sum_{\mu=1}^{\cP}\da_{\mu}^2= \sum_{\mu=2}^{\cP}\da_{\mu}^2+\da_1 ^{'2}+y_1 ^2-2y_1 \da'_1 \leq\en^2\Ra\da_1 ^{'2}+\sum_{\mu=2}^{\cP}\da_{\mu}^2\le\en^2-y_1 ^2+2y_1 \da'_1 \leq \en^2-y_1 ^2\ ,
\label{noisebound}
\ee
since $y_1 \da'_1 <0$. Or, 
\be
\da_1 ^{'2}+\sum_{\mu=1}^{\cP}\da_{\mu}^2\le\en^{'2}\equiv \en^2-y_1 ^2\ ,
\label{newerror}
\ee
Let us now introduce a new reduced response vector whose response to the $1^{st}$ pattern is set to zero:
\be
\yv ^{\;'}\equiv\sum_{\mu=2}^{\cR}y_{\mu}\env_{\mu}\ .
\label{reduced}
\ee
We can then identify $\da'_1 $ to be the noise associated with the $\mu=1$ response in this new feedforward problem, while the other $\da_{\mu}$'s can continue to represent the noise associated with all the other responses. Thus, a sufficient condition for a given synapse to be certain is
\be
|\yv ^{\;'}|>\ycmax(\yh ',\en')\ .
\label{newcondition}
\ee

A very similar condition arises if $y_1$ is positive but small enough to admit a topological transition. To see how, let us first remember that in order for a synapse to be certain, the solution space should not intersect with the $w=0$ hyperplane. Now, let us look at the solution space coming from denoised $\ytv$'s that have $\w{y}_1 >0$. Since, the solution space corresponding to these $\ytv$ 's do not have any additional semi-constrained dimension as compared to the observed response, $\yv$, the condition for no intersection with this part of the solution space  is simply given by 
\be
|\yv |>\ycmax(\yh ,\en)\ ,
\label{original}
\ee
a condition that guarantees a certain synapse when no topological transitions are considered. Next consider the solution space for denoised $\ytv$ 's with $\w{y}_1 =0$. The solution space for these $\ytv$ 's have an additional semi-constrained dimension corresponding to the $1^{st}$ pattern. We can therefore use the reduced response vector, $\yv ^{\;'}$ (\ref{reduced}), so that the solution space corresponding this new response vector with the error bound, $\en'$ (\ref{newerror}), along with the solution space with $\w{y}_1 >0$ accounts for the full solution space of $\yv $ with error $\en$. Note, that the noise budget is again reduced according to (\ref{newerror}) since we are committed to making at least an error of $y_1 $ to convert the $1^{st}$ response to a semi-constrained dimension. To ensure that there is no intersection of the $w=0$ hyperplane with the $\yv ^{\;'}$ solution space we must therefore also satisfy (\ref{newcondition}). We note that to calculate the right hand side using (\ref{ycmax}), the various projections have to be recalculated according to 
\ba
e_{y'}=
 {\sum_{\mu=2}^{\cR}y_\mu e_\mu \over \sqrt{\sum_{\mu=2}^{\cR}y_\mu^2 }} \ ,\mand \ e_{s'\x}=   \sqrt{\sum_{\mu\in A_-}e_\mu^2 + \Ta(\sgn(e_{y'}) e_{1} )e_{1} ^2}\ .
\label{trans-projections}
\ea
The condition (\ref{newcondition}) on $|\yv ^{\;'}|$ translates to a condition on $|\yv |$:
\be
|\yv |>\sqrt{\ycmax^2(\yh ',\en')+y_1 ^2}\Ra |\yv |>{\ycmax(\yh ',\en')\over \sqrt{1-\yh_{1}^2}}\ ,
\label{aftertrans}
\ee
where we have defined $\widehat{y}_{\mu}$'s to be the $\mu^{th}$ component of $\yh $. 

We note that this is a nonlinear inequality as the right hand side depends on $|\yv |$ through its implicit dependence on $\en'$. When we have a negative $y_1 $, only (\ref{aftertrans}) needs to be satisfied to guarantee a certain synapse, but if $y_1 $ is positive, both (\ref{original}) and (\ref{aftertrans}) have to be satisfied. It is not hard to see how this process should be continued if one has more than one topological transition within the allowed error. Since we know the precise sequence of topological transitions, all the sequential certainty-conditions can in principle be obtained. A synapse is certain if all of its certainty-conditions are satisfied. 

So far, we have obtained a way to check whether a synapse is certain given the response data, $\yv $. We also have an upper bound  of $y$-critical, $\ycmax$,  ignoring effects from topological transitions when all the observed responses are non-negative. We will now investigate how topological transitions can change this upper bound. We will start by quantifying  effects from a single topological transition by finding potentially a new  upper bound  for $y$-critical, $\ycmaxx$, such that we can say that if $|\yv |>\ycmaxx$, then the synapse is certain. Suppose we start out with a data vector whose norm is so large that there are no topological transitions. Then as we decrease the norm, but keep its orientation, $\yh $, fixed,  eventually a semi-constrained dimension will open up in the solution space, in our example, the $1^{st}$ direction. If we keep decreasing further, then at some point another response dimension will become semi-constrained due to the presence of noise. Let us however consider the situation where $\ycmaxx$ (that is yet to be computed) is going to turn out to be larger than the norm when the second transition occurs. In this case, we don't have to consider this possibility (and any other transitions) because then if $|\yv|>\ycmaxx$ the second transition cannot occur. We will later find a condition that guarantees this. Since we are trying to find the smallest value of $\ycmaxx$ that we can find, what all this means  is that at $\yv_{\mt{cr,max}}=\ycmaxx\yh $, one of the two inequalities (\ref{original} and \ref{newcondition}) becomes an equality. While the first equality is trivial to solve as the right hand side doesn't depend on $\ycmaxx$, the second equation is highly nonlinear\footnote{Here the ``1'' in the subscript indicates that this possibility for $y$-critical is computed by only considering the first topological transition.}:
\be
\yct={\ycmax(\yh ',\en')\over \sqrt{1-\yh_{1}^2}}\ ,
\label{ycttrans}
\ee
where
\be
\en'=\sqrt{\en^2-\yct^2\yh_{1}^2}\ .
\label{conserror}
\ee
In particular, we notice that there are two competing effects that ultimately determine $\yct$. The numerator  depends on $\en'$, which decreases as $\yct$ increases and therefore has an overall effect of decreasing $\yct$. On the other hand, the presence of $\yh_{1}$ in the denominator  within the square root tends to increase $\yct$. To determine the correct upper bound  for $y$-critical one has to compare the $\yct(\yh ' ,\yh_{1},\en')$ obtained from (\ref{ycttrans}) with $\ycmax(\yh ,\en)$, and then choose the maximum because then both the inequalities (\ref{original} and \ref{newcondition}) will be satisfied. For the negative response case, we simply need to solve (\ref{ycttrans}) to obtain $\yct$. 

Now, determining $\yct$ from (\ref{aftertrans})  involves solving a quartic equation leading to expressions that are not particularly insightful. However, we can obtain a relatively simple conservative estimate bypassing the nonlinearity if we have a lower bound  on $y$-critical, $\ycminn$ because we can use this bound to overestimate $\en'$:
\be
\ycmaxx\equiv \max\left\{ {\ycmax(\yh ',\en'')\over \sqrt{1-\yh_{1}^2}},\ycmax(\yh ,\en)\right\} \where \en''\equiv\sqrt{\en^2-\ycminn^2\yh_{1}^2}\geq \en'\ .
\label{ymaxcons}
\ee
Before we describe how we can obtain $\ycminn$, let us  note that  if
\be 
\ycminn\yh_{\mu}>\en \ ,\ \forall\ \mu>1\ ,
\label{nosecond}
\ee
then the second transition occurs at a magnitude that is lower than $\yct$, and therefore doesn't need to be incorporated in the $\yct$ calculation. Indeed, (\ref{nosecond}) is a sufficient condition but not a necessary one. 
\vs
{\it $\ycminn$, when one and only one non-zero response is  smaller than noise:} 
When we have a small response, $|y_1 |<\en$, we have seen that we have to consider solution space around a reduced response vector, $\yv ^{\;'}$ (\ref{reduced}), with a smaller error budget, $\en'$ (\ref{newerror}).   Accordingly, we can obtain an equation for a lower bound  on $y$-critical using (\ref{ycmin})~\footnote{To remind the readers, the expression for $\ycmin$ was obtained by computing $\ycr$ for   $\ytv=(1-\en)\yv $, a denoised point that is allowed because of the noise. In this case, the corresponding point is $\ytv=(1-\en')\yv ^{\;'}$.}, 
\be
\ycmint=\ycr(\yh ')+\en'\ ,
\ee
where $\mu=1$ along with $\mu=\cR+1\dots \cP$ are all treated as semi-constrained.
As before, since the norms along $\yv $ and $\yv ^{\;'}$ are related via 
\be
|\yv |\sqrt{1-\yh_{1}}=|\yv ^{\;'}|\ ,
\ee
we get an equation for $\ycmint$ very similar to (\ref{ycttrans}) for $\yct$:
\be
\ycmint={\ycr(\yh')+\en''\over \sqrt{1-\yh_{1}^2}}\ ,\where \en''=\sqrt{\en^2- \ycmint^2\yh_{1}^2}\ .
\ee
Although nonlinear, the above equation reduces to a quadratic equation for $\ycmint$,
\be
\ycmint^2-2\ycmint\ycr\sqrt{1-\yh_{1}^2}+\ycr^2-\en^2=0\ ,
\ee
solving which we get~\footnote{The second root gives a negative result, and accordingly doesn't reduce to the correct $\en\ra 0$ limit.},
\be
\ycmint=\ycr(\yh')\sqrt{1-\yh_{1}^2}+\sqrt{\en^2-\ycr^2(\yh')\yh_{1}^2}\ .
\ee
For positive $y_1 $ the above expression provides another lower bound  along with the one obtained without the transition, (\ref{ycmin}). To ensure we have the tightest possible lower bound we thus maximize:
\be
\ycminn\equiv \max\{\ycmin,\ycmint \}\ .
\ee
\begin{figure}[!thbp]
	\includegraphics[width=0.96\textwidth,angle=0]{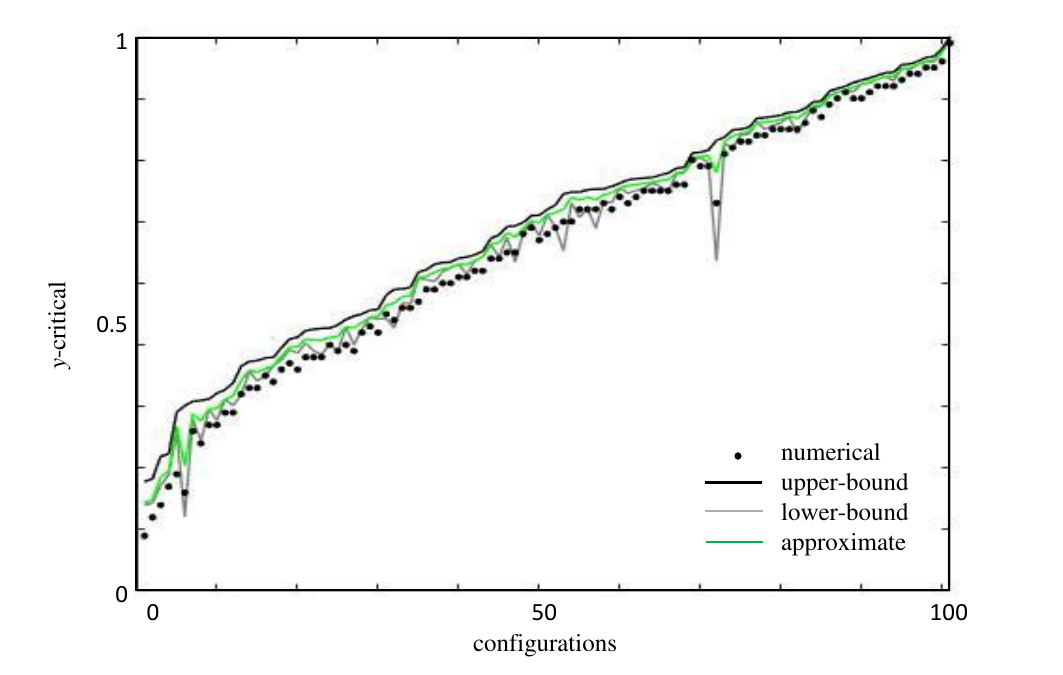}
	\caption{{\small {\bf Testing bounds on $y$-critical for solutions with error.} We show the same 102 random configurations of input-output activity as Fig.~\ref{fig:highDsims}B. The bold black, green, and gray curves represent the upper bound $\ycmaxx$, approximate $y_{\mt{cr,appr}}$, and lower bound $y$-critical $\ycminn$, values, respectively. The black dots correspond to the maximum value of $y$ in our simulations that resulted in mixed signs for the synaptic weights under consideration. 
			\label{fig:errorcurves}}}
\end{figure}
\vs
{\it When more than one non-zero responses are smaller than allowed error:} 
It is not difficult to see how the arguments above generalize if we have more than one small ($<\en$) observed response. We have to consider cases where all the negative observed responses, and different possible combinations of the positive responses, are set to zero. Let us denote $T$ to be one such possible set of $\mu$ indices. As before, we define a reduced response vector, which is now indexed by $T$:
\be
\yv_T\equiv \sum_{\mu\not\in T}y_{\mu}\env_\mu\ ,
\ee 
so $y_{T,\mu}=0$ for all $\mu\in T$. Then, essentially following the same algebraic manipulations as above we obtain a lower bound  according to
\be
y_{\mt{cr,min},T}=\ycr(\yh_T)\sqrt{1-\sum_{\mu\in T}\yh^2_{\mu}}+\sqrt{\en^2-\ycr^2(\yh_T)\sum_{\mu\in T}\yh^2_{\mu}}\ .
\ee
To reiterate, whenever an observed response is negative, which is inconsistent with a threshold linear transfer function, this means that some of the noise budget has to be used up to bring this response up to zero, and the same noise reduction occurs if one wants to consider topological transitions. Each $y_{\mt{cr,min},T}$ evaluated this way provides us with a lower bound, and hence we have to take a maximum over all these to find the tightest lower bound, $\ycminn$. To make things explicit, let us also enumerate the new expressions for the various projections of $\eh$ that one needs to calculate $\ycr(\yh_T)$:
\ba
e_{y,T}=
 \Bigg(\sum_{\mu\not\in T}y_\mu e_\mu\Bigg)\Bigg/ \sqrt{\sum_{\mu\not\in T}y_\mu^2 } \ ,\mand \ e_{s\x,T}=   \sqrt{\sum_{\mu\in T}\Ta(e_{\mu} e_{y,T} )e_\mu^2 }\ .
\label{gen-projections}
\ea

Once we have a lower bound, we can obtain conservative upper bounds analogous to (\ref{ymaxcons}) for each $T$:
\be
y_{\mt{cr,max},T}\equiv  {y_{\mt{cr,max},0}(\yh_T,\en_T)\over \sqrt{1-\sum_{\mu\in T}\yh^2_{\mu}}}\ ,\where \en_T\equiv\sqrt{\en^2-\ycminn^2\sum_{\mu\in T}\yh^2_{\mu}}\ .
\label{ycrmaxT}
\ee
As before, for a consistent upper bound  for $y$-critical, we need to take the maximum over all $y_{\mt{cr,max},T}$'s.

Finally, we note that we do not need to consider all possible transitions. While going through the sequence of transitions, as soon as we find a $T$ such that   $y_{\mt{cr,min},T}\yh_{\mu}>\en$ for all $\mu\not\in T$ we can stop as this means that by the time $|\yv|$ is small enough that any additional $y_\mu$'s can be set to zero, the synapse is already uncertain.  
\vs
{\it Numerical Simulation:} 
To illustrate the behavior of the various $y$-critical functions and check their utility, in Fig.~\ref{fig:errorcurves} we  have plotted $\ycminn$ (light grey curve) and $\ycmaxx$ (black curve) for the same 102 configurations as the ones depicted in Fig.~\ref{fig:highDsims}B involving a feedforward simulation with $\cN=6,\ \cP=5$, $\cR=2$, and $\cE<\en=0.1$. We also defined, $y_{\mt{cr,appr}}$, as a maximum over different approximations, $y_{\mt{cr,appr,T}}$'s, that incorporate topological transitions  and are defined as natural generalizations of (\ref{ycr-lead-en}):
\be
y_{\mt{cr,appr,T}}=W\sqrt{{e_{s\x,T}^2+e_u^2 \over e_{y,T}^2+ e_{s\x,T}^2+e_u^2 }}+ \en_T\sqrt{1+{e_{y,T}^2(e_p^2-e_{s\x,T}^2-e_{y,T}^2)\over (e_{y,T}^2+e_{s\x,T}^2+e_u^2)^2}}\ .
\ee
We have plotted $y_{\mt{cr,appr}}$, the approximation of $y$-critical, in green in Fig.~\ref{fig:errorcurves}. As in Fig.~\ref{fig:highDsims}B, the black dots here denote the maximum value of $y$ in our simulations that still admitted mixed signs for the synapse under consideration, for details on the simulations, please see Appendix F. As one can see, most of the black dots seem to  closely track the  $\ycminn$-curve, but some of the dots lie between the  $y_{\mt{cr,appr}}$ and $\ycminn$-curves. 
\vs
{\bf New sources of corrections in recurrent neural networks:} 
\vs
It is clear that recurrent neural networks inherit error corrections to $y$-critical that were already present in the feedforward case. There are two additional sources of error that one could consider as one moves from feedforward to recurrent networks. However, our numerical simulations of recurrent networks suggest that these are sometimes small effects, and we leave their systematic study for the future.

Firstly, we could account for the fact that the $\env_\mu$-directions themselves can change. This is because the inputs driving any given driven neuron can no longer be assumed to be fixed at $z_{\mu m}$ if the other driven neurons suffer from noise. However, these activity patterns define the $\env_\mu$-directions and $\eta_\mu$-coordinates. Allowing noise in input neurons would lead to similar corrections. 

\begin{figure}[thbp]
	\includegraphics[width=0.95\textwidth,angle=0]{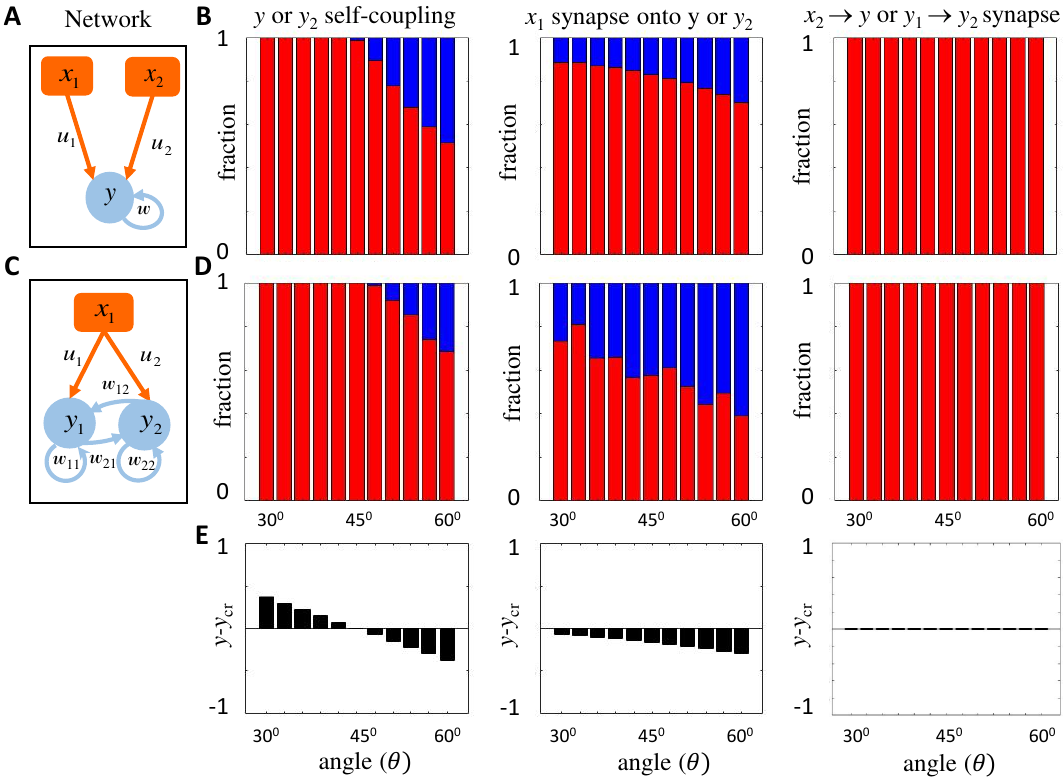}
	\caption{{\small {\bf Comparing simulation and theoretical results in $\cN=3$ recurrent network.} ({\bf A}) A simple $\cN=3$ recurrent neural network with one driven and two input neurons. Note that the $y_1$ neuron shown here maps onto the $y_3$ neuron in Fig.~\ref{fig:zeroerror}A by interpreting the $x_1$ and $x_2$ neurons shown here as the $x_3$ and $y_2$ neurons shown in Fig.~\ref{fig:zeroerror}A. ({\bf B}) Bar graphs depicting the fraction of positive (red) and negative (blue) weights from the network depicted in (A). 
	({\bf C}) Another $\cN=3$ recurrent neural network, this time with two driven and one input neuron. ({\bf D}) Bar graphs depicting the fraction of positive (red) and negative (blue) weights from the network depicted in (C). 
	({\bf E}) The black bars depict $y-\ycr$  for the corresponding synapses. 
			\label{fig:RCN3}}}
\end{figure}

Secondly, the total error in (\ref{error}) may be unevenly distributed across the driven neurons. If the total squared error summed over all responses and neurons is $\en_\tot^2$, then on average, the root mean square error associated with each driven neuron is $\en_\tot/\sqrt{\cD}$. We can thus hope that a substitution of  $\en=\en_\tot/\sqrt{\cD}$ in the various $y$-critical formulas will provide a good approximation. However, it's also possible that a few neurons will incur most of the error (up to $\en_\tot$), potentially leading to violation of the certainty conditions computed from the root mean square error over neurons. 
\vspace{5mm}\\
\renewcommand{\theequation}{D.\arabic{equation}}
\setcounter{equation}{0}
{\bf D. Beyond threshold-linear transfer functions}
\vs
So far, we have always modeled the firing rate as a threshold-linear function applied to the input drive. Here we will explain how our analyses of $y$-critical with noise also provide a formalism to analyze a much more general class of nonlinear transfer functions. \vs
{\it Bounded deviations from the threshold-linear function:} Let us start by considering transfer functions with bounded differences from the threshold-linear function:
\be
\Psi(x)=\Phi(x)+\Da(x), \ \mathrm{with}\ |\Da(x)|< \Da_0\ \forall\ x.
\ee 
In this case, the fixed-point equations become
\be
y_{\mu}=\Psi\LF\sum_{m=1}^{\cN}z_{\mu m}w_m\RF=\Phi\LF\sum_{m=1}^{\cN}z_{\mu m}w_m\RF+\da_\mu\ ,\where \da_\mu\equiv \Da\LF\sum_{m=1}^{\cN}z_{\mu m}w_m\RF\ .
\ee
Since $|\Da(x)|$ is bounded by $\Da_0$, we have a bound on the squared norm of $\vec{\da}$:
\be
|\vec{\da}|^2<\cP\Da_0^2\ .
\ee
It is therefore clear that we can estimate $y$-critical for the $\Psi$ nonlinearity with exactly the same formalism that we used to estimate $y$-critical for the threshold nonlinearity in the presence of noise. In particular, all the $y$-critical estimates (\ref{ycmaxM}, \ref{ycminM}, \ref{ycr-lead-en}) are valid with the substitution, $\en=\sqrt{\cP}\Da_0$. Moreover, one can account for other sources of noise (bounded by $\en_0$) by instead substituting
\be
\en=\sqrt{\en_0^2+\cP\Da^2_0} \ 
\ee
to obtain estimates and bounds on $y$-critical.\vs
{\it Bounded departures from any threshold-monotonic nonlinearity:} Let us now consider transfer functions, $\Psi(x)$, that are close to a function, $\Xi(x)$, that monotonically increases above a threshold, $x_T$:
\ba
\Xi(x)&=&0\ , \mx{if }x\leq x_T, \non\Xi(x)&>&\Xi(y)\ >\ 0, \mx{if }x>y> x_T, \mand\ \non
\Psi(x) &=& \Xi(x) + \Delta(x),\ \mathrm{with}\ |\Delta(x)| < \Delta_0\ \forall\ x.
\ea
Accordingly,  we find
\be
y_{\mu}=\Psi\LF\sum_{m=1}^{\cN}z_{\mu m}w_m\RF=\Xi(\eta_\mu)+\Da(\eta_\mu)\Ra \Xi(\eta_\mu)=y_\mu-\Delta(\eta_\mu)\ .
\ee 
Since the monotonicity condition ensures that $\Xi^{-1}$ is well defined above threshold, and $|\Da(\eta_\mu)|< \Da_0$,   we then have the upper bound,
\be
0\leq\Xi(\eta_\mu)< y_\mu+\Da_0\Ra \eta_\mu < \Xi^{-1}(y_\mu+\Da_0)\ .
\label{upper bound }
\ee
Additionally, if $y_{\mu}>\Da_0$, we also have a lower bound:
\be
0< y_\mu-\Da_0<\Xi(\eta_\mu)\Ra \eta_\mu > \Xi^{-1}(y_\mu-\Da_0)\ .
\ee 
Thus combining the upper and lower bounds, we find
\be
\Xi^{-1}(y_\mu+\Da_0)>\eta_\mu >\Xi^{-1}(y_\mu-\Da_0)>0
\ee
On the other hand, if $y_{\mu}\leq \Da_0$, then there is no lower bound, and any $\eta_\mu$ satisfying the upper bound  (\ref{upper bound }) is allowed. Now, we can introduce effective responses, representing the midpoint of possible super-threshold input drives, 
\be
\bar{y}_\mu\equiv\left\{
\begin{array}{cl} 
\2\LT\Xi^{-1}(y_\mu+\Da_0)+\Xi^{-1}(y_\mu-\Da_0)\RT&\mx{ if } y_{\mu}>\Da_0\\
\2\Xi^{-1}(y_\mu+\Da_0)&\mx{ if } y_{\mu}\leq\Da_0
\end{array}
\Rd\ ,
\ee
and effective noise limits,
\be
\bar{\en}_\mu\equiv\left\{
\begin{array}{cl} 
\2\LT\Xi^{-1}(y_\mu+\Da_0)-\Xi^{-1}(y_\mu-\Da_0)\RT&\mx{ if } y_{\mu}>\Da_0\\
\2\Xi^{-1}(y_\mu+\Da_0)&\mx{ if } y_{\mu}\leq\Da_0
\end{array}
\Rd\ ,
\ee
which allow $\eta_\mu$ to span the full allowed range. By inspection, we now see that the solution space is equivalent to the solution space of a threshold-linear problem:
\be
\Phi(\eta_\mu)=\bar{y}_{\mu}+\bar{\da}_{\mu},\with |\bar{\da}_{\mu}|\leq \bar{\en}_{\mu}.
\ee
Thus, again all the $y$-critical estimates (\ref{ycmaxM}, \ref{ycminM}, \ref{ycr-lead-en}) will be valid with the substitution  $y_{\mu}\ra\bar{y}_{\mu}$, and a conservative error bound
\be
\en^2= \sum_{\mu=1}^{\cP}\bar{\en}_{\mu}^2\ .
\ee
\vs
\renewcommand{\theequation}{E.\arabic{equation}}
\setcounter{equation}{0}
{\bf E. Certain synapses in low-dimensional recurrent networks with self-connections}
\vs
As discussed in Appendix A, when one moves from feedforward to recurrent neural networks with self-synapses, the input patterns  can no longer be considered independent from the target neuron responses. How then does one assess synapse certainty for driven neurons with self-synapses, such as $y_3$ in Fig.~\ref{fig:zeroerror}A, $y$ in Fig.~\ref{fig:RCN3}A, $y_1$ in Fig.~\ref{fig:RCN3}C, and $y_2$ in Fig.~\ref{fig:RCN3}C?

We begin by concretely analyzing neuron $y$ in Fig.~\ref{fig:RCN3}A, because this is the conceptually simplest example, and  fundamentally the mathematical analyses are the same for the other  examples. In particular, to test our formalism and analytical results using this neuron, we performed low-dimensional simulations where the $\cN\times\cN$ extended input pattern matrix was
\ba
& &\qquad\qquad x_1\qquad\qquad x_2\qquad\qquad y\non
Z&=&\LF\by{rrr}
-\sin\psi \cos\chi& \cos\psi \cos\chi&\sin\chi\\
\cos\psi        & \sin\psi&       0  \\
\sin\psi \sin\chi& -\cos\psi \sin\chi& \cos\chi
\ey\RF\ ,
\label{rec-input}
\ea
which is the same as $X$ in Eq. (\ref{extendedinputpatterns}), except that the role of $x_3$ is now played by $y$ itself.\footnote{In the context of Fig.~\ref{fig:zeroerror}A,  $y$, $x_1$, and $x_2$ can be identified with $y_3$, $y_2$, and $x_2$, respectively.} The third column of Eq. (\ref{rec-input}) corresponds to the responses of the driven neuron, but it also provides a self-input. The two input neuron responses are given by the first two columns. Eq. (\ref{rec-input}) is meant to correspond to the case where $\cP=2$ and $\cR=1$, such that $\mu=1,2,3$ correspond to the constrained, semi-constrained, and unconstrained response patterns, respectively. For the purpose of numerical testing, we assumed that $\chi\in (0^\circ,90^\circ)$ and $\psi\in (-90^\circ,0^\circ)$, as this range of angles ensures that driven neuron responses were non-negative.\footnote{This range also allowed us to ensure non-negativity for the other example recurrent circuits in Fig.~\ref{fig:zeroerror}A and Fig.~\ref{fig:RCN3}C.} We set $W=1$. See Appendix F for numerical simulation details.

This example problem has one  self-coupling, $w$, and two feedforward couplings, $u_1$ and $u_2$. For this response structure, one can use Eq. (\ref{ycritical}) to calculate $\ycr$ for each of these three couplings. We find
\ba
y_{\mt{cr},w}&=& W\cos\chi\label{w1cc}\non
y_{\mt{cr},u_1}&=& W\sqrt{\cos^2\psi+\cos^2\chi\sin^2\psi}\non
y_{\mt{cr},u_2}&=& W\sin\chi.
\label{rec-conds}
\ea
We assess synapse certainty by checking whether these formulae for $\ycr$ are smaller than the magnitude of $\vec{y}$,
\be
y=\sin\chi\ .
\ee
Since $W=1$, we see that the self-coupling becomes certain only if
\be
\sin\chi >\cos \chi\ , \mx{ or, }\  0^\circ<\ta\equiv\pi/2-\chi<45^\circ\ 
\label{rcn3}
\ee
(Fig.~\ref{fig:RCN3}B, \emph{left}), where $\ta$ has been defined to be the angle between $\eh_{yy}$ and $\ch=\yh$~\footnote{Note that
\be
\eh_{yy}=\sin\chi\env_1+\cos\chi\env_3=\cos\ta\env_1+\sin\ta\env_3\ ,
\ee
where we have  identified $\ch$ and $\uh$ with $\env_1$ and $\env_3$, respectively.}, consistent with the conventions of Eq. (\ref{Eq: SubspaceDecompEm}). Next, basic trigonometric manipulations tell us that the certainty condition can never be satisfied for $u_1$ (Fig.~\ref{fig:RCN3}B, \emph{middle}). Finally, we see that the condition for certainty is always {\it just} not satisfied for $u_2$. Here this implies that all solutions have $u_2\ge0$ (Fig.~\ref{fig:RCN3}B, \emph{right}). This is because it is a  very special case where $\cos\ga=e_{s\x}=0$  and  $\yh_\perp$ in (\ref{ei-decomposition}) is aligned with $\nh$, so that both the inequalities in (\ref{inequality}) turn into equalities. In each panel of Fig.~\ref{fig:RCN3}B, we tracked the fraction of positive and negative synapse signs across the simulations, as we varied $\ta$. In particular, we see that $w$ had a unique sign as long as $\ta<  45^\circ$, $u_1$ always had mixed signs, and $u_2$ was non-negative\footnote{As we explained before, $y_{\mt{cr},u_2}=y$ for all values of $\ta$. Hence the certainty condition is not satisfied because $u_2$ may be zero. The fact that $u_2$ can vanish is not discernible from our simulations because the weight magnitudes were generated randomly, and weights where $u_2 = 0$ comprise a zero measure set.}.

The same exact response matrix (\ref{rec-input}) can also be used to consider neurons $y_3$ in Fig.~\ref{fig:zeroerror}A and $y_2$ in Fig.~\ref{fig:RCN3}C. The only difference is that the three columns respectively encode: the responses of the second driven neuron, second input neuron, and third driven neuron in Fig.~\ref{fig:zeroerror}A; and the responses of the single input neuron and two driven neurons, $y_1$ and $y_2$, in Fig.~\ref{fig:RCN3}C. This correspondence can be seen by comparing the numerical results in Figs.~\ref{fig:RCN3}B and \ref{fig:RCN3}D. We use this correspondence to avoid having to simulate the full network in Fig.~\ref{fig:zeroerror}A, and the numerical results in Fig.~\ref{fig:zeroerror}C are the same as those in Fig.~\ref{fig:RCN3}B. 
\vs
\renewcommand{\theequation}{F.\arabic{equation}}
\setcounter{equation}{0}
{\bf F. Numerical Methods}
\vs
{\bf Low-dimensional numerical methods:} Here we detail the numerical methods relevant for Fig.~\ref{fig:zeroerror} and Fig.~\ref{fig:RCN3}.
\vs
{\it Feedforward analysis:} To test the analytic dependence in Fig.~\ref{fig:zeroerror}B, we wanted to simulate solutions without biasing ourselves by the particular search algorithm used to find solutions. Accordingly, to find solutions to the fixed point equations (\ref{eqn: SingleSteadyStateEqn}) with very small error ($\cE<\en=0.01$) we performed a random screen where each weight was chosen randomly from a uniform distribution between $-1$ and $+1$.  For feedforward circuits, given the synaptic weights, one can obtain the fixed point responses of the target neuron by direct substitution of the known input responses in (\ref{eqn: SingleSteadyStateEqn}) and then comparing these simulated target responses with the known target responses. We varied $\psi, \chi$ in the response data (\ref{rec-input}) systematically in steps of $6^\circ$~\footnote{Since here we were primarily interested in the zero-error result, we restricted ourselves to a range of $\psi,\chi$ where no topological transitions can occur due to the small but finite error we had to allow for numerical simulations.}. For the light and dark green curves, $\psi$ was fixed at $45^\circ$, and $\chi$ was varied between $(0^\circ,90^\circ)$ and $(90^\circ,180^\circ)$ respectively, while for the pink and purple curves $\chi$ was fixed at $45^\circ, 135^\circ$ respectively, and $\psi$ was varied between $(0^\circ,90^\circ)$. Finally, for a given choice of $\psi, \chi$, we systematically varied $y$ between 0 and 1, in intervals of $\Da y=0.01$. For each value of $\psi, \chi$, and $y$, we obtained  $\sim\cO(10^2-10^4)$ solutions\footnote{The number of solutions varied between 200 and 40,000 depending primarily on the value of $y$, the higher the value, typically the more difficult it was to find solutions.}  satisfying the error and the biological bound (\ref{sphere}) from five to ten million different trial weight vectors. We then identified the maximal value of $y$ for which the solutions had both positive and negative $w_1$'s. This simulation point should lie beneath the theoretical $\ycr$ if  no error is allowed. However, since the error is small but nonzero, occasionally the $y$-criticals determined from simulations did slightly exceed the theoretical value. Also, since we vary $y$ by small amounts $\Da y=0.01$, we expected the simulated $y$-criticals to be discrete but close to the theoretical predictions, which is exactly what we found in  Fig.~\ref{fig:zeroerror}B.    
\vs
{\it Recurrent analysis:} Because the recurrent network solution space separates into several feedforward solution spaces at zero error, we numerically treated the driven neurons one at a time. To find solutions for the recurrent neurons in Fig.~\ref{fig:zeroerror}A, Fig.~\ref{fig:RCN3}A, and Fig.~\ref{fig:RCN3}C, we  fixed $\chi,\psi$ and then performed screens with random weights, selected in the same manner as the feedforward simulations discussed earlier. For each set of weights, and for each $\mu=1,2$, we obtained the late time values of $y$ by solving the time evolution equation (Eq. (\ref{eqn: RateEqn}) with $\tau_i$ = 20ms) using Euler's method starting with initial conditions $y_i(0)=y_{\mu i}$, for $ \mu=1,2$. We used a time step of $\Da t =0.2$ ms. The $\w{y}_{\mu }$'s obtained from the simulation at late times, $t\sim 600$ ms, were then compared with $ y_{\mu }$ to obtain $\cE$. If the weights satisfied, $\cE<0.05\sqrt{\cD}$ and the biological bound (\ref{sphere}), then we considered the weights as solutions and checked the sign of the synaptic weights. For every value of $\psi, \chi$, we found at least  50 solutions\footnote{The number of solutions varied between $10^4$ and $10^5$ for the $\cD=1$ simulation in Fig.~\ref{fig:RCN3}B and between 56 and 110 for the $\cD=2$ simulation in Fig.~\ref{fig:RCN3}D. Note that for the $\cD=2$ simulation we have a six-dimensional weight-space, which makes it a lot harder to find solutions through random scanning. Also, for this latter case we only checked that the biological constraint is satisfied by the incoming weights to $y_1$.} to test the certainty predictions. 
\vs
{\bf High-dimensional numerical methods:} Here we detail the numerical methods relevant for Fig.~\ref{fig:highDsims} and Fig.~\ref{fig:errorcurves}.
\vs
{\it Generating random orthogonal matrices:} In several simulations we had to generate orthogonal response matrices. This meant that we had to obtain $\cP$ orthonormal $\cN$-dimensional vectors. This was done by first generating an $\cN\times \cN$ matrix, $G$, where each of its entries was randomly selected  from a uniform distribution between $-1$ and $+1$. We then antisymmetrized the matrix, $G\ra (G-G^T)/2$, and a random orthogonal response matrix was then obtained via matrix exponentiation, $Z=e^{G}$, where the matrix exponential is defined by substituting the matrix $G$ into the power series expansion of the exponential function and is distinct from the simple exponentiation of individual matrix elements. The first $\cP$ rows of $Z$ could then correspond to the $\cP$ orthogonal patterns, and $Z$ can be interpreted as the orthogonal extension of $z$ as discussed before. 
\vs
{\it Generating random orthogonal matrices with non-negativity constraints}. In recurrent networks, all driven neurons must have non-negative responses for all patterns. Accordingly, when the input response pattern includes responses of driven neurons, we follow a different procedure for generating the response matrix, which works as long as $\cI\geq \cP-1$. We started by choosing a $(\cP\times \cN)$-dimensional matrix, $z$, containing the responses of $\cD$ driven neurons and $\cI\geq \cP-1$ input neurons. The first columns of $z$ corresponded to driven neuron responses, and the last columns to input neuron responses, such that $z= \left(\begin{matrix}y & x\end{matrix}\right)$. We made sure that the responses of the driven neurons were all non-negative, as the threshold nonlinearity dictates, by choosing them to lie randomly between 0 and 1. To mimic a sparse response pattern, we  set  driven responses to 0 with 50\% probability. The feedforward inputs, on the other hand, were randomly selected between $-1$ and $1$. We then orthogonalized the input responses to the target neuron as follows. We start by normalizing the $\nu=1$ pattern:
\be
z_{1 m}\ra {z_{1 m}\over \sqrt{\sum_{n=1}^{\cN} z_{1 n}^2}}\ .
\ee
Then, for each row, $\nu=2\dots \cP$, in a sequential order we performed the following operations:  
\bi
\item We started by defining a ($\nu-1$)-dimensional square matrix, $x'$:
\be
x'_{\mu m}\equiv x_{\mu m}\for \mu,m=1\dots (\nu-1).
\ee
\item We next changed the first $m=1\dots \nu-1$ elements of the $\nu^{th}$ row of $x$
\be
x_{\nu m}\equiv-\sum_{\mu=1}^{\nu-1}x'^{-1}_{m\mu}\LF\sum_{i=1}^{\cD}y_{\mu i}y_{\nu i}+\sum_{n=\nu}^{\cI}x_{\mu n}x_{\nu n}\RF,
\ee
and thus $z$. The other elements of the $\nu^{th}$ row of $z$ were left unchanged. In particular, none of the driven neuron responses changed during this step. 
\item Finally, we rescaled all the elements of the  $\nu^{th}$ row of $z$ for normalization:
\be
z_{\nu m}\ra {z_{\nu m}\over \sqrt{\sum_{n=1}^{\cN} z_{\nu n}^2}}\ .
\ee
\ei
This algorithm essentially uses the responses of the input neurons to the $\nu$th stimulus to ensure that the full $\nu$th response pattern involving both the driven and input neurons is orthogonal to all $\mu\leq\nu-1$ patterns. 
\vs
{\it Generating target responses and response directions:} To generate  $\cP$ target responses with $\cS$ null responses, we simply randomly selected numbers between 0 and 1 for the $\cR=\cP-\cS$ nonzero responses. 

In some simulations, we wanted to consider situations where one has to account for a single  topological transition to compute $y$-critical. Accordingly, we tailored the responses as follows. First, we set one nonzero response of the target neuron  to a small value,  $0.1\en$. $\yh$ was then obtained by dividing the response vector by its norm. We then only considered those $\yh$'s whose other entries were large enough to prevent additional topological transitions from affecting $y$-critical. This was done by: evaluating $y_{\mt{cr},\min}$, the theoretical lower bound for $y$-critical that includes the first topological transition (Appendix C); constructing $\vec{y} = y_{\mt{cr},\min} \hat y$, which approximates the activity vector right below $y$-critical; and ensuring that all the other entries of $\vec{y}$ were greater than the allowed error, which guarantees that no other constrained dimensions can become semi-constrained in between $y_{\mt{cr},\min}$ and the true $y$-critical. This way, typically one and only one constrained direction became semi-constrained when we allowed solutions with errors $\lesssim \en$. 
\vs
{\it Finding solutions using gradient descent learning in feedforward networks:} In all the high-dimensional simulations, we had to find solutions to the fixed point equations (\ref{eqn: SingleSteadyStateEqn}). Since scanning a high-dimensional synaptic weight-space randomly is not numerically efficient, we applied gradient descent learning\footnote{Typically with learning rate $\sim 0.01$.} to obtain   solutions. For feedforward networks, this meant using the loss function
\ba
\cE^2= \sum_{\mu=1}^{\cP}\LF \w{y}_{\mu }- y_{\mu }\RF^2, \where \ 
\w{y}_{\mu}= \Phi\LF\sum_{m=1}^{\cN} z_{\mu m} w_{m}\RF\ .
\ea
We performed gradient descent optimization until we reached the desired error bound, $\cE<\en$. The  initial weights were first chosen randomly from a uniform distribution between $-1$ and $1$. The initial weight vector was then rescaled to have a norm between 0 and $W=1$, chosen uniformly.
\vs
{\it Finding solutions using gradient descent learning in recurrent networks:}
To find solutions for recurrent neural networks, we used the modified loss function,
\ba
\bar{\cE}^2\equiv\sum_{i=1}^{\cD} \sum_{\mu=1}^{\cP}\LF \w{y}_{\mu i }- y_{\mu i}\RF^2 \equiv \sum_{i=1}^{\cD}\bar{\cE}_i^2\ ,\where \ 
\w{y}_{\mu i}= \Phi\LF\sum_{m=1}^{\cN} z_{\mu m} w_{im}\RF\ ,
\ea
to perform gradient descent, instead of (\ref{error}). Since the responses of the driven neurons can vary for nonzero errors, the two loss functions, $\cE$ and $\bar{\cE}$, differ. However, it is numerically a lot quicker to obtain solutions via gradient descent with $\bar{\cE}$ as compared to using back-propagation through time to consider the entire time evolution of the network. Thus, the strategy we adopted to find solutions with $\cE\lesssim\en$ was to first find weights satisfying $\bar{\cE}\lesssim\bar{\en}=\en/10$. Also, the gradient descent was done in two stages. In the first stage we minimized the error associated with each individual driven neuron, $\bar{\cE}_i$ treating it as a feedforward problem. Once each of these errors were less than $\bar{\en}$, we performed a second stage of gradient descent to minimize $\bar{\cE}$ down to $\bar{\en}$. Next, we obtained the late time values of $y_i$'s by solving the time evolution equations (\ref{eqn: RateEqn}) with $\tau_i$ = 20 ms) using Euler's method with step time $\Da t=0.2$ ms for the weights obtained via gradient descent and starting with initial conditions $y_i(0)=y_{\mu i}$, $\forall \mu, i$. The $\w{y}_{\mu i}$'s obtained at late times, $t\sim 600$ ms, this way were compared with $  y_{\mu i}$ to obtain $\cE$. Finally, we checked that the weights satisfied the biological bound\footnote{We only checked that the target weights satisfied the weight bound, as that's what matters for the certainty conditions. Since we initialized weights amongst the non-target neurons to be between -7 and 7, it's likely that other components of the weight matrix were large.} (\ref{sphere}).
\vs
{\it Other minor simulation details:} In Fig.~\ref{fig:highDsims}B, we show results from a simulation with $\cN=6,\ \cP=5$, $\cR=2$, and $\cE<\en=0.1$. The solutions were obtained using a gradient descent learning rate of 0.02. We  varied the norm of the target response vector, $\vec{y}=y \yh$, systematically by $\Da y=0.01$ in a manner similar to the low dimensional simulations.

In Fig.~\ref{fig:highDsims}C, we considered a single input-output configuration for a given value of $\cN$  and $\cP$, and we found a single solution with $\cE < 0.001\sqrt{\cP}$ using a gradient descent learning rate of 0.005. 

In Figs.~\ref{fig:highDsims}D-E, we show results of a simulation for a $\cN=10, \cD=4, \cI=7, \cP=8$, and $\cR=3$ network where the norm of $\vec y$ was fixed to 0.79, which approximates the median value of $y$ for the given values of $\cN, \cP$ and $\cR$ (\ref{avg-median}). Our solutions were obtained using a gradient descent learning rate of 0.004 and the  overall error satisfied $0.017\sqrt{\cD}\lesssim\cE\lesssim 0.25\sqrt{\cD}$. 
\vs
\nopagebreak
\twocolumngrid

\end{document}